\documentclass[]{tandf2}

\usepackage{epstopdf}
\usepackage{subfigure}
\usepackage{amsmath}

\usepackage{physics}
\usepackage{cancel}

\usepackage[numbers,sort&compress]{natbib}
\bibpunct[, ]{[}{]}{,}{n}{,}{,}
\renewcommand\bibfont{\fontsize{10}{12}\selectfont}
\makeatletter
\def\NAT@def@citea{\def\@citea{\NAT@separator}}
\makeatother

\theoremstyle{plain}

\theoremstyle{definition}

\theoremstyle{remark}

\usepackage{amsmath}
\usepackage{bm}
\usepackage{mathrsfs}
\usepackage{color}
\definecolor{darkblue}{rgb}{0,0,0.6}
\definecolor{darkred}{rgb}{0.6,0,0}
\definecolor{darkgrey}{rgb}{0.6,0.6,0.6}

\usepackage[colorlinks=true,urlcolor=darkblue,citecolor=darkblue,linkcolor=darkred,hyperfootnotes=false]{hyperref}

\usepackage{ulem}
\usepackage{nameref}
\usepackage{stmaryrd}
\usepackage{cleveref}

\newcommand{\overbar}[1]{\mkern 1.5mu\overline{\mkern-1.5mu#1\mkern-1.5mu}\mkern 1.5mu}

\newcommand{\B}{B}           
\newcommand{\K}{\mathbb K}   

\newcommand{\beeta}{\boldsymbol{\eta}}

\newcommand{\ee}{\text{e}}

\newcommand{\p}{\partial}
\newcommand{\bx}{\text{\bf x}}

\newcommand{\bff}{\text{\bf f}}

\newcommand{\bG}{\text{\bf G}}
\newcommand{\bg}{\text{\bf g}}

\newcommand{\bu}{\text{\bf u}}
\newcommand{\bU}{\text{\bf U}}

\newcommand{\bF}{\text{\bf F}}

\definecolor{mycyan}{rgb}{.0,.5,.1}

\newcommand{\dW}{{\text{dW}}}

\newcommand{\tf}{t_{\text{f}}}
\newcommand{\tR}{t^{\text{R}}}
\newcommand{\vf}{v_{\text{f}}}
\newcommand{\Kf}{\mathcal K_{\text{f}}}
\newcommand{\xf}{x_{\text{f}}}
\newcommand{\bxf}{\bx_{\text{f}}}
\newcommand{\buf}{\bu_{\text{f}}}
\newcommand{\cl}{{\text{cl}}}

\newcommand{\betag}{\beta_g}

\usepackage[percent]{overpic}

\newcommand{\sms}{\kern.75pt}  
\newcommand{\sns}{\kern-.75pt}  

\begin{document}

\articletype{REVIEW}

\title{Path integrals and stochastic calculus}

\author{
\name{Thibaut~Arnoulx de Pirey\textsuperscript{a}$^{\ast}$\thanks{$^\ast$Corresponding author. Email: t.depirey@campus.technion.ac.il}, 
Leticia F.~Cugliandolo\textsuperscript{b,c}, \\
Vivien Lecomte\textsuperscript{d} and Frédéric~van~Wijland\textsuperscript{e}
}
\affil{
\textsuperscript{a}Department of Physics, Technion-Israel Institute of Technology, Haifa 32000, Israel
\\
\textsuperscript{b}Laboratoire de Physique Th\'eorique et Hautes \'Energies, Sorbonne Universit\'e \& \\
CNRS (UMR 7589), 4 place Jussieu, 75005 Paris, France
\\
\textsuperscript{c}Institut Universitaire de France, 1 rue Descartes, 75231  Paris Cedex 05 France
\\
\textsuperscript{d}Laboratoire Interdisciplinaire de Physique, Universit\'e Grenoble Alpes \& CNRS (UMR 5588), \\
140 avenue de la Physique, 38402 Saint-Martin d'H\`eres, France
\\
\textsuperscript{e}Laboratoire Mati\`ere et Syst\`emes Complexes, Universit\'e Paris Cité \& CNRS (UMR 7057), \\
10 rue Alice Domon et L\'eonie Duquet, 75205  Paris Cedex 13, France
}
}

\maketitle

\begin{abstract}
Path integrals are a ubiquitous tool in theoretical physics. However, their use is sometimes hindered by the lack of control 
on various manipulations---such as performing a change of the integration path---one would like to carry out in the light-hearted fashion that physicists enjoy. Similar issues arise in the field of stochastic calculus, which we review to prepare the ground for a proper construction of 
path integrals. At the level of path integration, and in arbitrary space dimension, we not only report on existing Riemannian geometry-based approaches that render path integrals amenable to the standard rules of calculus, but also bring forth new routes, based on a fully time-discretized approach, that achieve the same goal. We illustrate these various definitions of path integration on simple examples such as the diffusion of a particle on a sphere.  
\end{abstract}

\newpage

\tableofcontents

\newpage

\section{An introduction to the difficulties of manipulating path integrals}
\label{sec:intro}

The Markovian Langevin equation is successfully used to model stochastic phenomena, 
but its very definition comes with a quandary: it is a differential equation that describes the evolution of a non-differentiable process.
Stochastic calculus was built not only to make sense of such seemingly ill-defined processes,
but also to provide a set of rules that allow one to manipulate them (almost) as if they were differentiable functions. 
The implications of this well-established framework, such as the use of a (modified) chain rule when changing variables, are well known in theoretical physics;
what is less known, however, is that the path-integral representation of the probability of 
{such} Langevin processes suffers from even deeper pitfalls, which the usual stochastic calculus cannot cure.
This introduction aims at describing on simple illustrative examples the questions we address in this article, before reviewing and extending solutions that were proposed to solve them.

\subsection{Langevin equations with multiplicative noise}

Take a Langevin equation for a one-dimensional stochastic  process $x(t)$ with multiplicative noise, whose generic form is
\begin{equation}\label{eq:Langevindepart}
\frac{\dd x(t)}{\dd t}=f(x(t))+g(x(t)){\eta}(t)
\end{equation}
where $\eta$ is a Gaussian white noise with zero mean and correlations $\langle \eta(t)\eta(t')\rangle=\delta(t-t')$. 
Here, $f(x)$ represents a deterministic force (or `drift') and $g(x)$ the amplitude of the noise to which the process is subjected.
For example, $x(t)$ can be the position of a particle in a medium where temperature or friction depend 
on space through $g(x)$~\cite{lancon_drift_2001,lancon_brownian_2002}
or it could be a heterogeneous diffusion process~\cite{cherstvy_anomalous_2013,cherstvy_population_2013}.
In chemistry, $x(t)$ can describe the concentration of a chemical species subjected to population noise that goes to zero with  concentration itself~\cite{gillespie_chemical_2000}.
In biology, Eq.\ (\ref{eq:Langevindepart}) can model the stochastic level of expression of a gene subject to a noise depending on the level itself~\cite{kaern_stochasticity_2005},
or the population in ecosystems described by stochastic generalized Lotka--Volterra models~\cite{giuli_dynamical_2022}.
In finance, $x(t)$ can represent a stock price whose volatility depends on the stochastic value itself~\cite{hamao_correlations_1990}.
At larger scales, in cosmology, the inflaton field in  stochastic models for inflation follows an evolution equation of the form of 
Eq.~(\ref{eq:Langevindepart})
where the multiplicative noise arises from quantum fluctuations through a coarse-graining procedure~\cite{matacz_new_1997}.

Despite such a widespread use,
it is a well-known feature of equations of the form~(\ref{eq:Langevindepart}) that they must be considered with great care, as the process $x(t)$ is not differentiable~\cite{oksendal_stochastic_2013}. 
One way to endow Eq.~\eqref{eq:Langevindepart} with a well-defined mathematical meaning is to interpret it as an equation for the infinitesimal increment of $x$ between $t$ and $t+\Delta t$, for $\Delta t\to 0$,
\begin{equation}\label{eq:Langevindiscretalpha}
x(t+\Delta t)-x(t)=\Delta x=f(x(t)+\alpha \Delta x)\,\Delta t+g(x(t)+\alpha \Delta x)\,\Delta\eta
\end{equation}
with $0\leq \alpha\leq 1$ and where $\Delta\eta$ is a zero-mean Gaussian variable of variance $\Delta t$. This is called the $\alpha$-discretization scheme.
The $\alpha=0$ scheme is known as the Itō one, 
for $\alpha=1/2$ one defines a Stratonovich process, and with $\alpha=1$ the discretization is named after Hänggi--Klimontovich~\cite{Hanggi80,Hanggi78,Hanggi1982,Klimontovich}. 
Importantly, for fixed functions $f(x)$ and $g(x)$, each value of $\alpha$ generates a different process~\cite{gardiner_handbook_1994,kampen_stochastic_2007}.
The most direct way to understand this is to remark that an  $\alpha$-discretized  Langevin equation is equivalent to  
an $\alpha'$-discretized one with a different expression of the deterministic 
drift:\footnote{This is a classical fact of stochastic calculus which, for completeness, is explained in Sec.~\ref{subsubsec:equivalence}.
The equivalence in Eq.\ (\ref{eq:chaangealphaLa}) means that the distribution of the two processes is the same at all times --~when starting from the same initial condition.}
\begin{equation}
\label{eq:chaangealphaLa}
\frac{\dd x}{\dd t}\stackrel{\alpha}{=}f(x)+g(x)\eta\,\iff\,\frac{\dd x}{\dd t}\stackrel{\alpha'}{=}f(x)+(\alpha-\alpha')g'(x) g(x)+g(x)\eta
\end{equation}
where the  $\stackrel{\alpha}{=}$ symbol means that the continuous-time equation must be understood according to the $\alpha$-discretization of Eq.~\eqref{eq:Langevindiscretalpha}
(and where, from now on, we do not make the time dependencies of the processes explicit).
The conclusions drawn from this observation are simple:
\begin{itemize}
\item[(\textit{i})] A multiplicative Langevin equation of the form of Eq.\ (\ref{eq:Langevindepart}) must be endowed with a discretization scheme to present an unambiguous meaning.
%
\item[(\textit{ii})] There is no good or bad choice of discretization since one can always switch from one to another (at the price of changing the form of its drift) while still describing the \textit{same} process.
%
\item[(\textit{iii})] Although $x(t)$ is non-differentiable, a single parameter $\alpha$ lifts the ambiguity in the definition of the stochastic differential equation\ (\ref{eq:Langevindepart}) through the discretization rule~(\ref{eq:Langevindiscretalpha}).
%
\item[(\textit{iv})] Last and importantly, to claim that a phenomenon is well modelled by the multiplicative Langevin equation\ (\ref{eq:Langevindepart}), one needs more than simply asserting that the force is $f(x)$
and the noise amplitude is $g(x)$: one needs a procedure that \textit{provides} the discretization of Eq.\ (\ref{eq:Langevindepart}).
\end{itemize}

Equations like Eq.~\eqref{eq:Langevindepart} usually appear, in physics, after some coarse-graining procedure consisting in integrating out degrees of freedom of no direct interest~\cite{zwanzig_nonequilibrium_2001} (see App.~\ref{app:oscillators}). They also require the existence of a separation of 
time scales between the degree of freedom of interest and the surrounding environment. 
The Markov approximation, according to which the relaxation of the environment occurs over time scales much shorter than that of the degree of freedom of interest $x(t)$, is responsible for the noise $\eta(t)$ being $\delta$-correlated and, therefore, for the process $x(t)$ not being differentiable. When the Markov limit is carefully taken in an equation of the form of Eq.~\eqref{eq:Langevindepart} with a correlated noise, the increment of $x$ between $t$ and $t+\Delta t$ is shown to be given by the Stratonovich $\alpha=1/2$ scheme. 

Such physical approximations (coarse-graining and the Markov limit) need not be implemented at the level of the equations of motion. They can instead be applied to, say, a Liouville equation. In the Markov approximation, this results in a master equation, which, in the diffusive limit, is known as the Fokker--Planck (or as the Kolmogorov forward equation or the Smoluchowski equation). In the Fokker-Planck framework, instead of tracking individual fluctuating trajectories generated by Eq.~\eqref{eq:Langevindepart}, one focuses on the probability density $P(x,t)$ of the random process $x(t)$ and arrives at an equation of the form
\begin{equation}
\p_t P(x,t)=-\p_x(f(x) P(x,t))+\frac 12 \, \p_x^2\big(g^2(x) P(x,t) \big)
\end{equation}
which describes the same process as the one evolving according to Eq.~\eqref{eq:Langevindepart} understood in the Itō sense with $\alpha=0$. 
(The Fokker--Planck equation for generic $\alpha$ is derived in App.~\ref{app:FP}.)
Within the quantum mechanical setting in which randomness is intrinsic, \textit{i.e.}~not resulting from a loss of information, probability amplitudes are obtained from the Schrödinger equation. The latter, for a particle in a potential, also takes the form of a linear first-order in time, second-order in space, partial differential equation. This formal resemblance explains that tools developed in stochastic processes can be useful in quantum mechanics, and \textit{vice versa}. Interestingly, there have even been attempts to cast quantum mechanics within the Langevin language~\cite{grabert1979quantum, nelson_quantum_1985}.\\

In this review, we will not discuss the procedure to be followed to arrive at a well-defined Langevin equation (as it depends on the system at hand, see for instance~\cite{sokolov2010ito} for a discussion in the case of diffusion in a disordered medium, or~\cite{hanggi_nonlinear_1982} on the subject of characterizing the noise around the deterministic limits obtained from Markov processes).
We assume instead that this wearying work was already done by the reader and we will start directly from a multiplicative Langevin equation with a known discretization.
Our interest goes to the \textit{methodological} advantage of using a discretization scheme rather than another.
A simple situation that illustrates why this question matters is that of a change of variable. 
Given a smooth invertible function $U$, the process $u(t)=U(x(t))$ also evolves according to a Langevin equation
(see Sec.~\ref{subsubsec:differentiation}). However, it is only when Eq.~\eqref{eq:Langevindepart} 
is understood with the Stratonovich scheme that one can use the usual chain rule of differential calculus
to transform the Langevin equation on $x$ into a Langevin equation on $u$.
Namely, in the Stratonovich scheme the usual chain rule reads
\begin{equation}
\frac{\dd u(t)}{\dd t}=U'(x(t))\frac{\dd x(t)}{\dd t} 
\, , 
\end{equation} 
where the prime represents derivative with respect to the argument, 
and then
\begin{equation}\label{eq:Langevinstratochain}
\frac{\dd x}{\dd t}\stackrel{\frac 12}{=}f(x)+g(x)\eta\, \;\;\; \implies\, \;\;\; \frac{\dd u}{\dd t}\stackrel{\frac 12}{=}F(u)+G(u)\eta
\,,
\end{equation}
where $F(u)$ and $G(u)$ are defined from simple ``covariant'' 
relations $F(U(x))=U'(x) f(x)$ and $G(U(x))=U'(x) g(x)$.
%
%
 Instead, when Eq.~\eqref{eq:Langevindepart} is understood in the Itō scheme, one has to use the celebrated Itō formula~\cite{ito_stochastic_1944},
\begin{equation}\label{eq:LangevinItochain}
\frac{\dd x}{\dd t}\stackrel{0}{=}f(x)+g(x)\eta\,\;\;\; \implies\, \;\;\; \frac{\dd u}{\dd t}\stackrel{0}{=}U' \, \frac{\dd x}{\dd t}+\frac 12 \, U'' g^2\,,
\end{equation}
meaning that the drift of the Itō Langevin equation for $u(t)$ is now equal to $U'(x) f(x) + \frac 12 \, U''(x) g(x)^2$ (with $x=U^{-1}(u)$)
instead of being covariant as in the Stratonovich case.
The lesson we learn is that the discretization scheme affects the rules of computation when changing variables, and that a special choice, the Stratonovich one,
guarantees a form of covariance which allows one to manipulate the Langevin equation as if $x(t)$ were differentiable.
\\

This brings us to the topic of this work. There is a third description of random processes based on path integrals where the fundamental object is the probability distribution over random trajectories. Originally, Wiener~\cite{wiener_differential-space_1923,wiener_average_1924} 
built them to analyze the properties of Brownian motion, but they became a central tool of theoretical physics after 
Feynman~\cite{feynman_space-time_1948} reformulated quantum mechanics in terms of path integrals. 
Following  the work of Onsager and Machlup~\cite{onsager_fluctuations_1953,machlup_fluctuations_1953II}, they also became a cornerstone in the study of classical irreversible processes.
Manifold analytical calculations are easier to set within the  
path-integral representation of stochastic processes. A few examples 
are: perturbative expansions, instanton calculations used to evaluate escape times~\cite{Caroli81}, the identification 
and analysis of dynamic symmetries leading, for example, to fluctuation theorems~\cite{AronLeticia2010,seifert2012stochastic},  
or the derivation of mean-field dynamic treatments like the ones used to study 
glassy dynamics~\cite{Cugliandolo-Houches}, for instance. 
 
 In much the same way as with Langevin equations, in path integrals one manipulates non-differentiable trajectories, and this comes with its share of mathematical difficulties, as raised by Edwards and Gulyaev~\cite{edwards_path_1964} in 1964 
(see \textit{e.g.}~Refs.~\cite{gervais_point_1976,Sa77,deininghaus_nonlinear_1979,AlDa90,ApOr96,aron_dynamical_2016,Cugliandolo-Lecomte17a} for later discussions on such issues
and~\cite{cartier_functional_2006} for a mathematical review of the subtleties of functional integration). 
These are the ones we would like to examine now.
As will soon become apparent, path integrals are in fact more sensitive to discretization issues than Langevin equations 
--~and this in spite of the fact that the white noise has been integrated over and does not appear explicitly in
the path-integral action.
We begin by illustrating such difficulties on an example.
For simplicity, we will now use standard expressions of path integrals, say for the probability of a time realization of a Langevin process, and 
if the reader has not already encountered these, 
we actually derive them 
in later sections (see also Refs.~\cite{zinn-justin_quantum_2002,chaichian_stochastic_2001,kleinert_path_2009} for reviews).
 
\subsection{A Brownian particle}
\label{sec:brownian-particle}

Consider a large particle of mass $m$ and velocity $v$ in water, whose motion is modeled by the Langevin equation
\begin{equation}\label{eq:langevin_colloid}
m\frac{\dd v}{\dd t}=-\gamma v+\sqrt{2\gamma T}\eta
\, , \qquad\qquad 
\langle\eta(t)\eta(t')\rangle=\delta(t-t')
\, ,
\end{equation}
where $\gamma$ is the friction coefficient, $T$ is the temperature of the water bath 
and the Boltzmann constant is set to $k_{\sms\text{B}\!}=1$. The noise $\eta$ is again white, Gaussian and with zero mean.
For simplicity, we restrict here to the one-dimensional case. 
Starting from an initial condition in which 
$v(0)=0$, the probability of observing a velocity $\vf$ at time $\tf$ can be obtained from a summation over all velocity trajectories going from $v(0)=0$ to $v(\tf)=\vf$:
\begin{equation}\label{eq:PItrivialIto}
{\mathbb P}(\vf,\tf|0,0)
=
\int_{v(0)=0}^{v(\tf)=\vf}{\mathscr D} v \; \exp\left[-\frac{1}{4\gamma T}\int_0^{\tf}\dd t\left(m\frac{\dd v}{\dd t}+\gamma v\right)^2\right].
\end{equation}
This expression comes from the noise distribution being $\propto \text{exp}\: \big[-\frac 12 \int_0^{\tf} \dd t\, \eta^2\big]$ (encoding that $\eta$ is Gaussian and white)
and from remarking that\ (\ref{eq:langevin_colloid}) implies $\eta = \big(m \dd v/\dd t + \gamma v\big)/\sqrt{2\gamma T}$.
As for Langevin equations, expressions such as Eq.~\eqref{eq:PItrivialIto} acquire  an unequivocal meaning when a discretization scheme is provided
(see Sec.~\ref{subsec:pi_additive} for a complete derivation).
 Here we must understand Eq.~\eqref{eq:PItrivialIto} as the $N\to \infty$ limit of
\begin{equation}
\prod_{k=1}^{N-1}\left(\frac{m}{\sqrt{4\pi\gamma T\Delta t}}\right)\dd v_k\to {\mathscr D} v
\label{eq:measure-v}
\end{equation}
with $\Delta t = \tf/N$ a small time interval and $k=1, \dots, N$ labelling discrete time steps
along with a time-discretized action
\begin{equation}
\label{eq:actionitoexplicit}
\Delta t\sum_{k=0}^{{N}-1}\left(m\frac{v_{k+1}-v_{k}}{\Delta t}+\gamma v_{k}\right)^2\to \int_0^{\tf}\dd t\left(m\frac{\dd v}{\dd t}+\gamma v\right)^2
\! , 
\end{equation}
$v_0=0$ and $v_{{N}}=v_{\rm f}$. These discretized expressions are the direct analogs of the Itō discretized form of the Langevin equation.
Similarly, as we later show in Eq.~\eqref{eq:action_additive_alpha} for a generic one-dimensional additive process, other schemes could be used to discretize the action,  such as the Stratonovich one, leading to
\begin{equation}\label{eq:PItrivialStrato}
{\mathbb P}(\vf,\tf|0,0)=\int_{v(0)=0}^{v(\tf)=\vf}{\mathscr D} v \; \exp\left[-\frac{1}{4\gamma T}\int_0^{\tf}\dd t\left(m\frac{\dd v}{\dd t}+\gamma v\right)^2+\frac{\gamma}{2m} \, \tf\right]
\end{equation}
with the same measure $\mathscr D v$ as in Eq.~(\ref{eq:PItrivialIto}), but now
\begin{equation}
\label{eq:actionstratoexplicit}
\Delta t\sum_{k=0}^{{N}-1}\left(m\frac{v_{k+1}-v_{k}}{\Delta t}+\gamma \frac{v_{k}+v_{k+1}}{2}\right)^2\to \int_0^{\tf}\dd t\left(m\frac{\dd v}{\dd t}+\gamma v\right)^2.
\end{equation}
We stress that the two expressions~(\ref{eq:PItrivialIto}) and~(\ref{eq:PItrivialStrato}) of the action (together with their corresponding discretizations (\ref{eq:actionitoexplicit}) and (\ref{eq:actionstratoexplicit}))
describe the \textit{same} process defined by Eq.~(\ref{eq:langevin_colloid}). Accordingly, any $\alpha$-discretization scheme could be used to build an equivalent path-integral representation of the stochastic process defined by Eq.~(\ref{eq:langevin_colloid}) (see for instance~\cite{itami_universal_2017}). To each parameter $\alpha$ is associated a different continuous-time
expression of the action, but each of these describes the same process. We will cover this extensively in Sec.~\ref{sec:pathint}. Also, we remark that while the discretization scheme is unimportant for Langevin processes with additive noise such as $v$, it plays a manifest role in the corresponding path-integral action, as can be seen from the difference between the expressions of Eqs.~\eqref{eq:PItrivialIto} and~\eqref{eq:PItrivialStrato}. 
\\

\begin{figure}
\centerline{
\includegraphics[width=.75\columnwidth]{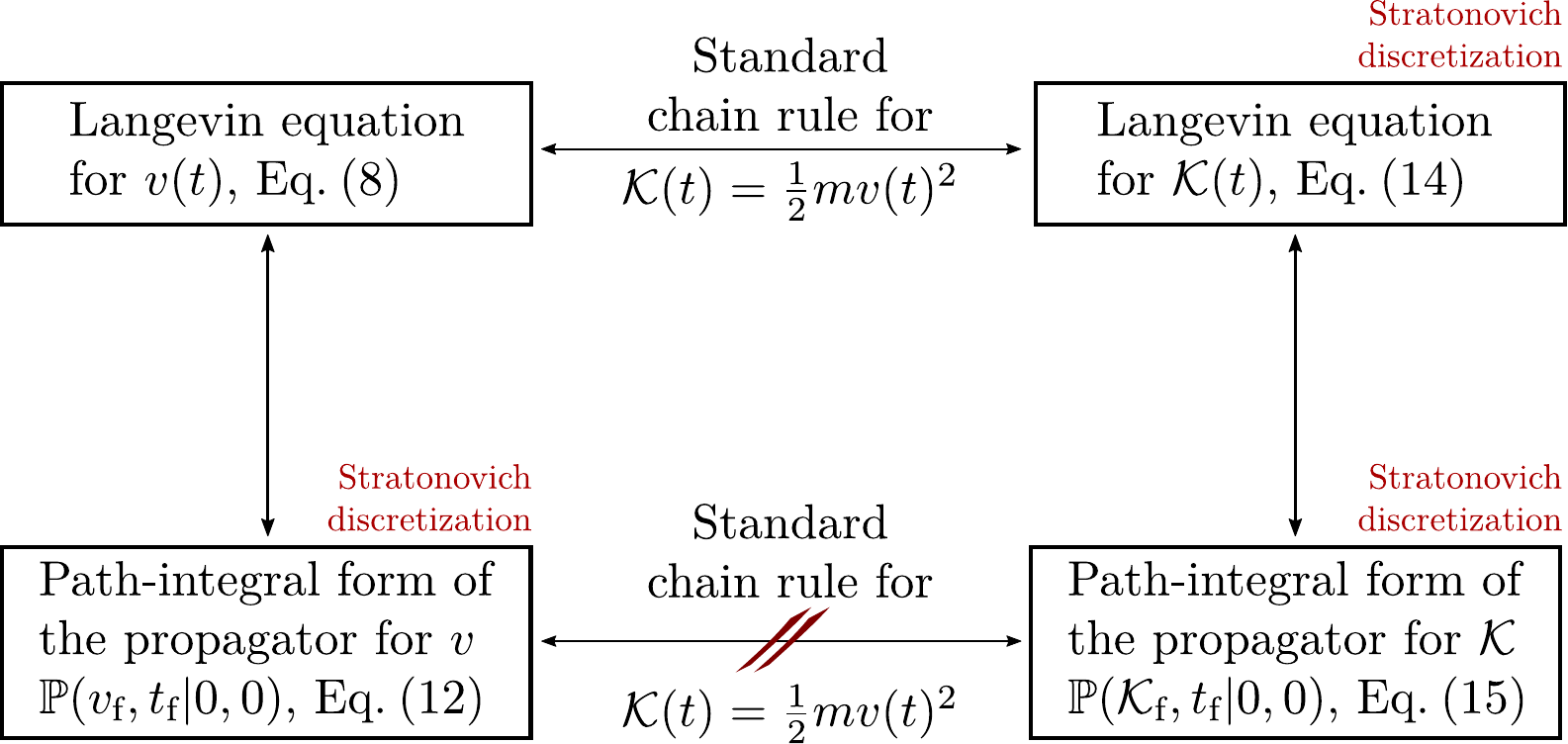}
}
\caption{%
  Schematic representation of the inapplicability of the standard chain rule in the path-integral representation of the propagator of Langevin processes,
  for the example studied in Sec.~\ref{sec:brownian-particle} 
  --~and this even though the Stratonovich discretization is applied (or any other linear $\alpha$-discretization scheme such as that of Eq.~(\ref{eq:Langevindiscretalpha})).
}\label{fig:scheme-example}
\end{figure}

We now ask about the statistics of the kinetic energy ${\mathcal K}=\frac{m}{2} v^2$ of the particle. 
At the level of Langevin equations, the rules of stochastic calculus allow us to deduce multiplicative noise 
Langevin equations for ${\mathcal K}$ in the Stratonovich discretizations:
\begin{align}
\label{eq:langevin_kineticS}
\frac{\dd {\mathcal K}}{\dd t}\stackrel{\frac 12}{=}&-\frac{2\gamma}{m}{\mathcal K}+\sqrt{\frac{4\gamma T {\mathcal K}}{m}} \, \eta
\; . 
\end{align}
Note that even though the transformation from $v$ to ${\mathcal K}$ is not invertible, it is possible to obtain a Langevin equation for the kinetic energy $\mathcal{K}$ because of the statistical invariance of the process in Eq.~\eqref{eq:langevin_colloid} under $\eta(t) \to -\eta(t)$ at any time step (which would not hold, \textit{e.g.} for a time-correlated noise). Regarding the corresponding path-integral formulation, the  kinetic energy
probability density at time $\tf$ knowing that ${\mathcal K}=0$ at 
time $t=0$, which we denote ${\mathbb P}({\Kf},\tf|0,0)$, reads, in the Stratonovich scheme
\begin{eqnarray}
\label{eq:PItrivialStratoK}
&& 
\!\!\!\!\!\!  
{\mathbb P}({\Kf},\tf|0,0)
\nonumber\\
&& 
\!\!\!\!\!\!   
= \!
\int_{\mathcal K(0)=0}^{\mathcal K(\tf)=\Kf} \!\!\!\!\! {\mathscr D}{\mathcal K}
\exp\left\{
-\frac{m}{8\gamma T}
\int_0^{\tf} \!\! \dd t \left[\frac{1}{ {\mathcal K}}\left(\frac{\dd {\mathcal K}}{\dd  t}+\frac{2\gamma}{m}{\mathcal K}\right)^2 \!\! +\frac{2\gamma T}{m{\mathcal K}}\frac{\dd {\mathcal K}}{\dd t}{+}\frac{\gamma^2 T^2}{m^2 {\mathcal K}}\right] \! +\frac{\gamma \tf}{2m}
\right\}
\qquad
\end{eqnarray}
where 
\begin{align}
& 
\prod_{k=1}^{{N}-1}\sqrt{\frac{m}{4\pi\gamma T \Delta t ({\mathcal K}_k+{\mathcal K}_{k+1})}}\,\dd {\mathcal K}_k
\to {\mathscr D} {\mathcal K}
\nonumber
\intertext{and}
& 
\Delta t\sum_{k=0}^{{N}-1}\left[\frac{2}{{\mathcal K}_k+{\mathcal K}_{k+1}}\left(\frac{{\mathcal K}_{k+1}-{\mathcal K}_{k}}{\Delta t}+\frac{\gamma}{m} ({\mathcal K}_{k}+{\mathcal K}_{k+1})\right)^2 
\right.
\nonumber\\
& 
\qquad 
\left.
\qquad\qquad 
+\frac{4\gamma T}{m({\mathcal K}_k+{\mathcal K}_{k+1})}\frac{{\mathcal K}_{k+1}-{\mathcal K}_k}{\Delta t}{+}\frac{2\gamma^2 T^2}{m^2({\mathcal K}_k+{\mathcal K}_{k+1})}
\right]
\nonumber\\[5pt]
&
\qquad\quad 
\to \int_0^{\tf}\dd t\left[\frac{1}{ {\mathcal K}}\left(\frac{\dd {\mathcal K}}{\dd  t}+\frac{2\gamma}{m}{\mathcal K}\right)^2 +\frac{2\gamma T}{m{\mathcal K}}\frac{\dd {\mathcal K}}{\dd t}{+}\frac{\gamma^2 T^2}{m^2 {\mathcal K}}\right]
.
\end{align}
At the Langevin level, the Stratonovich discretization is consistent with differential calculus and switching from $v$ to ${\mathcal K}$ can be done as if these  functions were differentiable. However, naively changing variables 
from $v$ to ${\mathcal K}$ starting from the path-integral probability~\eqref{eq:PItrivialStrato} for $v$ would not lead to the correct path-integral expression~\eqref{eq:PItrivialStratoK} for $\mathcal K$ (the last two terms in the time integral would be absent). This is illustrated in Fig.~\ref{fig:scheme-example}.

This simple example allows us to phrase the questions of interest throughout this work: 
\begin{itemize}
\item[(\textit{i})]
Starting from an action in the Itō (or Stratonovich, or any $\alpha$-discretized) form: Can we extend the modified chain rule of stochastic calculus to path-integral calculus (without using Langevin equations as intermediate steps)?
\item[(\textit{ii})]
Can one build a discretization scheme that allows one to deal with functions in path integrals as if these were differentiable (as is achieved by the Stratonovich discretization for Langevin equations)?
\end{itemize}
These are really the two sides of the same coin: either one sticks to a given discretization and then the rules of differential calculus have to be adapted, or one imposes differential calculus to hold, but this requires finding the appropriate discretization schemes. Such questions have already been addressed and {partially} answered in the past. We review the existing literature, and further bring to the fore alternative answers to these old questions.

\subsection{Motivations and outline}

We have just illustrated the core of the mathematical problem we want to address. These technical aspects of path integrals are of importance in a wide array of sciences. Indeed, stochastic processes are ubiquitous in mathematical descriptions of the physical world. In situations in which  one focuses on a subset of degrees of freedom of a deterministic dynamics, information is lost, and this results in effective randomness. This applies to inflationary cosmology~\cite{matacz_new_1997,vennin_correlation_2015,Renaux-Petel}, climate dynamics~\cite{gottwald_crommelin_franzke_2017}, colloidal particles in solvents~\cite{einstein1905molekularkinetischen}, Bose--Einstein condensates~\cite{duine_stochastic_2001}, to name but a few. Phenomena outside the realm of physics, whether option pricing~\cite{karatzas_methods_2016} or myosin dynamics~\cite{bustamante_physics_2001}, are also described by similar tools. 
Going down in scale one meets the quantum description of matter which is intrinsically random.
Common mathematical tools that pervade these areas of science are stochastic differential equations and their path-integral representation. 
It is thus of paramount importance to identify a sound mathematical framework that paves the way for 
their use.

\medskip

In this article, we review in a self-contained manner the pitfalls presented by different 
path-integral representations of the probability distribution of a Langevin equation,
and we explain the methods that allow one to properly manipulate the action and the measure upon a change of variables 
--~stressing that the degree of care one has to demonstrate goes one order beyond that of the usual discretization issues for Langevin equations.
The goals of this review are thus mainly  of methodological nature, but achieving them is  essential if one wants to manipulate path-integral representations in a consistent manner.

We begin in Sec.~\ref{sec:calcul_sto} by reviewing the role of time discretization in one and more dimensions:
we start by presenting the discretization schemes of Langevin equations in a self-contained manner, 
we then explain how integrals involving such processes are themselves discretized (pinpointing the need to go beyond the usual $\alpha$-discretization scheme)
and we detail finally the specifics of the multidimensional case.

In Sec.~\ref{sec:pathint}, we describe a first version of the path-integral construction based on the usual linear 
time-discretization procedure also used to discretize the Langevin equation,
and we show that it is not `covariant', \textit{i.e.}~that a naive use of the chain rule in the action at the continuous-time level would lead to incorrect results (even caring about the possible
changes induced in the measure).
The source of such a conundrum lies in the correct handling of all terms of relevant order after a change of variables --~as done when deriving Itō's lemma for Langevin equations.

In  Sec.~\ref{sec:ItocalculusPI}, we show that, while Langevin equations feature a term $\frac{\dd x}{\dd t}$ that already requires special care, 
extra care is needed to manipulate the term $(\frac{\dd x}{\dd t})^2$ that appears in the exponential weight of a path. 
This observation is at the root of the mathematical difficulties that arise when changing variables in the action. 
We show that,  in general,  it is not possible to use the modified chain rule of $\alpha$-discretized stochastic differential calculus
at the path-integral level 
--~hence the failure of the usual chain rule in Stratonovich-discretized path integrals~-- 
and we give the proper transformation rule of the $\alpha$-discretized path-integral weight. 
This is one way of addressing the problem of changing variables. Special cases are the Itō and Hänggi--Klimontovich discretized path integrals:
for suitably discretized path measures, the blind use of the modified chain rule does yield correct results~\cite{ding2021timeslicing}, potentially up to boundary terms.

Historically, DeWitt~\cite{dewitt_dynamical_1957} first proposed a covariant extension of Feynman's path-integral formulation of quantum mechanics to curved spaces. A similar construction was then used by Graham~\cite{graham_covariant_1977,graham_path_1977} for classical diffusive processes. In their formulation, the propagator of the process between two infinitesimally close times is expressed by means of the continuous-time action evaluated at the least-action trajectory. This requires solving the classical equation of motion over an infinitesimal time window with boundary conditions at $x$ and $x+\Delta x$. We will henceforth refer to this discretization as being of implicit nature in the increment $\Delta x$. We review these approaches in Sec.~\ref{sec:DWG}. 

Another  solution to the lack of covariance  
consists in altering the discretization scheme of the path integral so as to make the continuous-time expression consistent with differential calculus. 
Indeed, the alternative construction that we propose in Secs.~\ref{sec:explicit-covariant} and~\ref{sec:covLangevin} is based on higher-order 
extensions of the Stratonovich discretization of Langevin equations,
generalizing the 1D approach of~\cite{cugliandolo2019building} to the case of an arbitrary number of dimensions.
The continuous-time expressions that we obtain are compatible with differential calculus, as already achieved by DeWitt and Graham, but this covariance property extends to the fully discretized level. Our scheme will appear more familiar in spirit to statistical physicists.

We close the paper with a concluding section. Four Appendices provide further details and the last one summarizes the most 
relevant mathematical expressions.

\section{Stochastic calculus}\label{sec:calcul_sto}

This section reviews stochastic calculus at the level of the Langevin and Fokker--Planck equations without referring just yet to path integrals. The difficulties intrinsic to working in more than one space dimension are discussed.
 
\subsection{Linear discretization of stochastic differential equations} 

As we shall review below, a stochastic differential equation involving a multiplicative noise (one in which the noise $\eta$ appears to be multiplied by a state-dependent function $g$, as in Eq.~\eqref{eq:langevin_mult} below), acquires a well-defined mathematical meaning once endowed with a discretization rule. Such equations with multiplicative noise are by no means a rarity. For instance, the mobility of a Brownian colloid diffusing in the vicinity of a wall depends on its distance to the wall~\cite{PhysRevLett.57.17,ANDARWA201484,PhysRevE.91.052305,doi:10.1063/1.4964935,marbach2018transport}. The description of rotational Brownian motion~\cite{coffey2012langevin} (with applications to dielectrics~\cite{10.2307/20489032}, magnetism~\cite{brown1963thermal}, and active matter~\cite{cates2015motility}) also involves, in order to enforce a spherical constraint, a multiplicative noise. 
The evolution of the concentration of species, both in ecology and chemistry, involves a population noise that depends on the concentrations~\cite{gillespie_chemical_2000,spagnolo_noise_2004} (simply because the concentrations have to remain positive).
Another celebrated example outside of the realm of physics is the Black and Scholes equation~\cite{hull2009options} proposed to model the evolution of some specific financial assets. Of course, any non-linear transformation of a stochastic variable evolving according to a Langevin equation with additive noise is governed by a Langevin equation with multiplicative noise, as illustrated by our opening example, Eqs.~\eqref{eq:langevin_colloid} and~\eqref{eq:langevin_kineticS}. 

In the physical sciences, there are mostly two channels through which such Langevin first-order differential equations arise~\cite{kampen_ito_1981}. In the first one, a large physical system is described by dynamical equations that couple the degrees of freedom of interest, hereafter denoted by $x(t)$, to some other external degrees of freedom referred to as a bath or an environment. Integrating out the latter generically yields a dynamical equation for $x(t)$ which features both colored noise and colored friction~\cite{zwanzig_nonequilibrium_2001}. It is then the Markov limit, in which the relaxation time of the external degrees of freedom is assumed to be much smaller than the typical timescale associated with the dynamics of $x(t)$, that defines the correct limiting stochastic differential equation and that specifies the associated discretization rule (in most cases, this is how the Stratonovich discretization emerges). In the second situation, physics is fundamentally described by master equations (derived \textit{e.g.}~from some Liouville equation). The correspondence between such a 
description with a stochastic differential equation also fixes the proper discretization scheme used to represent the stochastic process. Most of the discussion that follows can be found in classic textbooks such as  Gardiner's~\cite{gardiner_handbook_1994} or Van Kampen's~\cite{kampen_stochastic_2007}.

\subsubsection{The stochastic equation}
\label{sec:stochastic-equation}

\begin{figure}[t!]
~\\[6mm]
\centerline{
\includegraphics[width=.66\columnwidth]{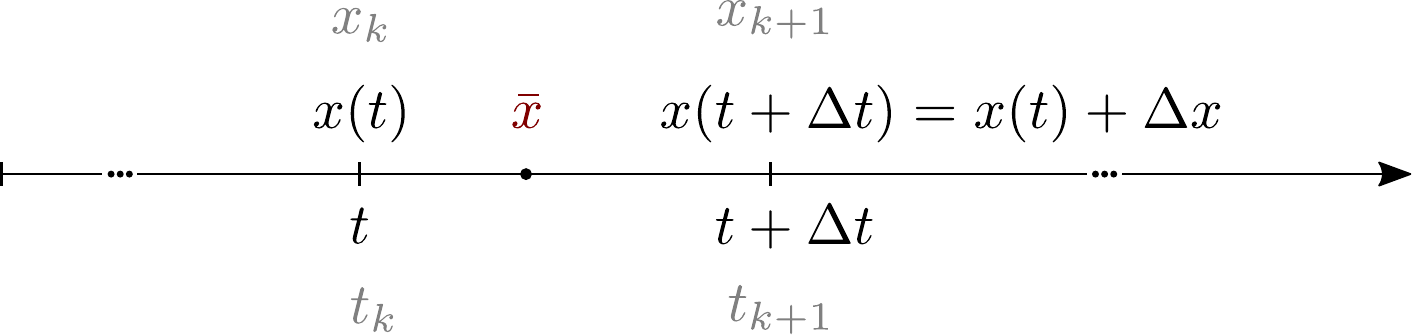}
}
~\\[-5mm]
\caption{The discretization point, $\bar x$, is the point at which functions of $x$ are evaluated in continuous-time writings of the Langevin equation (or in the path-integral action). 
It is expressed as a function of $x(t)$ and $\Delta x = x(t+\Delta t)-x(t)$. 
Discrete-time notations are displayed in gray.
The most standard discretization scheme is  linear, see Eq.~\eqref{eq:lineardiscretizationscheme}.
In this paper, we also consider quadratic discretization schemes (see Eq.~(\ref{eq:higher_order}) in 1D and~(\ref{eq:defxbark_ddim}) in any dimension).
}
\label{fig:discretisationpoint}
\end{figure}

We now consider a dynamical variable $x(t)$, the evolution of which is assumed to be given by a stochastic 
differential equation with multiplicative noise
\begin{align} \label{eq:langevin_mult}
\frac{\dd x}{\dd t} \stackrel{\mathfrak{d}}{=} f(x) + g(x) \eta(t)
\end{align}
with $x=x(t)$ and where, for the sake of clarity, we wrote the time dependence of the noise $\eta$ explicitly. 
This noise is Gaussian and white with zero mean, that is
\begin{eqnarray}
\langle \eta(t) \rangle = 0 
\; , 
\qquad\qquad
  \langle\eta(t)\eta(t')\rangle=\, \delta(t-t')
\; .
\end{eqnarray}
The label $\mathfrak d$ above the equal sign stands for a reminder that Eq.~\eqref{eq:langevin_mult} comes hand-in-hand with an accompanying discretization scheme (Van Kampen~\cite{kampen_ito_1981} refers to Eq.~\eqref{eq:langevin_mult} as a pre-equation). Concretely, Eq.~\eqref{eq:langevin_mult} should be understood as the continuous-time limit, \textit{i.e.}~the limit in which the time step $\Delta t$ goes to zero, of the discrete companion evolution rule 
\begin{align} \label{eq:discrete_bar}
\Delta x(t) = x(t + \Delta t) - x(t) = f(\bar{x})\Delta t + g(\bar{x})\Delta \eta(t) \, ,
\end{align}
where
$\bar x$ is the discretization point (see Fig.~\ref{fig:discretisationpoint}), and 
the $\Delta \eta(t)$'s 
are independent and identically distributed Gaussian variables with zero mean and variance $\Delta t$:
\begin{eqnarray}
\langle \Delta\eta(t) \rangle =0 
\; , 
\qquad\qquad
  \langle \Delta\eta(t) \Delta\eta(t')  \rangle= \Delta t \, \delta_{tt'}
\; ,
\label{eq:Delta-noise-corr}
\end{eqnarray}%
with $\delta_{tt'}$ denoting the Kronecker delta.
Notation-wise, we shall also use the discrete sequence of steps $x_k=x(k\Delta t)$, with $k=0,\ldots,N$
and $N={\tf}/\Delta t$, in particular, to write sums of functions of the discrete values of the variable.
The independent identically distributed Gaussian variables 
$\Delta\eta_k$ have variance $\Delta t$ and zero mean. Here, $[0,{\tf}]$ refers to the time window over which 
we sample the random process. In this notation,  Eq.~(\ref{eq:discrete_bar}) becomes
\begin{align} \label{eq:discrete_bar-more}
\Delta x_k = x_{k+1}-x_k= f(\bar{x}_k)\Delta t + g(\bar{x}_k)\Delta \eta_k \, .
\end{align}
In Eq.~\eqref{eq:discrete_bar}, $f$ and $g$ are evaluated at $\bar{x}$, a function of $x(t + \Delta t)$ and $x(t)$, the choice of which fully determines the discretization scheme. (In Eq.~(\ref{eq:discrete_bar-more}) $\bar{x}$ has been replaced by 
$\bar{x}_k$ where $x(t)$ and $x(t+\Delta t)$ have in turn been replaced by $x_k$ and $x_{k+1}$.) 
It is of paramount importance to notice that as a consequence of Eq.~\eqref{eq:discrete_bar} and the 
statistical properties of the noise, the increment of the process scales as
\begin{align} \label{eq:scaling}
\Delta x(t) = O\big(\sqrt{\Delta t}\,\big) 
\end{align} 
as $\Delta t\to 0$.
Therefore, the trajectories $x(t)$ obtained in the continuous-time limit $\Delta t \to 0$ are (almost surely) nowhere differentiable. This explains why, as we will see later, the discretization scheme of first-order 
stochastic differential equations matters while it does not in the $\Delta t \to 0$ limit when discretizing first-order 
ordinary differential equations.

\subsubsection{Linear discretization prescriptions}

A common discretization scheme~\cite{janssen_renormalized_1992} is the so-called $\alpha$-discretization prescription
\begin{align}
\bar{x} = x(t) + \alpha \Delta x(t) \, ,
\label{eq:lineardiscretizationscheme}
\end{align}
with $\alpha \in [0,1]$, that is, a \textit{linear} function of the increment $\Delta x$. 
The Itō or pre-point convention corresponds to $\alpha = 0$ for which Eq.~\eqref{eq:discrete_bar} provides an explicit expression for the increment $\Delta x(t)$. It moreover guarantees the statistical independence of $x(t)$ with respect to $\Delta \eta(t)$. The $\alpha = 1/2$ case corresponds to the Stratonovich or mid-point convention. The Stratonovich scheme is time-symmetric and, as we shall see further down, it allows for the usual chain rule of differential calculus to hold (at the Langevin equation level). Finally, the $\alpha = 1$ discretization scheme is called the Hänggi--Klimontovich or the post-point one, and it has proved convenient in the study of relativistic Brownian motion~\cite{dunkel2005theory}. In view of performing efficient numerical simulations, the question of finding the ``best" discretization scheme is a very active one that goes well beyond the present discussion. We refer the interested reader to recent reviews in this area~\cite{milshtejn_approximate_1974,milshtejn_approximate_1975,
kloeden_numerical_1995,mannella_integration_2002, kloeden2012numerical, farago2014langevin}.

We emphasize that, contrary to what holds for ordinary differential equations, different $\alpha$-discretized companion processes sharing the same $f$ and $g$ functions lead to different stochastic processes in the $\Delta t \to 0$ limit, and are thus characterized by different distributions. This is simply proven by the calculation of the difference between the $\Delta x(t)$ generated with 
$\alpha \neq \alpha'$ with the help of Eq.~\eqref{eq:scaling}: 
\begin{align}
& 
\left[g(x(t) + \alpha \Delta x(t)) - g(x(t) + \alpha' \Delta x(t))\right]\Delta \eta(t) 
\nonumber\\
& \qquad\qquad\qquad
= (\alpha - \alpha') g'(x(t))\, \Delta x(t)\, \Delta \eta(t) + O\big(\Delta t^{3/2}\big) 
\nonumber \\ 
& \qquad\qquad\qquad = O(\Delta t) \, .
\end{align}
Therefore, for two different $\alpha$-discretization schemes, the difference in the increments $\Delta x(t)$ is of order $O(\Delta t)$, as is the contribution to the increments of the deterministic term, and 
cannot be neglected in Eq.~(\ref{eq:discrete_bar}) in the $\Delta t \to 0$ limit. Note that, by contrast,
\begin{align}
\left[f(x(t) + \alpha \Delta x(t)) - f(x(t) + \alpha' \Delta x(t))\right]\Delta t & = O\big(\Delta t^{3/2}\big) \, ,
\end{align}
which expresses that the way in which 
the deterministic term is discretized does not bear any influence in the $\Delta t \to 0$ limit,
in the Langevin equation. 

The sensitivity to the discretization scheme also reflects on the Fokker--Planck equation associated 
to Eq.~\eqref{eq:langevin_mult} (read in $\alpha$-discretization), which reads
(App.~\ref{app:FP})~\cite{gardiner_handbook_1994,kampen_stochastic_2007,Langouche79,arenas2012}
\begin{eqnarray}
\partial_t P(x,t)
 = 
- \partial_x[ (f(x) +\alpha g(x) g'(x)) P(x,t) ] 
+ \frac{1}{2} \, \partial_x^2 [ g^2(x) P(x,t) ] \, .
\label{eq:FP-without-drift}
\end{eqnarray}
This  is a deterministic partial differential equation and it is thus immune to any discretization issue.
Once supplemented with an initial condition $P(x,0)$, 
it describes the deterministic evolution of the probability density $P(x,t)$ of finding $x$ at 
time $t$. Equation~(\ref{eq:FP-without-drift})
can be written in the form of a continuity equation 
$\partial_t P + \partial_x J = 0$ and its stationary solution with vanishing current, $J=0$, is
\begin{equation}
P_{\rm st}(x) = Z^{-1} \ [g(x)]^{2(\alpha-1)} \ 
\exp \left[ \, 2 \int^x \!\! \dd x' \; \dfrac{f(x')}{g^2(x')} \, \right]
\label{eq:stat-sol-no-drift}
\end{equation}
where $\int^x$ represents the indefinite integral over $x'$ and $Z$ is a normalization 
constant~\cite{gardiner_handbook_1994,kampen_stochastic_2007}. 
The  approach to this asymptotic form can be proven with the construction of an $H$-function
or with the mapping of the Fokker--Planck operator onto a Schrödinger operator and the analysis of its eigenvalue 
problem~\cite{Parisi}. That $P_{\rm st}$ depends on $\alpha$ and $g$ explicitly shows that these ingredients affect the stationary properties of the system~\cite{Horsthemke84,Sagues07,Stratonovich92}. However, if we allow ourselves to consider the special ``drift force''~\cite{Klimontovich1990}
\begin{equation}
f(x) = - \frac{1}{2} g^2(x) \, \beta V'(x) + (1-\alpha)  g(x) g'(x) \; , 
\label{eq:equil-choice-f}
\end{equation} 
 $\alpha$ is eliminated from the Fokker--Planck equation,
\begin{eqnarray}
&&
\partial_t P(x,t)
=
 \partial_x 
\Big\{ 
g^2(x) \Big[ \frac 12 \, \beta V'(x)  P(x,t) + \frac 12 \, \partial_x P(x,t) \Big]
\Big\}
,
\label{eq:FP-correct}
\end{eqnarray}
and no observable depends on this parameter either. The asymptotic solution to this new equation reads
\begin{equation}
P_{\rm st}(x) =  Z^{-1} \ \ee^{-  \beta V(x)} = P_{\rm GB}(x)
\; , 
\end{equation}
which, independently of $\alpha$ and $g$, is the standard Gibbs--Boltzmann distribution 
in the canonical ensemble of a system with potential energy $V(x)$.

Therefore, in order to describe the equilibrium Langevin dynamics of a  multiplicative white noise system 
that samples the standard Gibbs--Boltzmann distribution, one needs to work with the equation
\begin{equation}
\frac{\dd x}{\dd t} \stackrel{\alpha}{=} - \frac 1 2 g^2(x) \, \beta V'(x) + (1-\alpha) g(x) g'(x) + g(x) \eta  \, ,
 \label{eq:x-eom-drifted}
\end{equation}
using an $\alpha$-prescription. Note the presence of a non-trivial additional drift force
even in the Stratonovich ($\alpha=1/2$) scheme. It is only with a post-point (Hänggi--Klimontovich) discretization scheme $\alpha=1$~\cite{Hanggi80,Hanggi78,Hanggi1982,Klimontovich} that this additional term vanishes.

\subsubsection{Equivalence between differently discretized processes}
\label{subsubsec:equivalence}

While distinct discretizations of the same continuous-time expression lead  to different processes, there are different yet equivalent ways to describe the same physical process using a Langevin equation. To be more explicit, we consider an $\alpha$-discretized stochastic differential equation of the form
\begin{align}
\frac{\dd x}{\dd t} \stackrel{\alpha}{=} f(x) + g(x) \eta \, ,
\end{align}
which yields in discrete time
\begin{align} \label{eq:langevin_discrete_change}
\Delta x = f(x) \Delta t + g(x + \alpha \Delta x) \Delta \eta \, .
\end{align}
The latter evolution rule can be rewritten for any $\alpha' \in [0,1]$ as
\begin{align} \label{eq:change_discrete}
\Delta x & = f(x) \Delta t + g(x + \alpha' \Delta x + (\alpha - \alpha') \Delta x) \Delta \eta 
\nonumber \\ 
& 
= f(x) \Delta t + g( x + \alpha' \Delta x) \Delta \eta + (\alpha - \alpha') g'( x + \alpha' \Delta x)\Delta x \Delta \eta + O(\Delta t^{3/2}) \nonumber \\ 
& 
= f(x) \Delta t + g( x + \alpha' \Delta x) \Delta \eta + (\alpha - \alpha') g(x)g'(x)\Delta \eta^2 + O(\Delta t^{3/2}) \, , 
\end{align}
and 
explicitly displays a $\Delta \eta^2$ contribution, at odds with the original dynamics in Eq.~\eqref{eq:langevin_discrete_change}. However, when computing the continuous-time Fokker--Planck equation associated with the $\Delta t \to 0$ limit of Eq.~\eqref{eq:change_discrete}, one realizes that the $\Delta \eta^2$ only contributes through its first moment $\langle \Delta \eta^2\rangle = \Delta t$. One can thus rewrite
\begin{align} \label{eq:change_discrete_subs}
\Delta x &\doteq f(x) \Delta t + g( x + \alpha' \Delta x) \Delta \eta + (\alpha - \alpha') g(x)g'(x)\Delta t \, . 
\end{align}
The $\doteq $ sign does not mean there is a point-wise equality between Eq.~\eqref{eq:langevin_discrete_change} and Eq.~\eqref{eq:change_discrete_subs} but rather that these two discrete-time evolution rules generate the same random process in the continuous-time limit. This is the first example of a \textit{{substitution rule}}, 
a notion that we will shortly clarify. Therefore, in the continuous-time limit, we can assert that
\begin{align} \label{eq:change_discrete_cont}
\frac{\dd x}{\dd t} & \stackrel{\alpha}{=} f(x) + g(x) \eta  
\nonumber \\ & \stackrel{\alpha'}{=} f(x) + (\alpha - \alpha') g(x) g'(x)  + g(x) \eta \, ,
\end{align}
keeping in mind that these two Langevin equations are not equal point-by-point, but yield processes with the same distribution at all times. 

\smallskip

This concludes our review of the most common linear schemes used to discretize Langevin equations and their main 
properties. 
We next investigate how the rules of calculus --~integration and differentiation~-- are affected by the singular nature of the paths generated by Langevin equations with white noise.

\subsection{Integration}\label{sec:integration}

We now consider a random process $x(t)$ which evolves according to the Langevin equation
\begin{align} \label{eq:langevin_alpha}
\frac{\dd x}{\dd t} \stackrel{\alpha}{=} f(x) + g(x) \eta \, ,
\end{align}
understood as $\alpha$-discretized. First, we focus on observables of the form
\begin{equation} \label{eq:sum_riemann}
\mathcal{O}_0 = \int_0^{\tf} \dd t \, h(x(t)) \, ,
\end{equation}
where $h$ is a (smooth enough) arbitrary function, which, when expressed in terms of the discrete-time companion process, 
reads
\begin{align}
\mathcal{O}_0 = \lim_{\Delta t \to 0} \, \sum_{k = 0}^N \Delta t \, h(x_k) \, .
\end{align}
In the $\Delta t \to 0$ limit, one could have also written
\begin{align}
\mathcal{O}_0 = \lim_{\Delta t \to 0} \, \sum_{k = 0}^N \Delta t \, h(x_k + \alpha' \Delta x_k) \, , 
\end{align}
for any $\alpha' \in [0,1]$ with $\Delta x_k = x_{k+1}-x_k$. In other words, in the continuous-time limit, the specific discretization scheme of the integral in Eq.~\eqref{eq:sum_riemann} is irrelevant, as expected for such standard Riemann sums.

Other interesting observables that, for example, often arise in the field of stochastic thermodynamics~\cite{seifert2012stochastic,sekimoto2010stochastic} are of the form
\begin{align} \label{eq:integral_alpha}
\mathcal{O}_1 \stackrel{\alpha'}{=} \int_0^{\tf} \dd t \, \dot{x}(t) \, h(x(t)) \, ,  
\end{align}
which in terms of the discrete-time companion process reads
\begin{align}
\mathcal{O}_1 = \lim_{\Delta t \to 0} \, \sum_{k = 0}^{N-1} \Delta t \, \frac{\Delta x_k}{\Delta t} \,  h(x_k + \alpha' \Delta x_k) \, .
\end{align}
The discretization scheme has been made explicit above the equality sign by the $\alpha'$ label appearing in Eq.~\eqref{eq:integral_alpha}. Due to the scaling $\Delta x_k = O(\sqrt{\Delta t})$, the discretization of the integral, namely the point at which the function $h$ is evaluated, is relevant even in the $\Delta t \to 0$ limit. This statement is very similar to the fact that one needs to specify the discretization of $g$ in the discrete-time companion process of Eq.~\eqref{eq:langevin_alpha}. Note that $\alpha$ and $\alpha'$ are not necessarily related: $\alpha$ determines the evolution of the process $x$ while $\alpha'$ enters in the definition of the observable $\mathcal{O}_1$. Integrals of the form Eq.~\eqref{eq:integral_alpha} with $\alpha' = 0$ (respectively $\alpha' = 1/2$) are referred to as Itō integrals (respectively, Stratonovich integrals).

The analysis of the observable $\mathcal{O}_2$, defined by its discrete expression as
\begin{align}
\mathcal{O}_2 = \lim_{\Delta t \to 0} \, \sum_{k = 0}^{N-1} \Delta t \, \frac{\Delta x_k^2}{\Delta t} \, h(x_k) 
\, ,
\label{eq:O2-discrete}
\end{align}
which is finite in the $\Delta t \to 0$ limit, will allow us to identify \textit{substitution relations}.
The discretization of $h$ is irrelevant to define the continuous-time limit of $\mathcal{O}_2$
and this is the reason why we have not added a superscript to the equal sign in Eq.~(\ref{eq:O2-discrete})
and we simply evaluated $h$ at $x_k$. The finite character of $\mathcal{O}_2$ is proved by 
replacing $
\Delta x_k = f(x_k) \Delta t + g\left(x_k + \alpha \Delta x_k\right)\Delta \eta_k 
$
in~(\ref{eq:O2-discrete}):
\begin{align}\mathcal{O}_2 
& = \lim_{\Delta t \to 0} \sum_{k = 0}^{N-1} \, 
\Delta t \, \frac{\left[ f(x_k) \Delta t + g(x_k + \alpha \Delta x_k)\Delta \eta_k\right]^2}{\Delta t} \, h(x_k)  
\nonumber \\ & 
= \lim_{\Delta t \to 0} \sum_{k = 0}^{N-1} \, 
\Delta t \,\frac{\Delta \eta_k^2}{\Delta t} \, g^2(x_k + \alpha \Delta x_k) \, h(x_k) \nonumber  \\ & 
= \lim_{\Delta t \to 0} \sum_{k = 0}^{N-1} \, 
\Delta t \,\frac{\Delta \eta_k^2}{\Delta t} \, g^2(x_k) \, h(x_k) \, , 
\end{align}
which is manifestly finite.
Interestingly, in the $L^2$-norm sense, the statistical properties of $\mathcal{O}_2$ and $\mathcal{O}_2'$ defined by
\begin{align}
\mathcal{O}'_2 & =  \lim_{\Delta t \to 0} \, \sum_{k = 0}^{N-1} \Delta t \, g^2(x_k) \, h(x_k) 
\end{align}
are the same, as can be checked by the following calculation:
\begin{align} \label{eq:proof_convL2}
& \lim_{\Delta t \to 0} \left\langle \left[\sum_{k = 0}^{N-1} \Delta t \, g^2(x_k) \, h(x_k) \left( 1 - \frac{\Delta \eta_k^2}{\Delta t}\right)\right]^2 \right\rangle 
\nonumber \\ &  
\qquad 
= \lim_{\Delta t \to 0} \left\langle \sum_{k ,k' = 0}^{N-1} \Delta t^2 \, g^2(x_k) g^2(x_{k'})\, h(x_k)h(x_{k'}) 
\left( 1 - \frac{\Delta \eta_k^2}{\Delta t}\right) \left( 1 - \frac{\Delta {\eta_{k'}}^2}{\Delta t}\right) \right\rangle 
\nonumber  \\ & 
\qquad
= \lim_{\Delta t \to 0} \left\langle \sum_{k = 0}^{N-1} \Delta t^2 \, g^4(x_k)\, h^2(x_k) \left( 1 - \frac{\Delta \eta_k^2}{\Delta t}\right)^2\right\rangle  = 0 \, .
\end{align}
In the second line, for $k\neq k'$ we reordered $k'<k$, averaged first over $\Delta\eta_k$ (which is not correlated to $x_k$, $x_{k'}$ or $\Delta \eta_{k'}$) and used  that $\langle\Delta t (1-\Delta \eta_k^2/\Delta t)\rangle =0$ (see Eqs.~(B.8)-(B.13) in~\cite{Cugliandolo-Lecomte17a} for more details).
This justifies the substitution relation 
\begin{align} \label{eq:substitution_first}
\Delta x_k^2\doteq  g^2(x_k) \Delta t 
\end{align}
or, equivalently,
\begin{align} \label{eq:substitution_eta}
\Delta \eta_k^2\doteq  \Delta t \, .
\end{align}

Following the same line of reasoning one can show (see App.~B in~\cite{Cugliandolo-Lecomte17a})
\begin{align} \label{eq:subsitution_second}
\Delta x_k^4 \doteq   3 \, g^4(x_k) \, \Delta t^2 \, , 
\end{align}
and similar substitution relations for higher even powers of $\Delta x_k$ with the factors and multiplicities complying with Wick's rules.
Such substitution rules have been discussed in the context of path integrals in curved space~\cite{mclaughlin_path_1971,weiss_operator_1978,gervais_point_1976,hirshfeld_canonical_1978,girotti_generalized_1983}
and we will come back to them later. 
As will be detailed in Sec.~\ref{subsubsec:third-order} and further on, odd-order substitution rules require a special care (see also Ref.~\cite{Cugliandolo-Lecomte17a} in one dimension).

These substitution rules allow one  to express  $\alpha'$-discretized $\mathcal{O}_1$ observables in terms of $\alpha''$-discretized ones. The relation works as follows:
\begin{align} \label{eq:integral_change_var}
\mathcal{O}_1 & \stackrel{\alpha'}{=} \int_0^{\tf}  \dd t  \; \dot{x}(t) \, h(x(t))  
= 
\lim_{\Delta t \to 0} \, \sum_{k = 0}^{N-1} \Delta t \, \frac{\Delta x_k}{\Delta t} \,  h(x_k + \alpha' \Delta x_k) 
\nonumber \\ 
& =  \lim_{\Delta t \to 0}  \, \sum_{k = 0}^{N-1} 
\left\{ \Delta t \,\frac{\Delta x_k}{\Delta t}  
\left[ h(x_k + \alpha'' \Delta x_k) + (\alpha' - \alpha'') h'(x_k) \Delta x_k \right] + O\big(\Delta t^{3/2}\big) \right\} 
 \nonumber \\ 
 & 
\stackrel{\alpha''}{=} \int_0^{\tf}  \dd t  \, \dot{x}(t) \, h(x(t)) + \left(\alpha' - \alpha''\right)\int_0^{\tf} \dd t \, g^2(x(t)) \, h'(x(t)) 
\end{align}
where we used~(\ref{eq:substitution_first}).

As a final comment, we stress that even though the point at which the function $h$ is evaluated is irrelevant to determine the observable ${\mathcal O}_2$ in the $\Delta t\to 0$ limit, finite $\Delta t$ corrections to ${\mathcal O}_2$, if needed (and they will be needed when we address path integration), do depend on the value at which $h$ is evaluated. Equivalently, this issue surfaces if we want to regularize integrals of the type 
\begin{equation}
\mathcal{O}_3 =  \int_0^{\tf}\dd t \; \dot{x}^2(t) \,  h(x(t))
\; .
\end{equation}
Indeed,  it is clear that any observable of the form
\begin{align}
\mathcal{O}_3 = \sum_{k=0}^{N-1} \Delta t \, \frac{\Delta x_k^2}{\Delta t^2} \,  h(x_k) 
\end{align}
only has infinite moments when $x_k$ is sampled from the companion process associated to Eq.~\eqref{eq:langevin_alpha} in the $\Delta t \to 0$ limit. Suppose, however, that our interest goes to  the observable defined by
\begin{align} \label{eq:O3}
\mathcal{O}_3' =  \sum_{k=0}^{N-1} \Delta t \, h(\bar{x}_k) 
\frac{\Delta x_k^2}{\Delta t^2}
- \sum_{k=0}^{N-1} \Delta t \, h(x_k) \frac{\Delta x_k^2}{\Delta t^2} \, ,
\end{align}
where $\bar{x}_k$ is again a function of $x_k$ and $x_{k+1}$ and where the diverging part in the continuous limit has been subtracted. Owing to the scaling of $\Delta x_k$, namely,
\begin{align}
\Delta x_k^2 \sim \Delta t \, ,
\end{align}
one needs to know $h(\bar{x}_k)$ up to order $O(\Delta t)$ in order to collect all finite terms in the continuous-time limit. This is in stark contrast with standard Itō or Stratonovich integrals (or any integral of the type $\mathcal{O}_1$) in which the function $h$ needs to be known up to order $O(\sqrt{\Delta t})$ only. In particular, observables such as $\mathcal{O}_3$ (that we will encounter when dealing with path integrals) are sensitive to higher-order terms in the discretization. To render this property more explicit, we introduce an $\alpha, \beta$ \textit{non-linear} quadratic discretization scheme defined by
\begin{align} \label{eq:higher_order}
\bar{x}_k = x_k + \alpha \Delta x_k + \beta(x_k) \Delta x_k^2 \, ,
\end{align}
where the notation stresses that $\beta(x_k)$ may depend on $x_k$ and which yields
\begin{align} \label{eq:quadratic_discrete}
\mathcal{O}_3' & = 
\sum_{k=0}^{N-1} \Delta t \, \Big[h\big(x_k + \alpha \Delta x_k + \beta(x_k) \Delta x_k^2\big) - h(x_k)\Big] 
\frac{\Delta x_k^2}{\Delta t^2}
\nonumber \\ 
& 
= \sum_{k=0}^{N-1} \Delta t \, \left[\alpha \, \Delta x_k \, h'(x_k) + \left(\beta(x_k) h'(x_k) + \frac{\alpha^2}{2}h''(x_k)\right) \Delta x_k^2 \right] 
\frac{\Delta x_k^2}{\Delta t^2}
\nonumber \\
& 
= \sum_{k=0}^{N-1} \alpha h'(x_k) \frac{\Delta x_k^3}{\Delta t} + \sum_{k=0}^{N-1} \left(\beta(x_k) h'(x_k) + \frac{\alpha^2}{2}h''(x_k)\right) 
\frac{\Delta x_k^4}{\Delta t}  
\nonumber \\ 
&\doteq  \sum_{k=0}^{N-1} \alpha h'(x_k) \frac{\Delta x_k^3}{\Delta t} 
+ 3 \sum_{k=0}^{N-1} g^4(x_k)\left(\beta(x_k) h'(x_k) + \frac{\alpha^2}{2}h''(x_k)\right)\Delta t  \, , 
\end{align}
after having used the substitution rule~\eqref{eq:subsitution_second}, valid in the $\Delta t \to 0$ limit, 
in the last line. The cubic term $\Delta x_k^3/\Delta t$, which has no obvious continuous-time limit, 
will be dealt with in Sec.~\ref{subsubsec:third-order}
in the more general higher dimensional case. It is of order $O(\Delta t^{1/2})$ and thus cannot be discarded in Eq.~(\ref{eq:quadratic_discrete}) in the $\Delta t\to 0$ limit.

The goal of this subsection was to provide the reader with a review of stochastic integration. We have discussed four types of integral observables, ${\mathcal O}_0$, ${\mathcal O}_1$, ${\mathcal O}_2$ and ${\mathcal O}_3$ that will each appear in the exponential weight of path integrals. We now turn to the differentiation of a stochastic path.

\subsection{Differentiation}
\label{subsubsec:differentiation}

The other operation that we wish to extend to stochastic paths is differentiation. Going back to our initial equation
\begin{align} \label{eq:lagevin_change_var}
\frac{\dd x}{\dd t} \stackrel{\alpha}{=} f(x) + g(x)\eta \, , 
\end{align}
we now define the process $u(t) = U(x(t))$ where $U$ is some smooth invertible function. 
Its distribution $P_u$ is governed by a Fokker--Planck equation of the form~(\ref{eq:FP-without-drift}) 
inferred from the distribution $P_x$ of the process $x$ and from the change of measure
\begin{equation}
  \label{eq:chvarPxPu}
  \dd x \, P_x (x,t) = \dd u \, P_u(u,t)
\end{equation}
(where we temporarily made explicit, as an index, the process of interest).
The statistics of $u(t)$ can also be directly studied within the Langevin framework, as we now explain.
To this end, we start with the discrete-time companion process $x$ which evolves according to
\begin{align}
\Delta x = f(x)\Delta t + g\left(x + \alpha \Delta x\right)\Delta \eta \, .
\end{align}
In discrete time, the evolution of the new variable $u$ is therefore given by 
\begin{align} \label{eq:discrete_U}
\Delta u & = u(t + \Delta t) - u(t) = U(x + \Delta x) - U(x) 
\nonumber \\
&
 = U'(x) \, \Delta x + \frac{1}{2}\, U''(x) \, \Delta x^2 + O\big(\Delta t^{3/2}\big) 
 \nonumber \\
& \stackrel{}{=}   U'(x) \, f(x) \,\Delta t + U'(x) \, g(x + \alpha \Delta x)  \, \Delta \eta + \frac{1}{2} \, U''(x) \, g^2(x) \Delta \eta^2 + O\big(\Delta t^{3/2}\big) 
\nonumber \\
& 
\stackrel{}{=} 
 f(x) \, U'(x) \Delta t + g(x + \alpha \Delta x) U'(x + \alpha \Delta x) \Delta \eta + \left(\frac{1}{2} - \alpha\right)g^2(x) U''(x) \Delta \eta^2  
\nonumber\\
& 
\quad + O\big(\Delta t^{3/2}\big)
\, . 
\end{align} 
In these expression, we could also replace the functions evaluated in $x=x(t)$ by the same functions evaluated in $x+\alpha\Delta x$, since this would only add terms of order $\Delta t^{3/2}$.
With the purpose of replacing the $x$ dependence still present in the right-hand-side by $u$ we use 
(see App.~\ref{app:inverse} for the analysis of the discretization of the inverse)
\begin{align}
x = U^{-1}(u) \, ,
\end{align}
and
\begin{align}
x + \alpha \Delta x = U^{-1}(u + \alpha \Delta u) + O(\Delta t) \, .
\end{align}
Equation~\eqref{eq:discrete_U} thus becomes
\begin{align} \label{eq:discrete_U_bis}
\Delta u & \stackrel{}{=}  F(u) \Delta t + \left(\frac{1}{2} - \alpha\right)g^2(U^{-1}(u))\, U''(U^{-1}(u)) \Delta \eta^2 + G(u + \alpha \Delta u) \Delta \eta + O\big(\Delta t^{3/2}\big) 
\nonumber \\
&\doteq  F(u) \Delta t + \left(\frac{1}{2} - \alpha\right)g^2(U^{-1}(u)) \, U''(U^{-1}(u)) \Delta t + G(u + \alpha \Delta u) \Delta \eta  \, ,
\end{align} 
with
\begin{align}
F = \left(f U'\right) \circ U^{-1} \, , \qquad\qquad
G = \left(g U'\right) \circ U^{-1} \, ,
\label{eq:FGoffg}
\end{align}
and where again the $\doteq $ notation expresses that the two discrete-time evolutions generate the same processes in the limit $\Delta t \to 0$ (this is not a point-wise equality). The $\circ$ symbol represents convolution between two functions, 
that is, $(f\circ g)(z) = f(g(z))$.
The expressions  (\ref{eq:discrete_U_bis})-(\ref{eq:FGoffg}) 
can be established either  by computing the Kramers--Moyal expansion of the process $u$ and determining the Fokker--Planck equation associated with Eq.~\eqref{eq:discrete_U_bis}  in the continuous-time limit, or by resorting to the substitution rule Eq.~\eqref{eq:substitution_first}. 
Again in Eq.~(\ref{eq:discrete_U_bis}) the functions of $u$ can be evaluated in $u+\alpha \Delta u$ since this would only add subdominant terms of order $O(\Delta t^{3/2})$.
In the continuous-time limit, we thus arrive at an $\alpha$-discretized Langevin equation for $u(t)$:
\begin{align} \label{eq:ito_lemma}
\frac{\dd u}{\dd t} & \stackrel{\alpha}{=} U' \, \frac{\dd x}{\dd t} + \left(\frac{1}{2} - \alpha\right)g^2(U^{-1}(u)) \, U''(U^{-1}(u)) 
\nonumber \\ 
& 
\stackrel{\alpha}{=} F(u) + \left(\frac{1}{2} - \alpha\right)g^2(U^{-1}(u)) \, U''(U^{-1}(u)) + G(u) \eta \, .
\end{align}

The formula above is at the core of stochastic calculus as it explicitly shows how working with paths generated by Eq.~\eqref{eq:langevin_alpha} modifies the chain rule of ordinary differential calculus. For $\alpha = 0$, Eq.~\eqref{eq:ito_lemma} is the celebrated Itō's lemma. For $\alpha = 1/2$, the standard chain rule holds and this is one of the most important assets of Stratonovich-discretized stochastic differential equations. 
Note also that, while a discussion of the discretization scheme is irrelevant at the level of Eq.\ (\ref{eq:lagevin_change_var}) whenever $g$ is a constant (namely for a process with additive  noise), 
it becomes a requirement when studying the evolution of a non-linear function of $x$, as seen in Eq.\ (\ref{eq:ito_lemma}).\\

An interesting consequence of formula~\eqref{eq:ito_lemma} is that it is always possible (for a one-dimensional process) to perform a non-linear change of variables so as to turn a process with multiplicative noise into one with additive noise. This is indeed easily achieved if one starts from a Stratonovich-discretized process (from Eq.~\eqref{eq:change_discrete_cont} we know that any $\alpha$-discretized stochastic differential equation can be turned into a Stratonovich-discretized one) of the form
\begin{align}
\frac{\dd x}{\dd t} \stackrel{\frac 12}{=} f(x) + g(x) \eta \, .
\end{align}
Assuming now that $g$ is a non vanishing function of $x$, we choose $U$ to be such that
\begin{align}
\left(g U'\right) \circ U^{-1} = \text{Id} \, ,
\end{align}
namely
\begin{align}
U(x) = \exp{ \int_{x_0}^x \dd x' \,  \frac{1}{g(x')}} \, ,
\end{align}
for some $x_0$. This immediately leads to
\begin{align}
\frac{\dd u}{\dd t} = F(u) + \eta \, ,
\end{align}
for $u(t) = U(x(t))$, which has additive noise and where $F$ is given in Eq.~\eqref{eq:FGoffg}. 
We refer the reader to Ref.~\cite{hanggi_derivations_1978} for a discussion on other transformations, in particular in the case in which the drift or diffusion depend explicitly on time.
Note that the conversion of a multiplicative noise process into an additive noise one 
is a peculiarity of processes living on the real axis which does not extend to higher dimensions in general.

\subsection{What changes in higher dimensions} \label{sec:multidim}

We now introduce a $d$-dimensional stochastic process $\bx(t)$ with components $x^\mu(t)$ for $\mu \in \llbracket 1, d\rrbracket$. Einstein summation convention is hereafter used throughout. 
The time evolution of the components of the stochastic vector $\bx(t)$ is governed by
\begin{align} \label{eq:multidim_langevin}
\frac{\dd x^\mu}{\dd t} \stackrel{\mathfrak{d}}{=} f^\mu(\bx) + g^{\mu i}(\bx) \eta_i \, ,
\end{align}
where the $\mathfrak{d}$ accounts for the underlying discretization scheme and where the index $i$ runs from $1$ to $n$ with 
$n\geq d$
%
%
 and the $n$-dimensional Gaussian white noise $\beeta(t)$ has mean and correlations
\begin{align}
& \left\langle \eta_i(t) \right\rangle = 0 \, , \qquad\qquad \left\langle \eta_i(t)\eta_j(t') \right\rangle = \delta_{ij}\delta(t-t') \, .
\end{align}
We shall later realize that the up or down position of the space index $\mu$ corresponds to a contravariant or covariant vector with respect to a given metric tensor. However, the position of the internal noise index $i$ bears no specific geometric meaning. In the $\alpha$-discretization scheme, we have
\begin{align}\label{eq:alphadiscrete}
\Delta x^\mu_k = x^{\mu}((k+1)\Delta t) - x^{\mu}(k\Delta t) = f^\mu(\bx_k) + g^{\mu i}\left(\bx_k + \alpha \Delta \bx_k\right)\Delta \eta_{i,k} \, ,
\end{align}
where the noise correlations are such that
\begin{align}
\left\langle \Delta \eta_{i,k} \right\rangle = 0 \, ,  \qquad\qquad \left\langle \Delta \eta_{i,k} \Delta \eta_{j,k'} \right\rangle = \Delta t \, \delta_{ij}\delta_{kk'} \, .
\end{align}
We furthermore introduce the $d \times d$ matrix with elements $\omega^{\mu\nu}(\bx)$ defined by
\begin{align}
\omega^{\mu\nu}(\bx) =  g^{\mu i}(\bx) g^{\nu j}(\bx) \delta_{ij} \,.
\label{eq:def-omega}
\end{align}
The substitution rules Eq.~\eqref{eq:substitution_first}-\eqref{eq:subsitution_second} are generalized as follows~\cite{mclaughlin_path_1971,weiss_operator_1978,gervais_point_1976,hirshfeld_canonical_1978,girotti_generalized_1983}:
\begin{align} \label{eq:subs_multi_first}
\Delta x^\mu \Delta x^\nu\doteq  \omega^{\mu\nu}(\bx) \, \Delta t \, ,  \qquad\qquad \Delta \eta_i \Delta \eta_j\doteq  \Delta t \, \delta_{ij} \, ,
\end{align}
and 
\begin{align}
\Delta x^\mu \Delta x^\nu \Delta x^\rho \Delta x^\sigma\doteq  
\left[\omega^{\mu\nu}(\bx)\omega^{\rho\sigma}(\bx) + \omega^{\mu\rho}(\bx)\omega^{\nu\sigma}(\bx) + \omega^{\mu\sigma}(\bx)\omega^{\nu\rho}(\bx)\right]
\Delta t^2 \, .
\label{eq:quarticsubstit}
\end{align}

\subsubsection{Changing linear discretization scheme or variables}
\label{sec:chang-line-discr}

From the rules that we have just identified, it is possible to derive the relations 
between $\alpha$-discretized and $\alpha'$-discretized stochastic differential equations, which read
\begin{align}
\frac{\dd x^\mu}{\dd t} & \stackrel{\alpha}{=} f^\mu + g^{\mu i} \eta_i  
\nonumber \\ & \stackrel{\alpha'}{=} f^\mu + \left(\alpha - \alpha'\right) g^{\nu i} \, \partial_\nu g^{\mu i} +  g^{\mu i} \eta_i \, , 
\end{align}
with $\partial_\nu  = \partial/\partial x^\nu $.

We now investigate the issue of differentiation and changes of variables. 
Let $\textbf{U}$ be a smooth invertible transformation between two open subsets of  $\mathbb{R}^d$ and define the process $\bu(t) = \textbf{U}(\bx(t))$. 
Assuming that $\bx$ evolves according to Eq.~\eqref{eq:multidim_langevin} understood as $\alpha$-discretized, we obtain
\begin{align} \label{eq:change_var}
\frac{\dd u^\mu}{\dd t} & 
\stackrel{\alpha}{=} \left[f^\rho \, \partial_\rho U^\mu\right]\left(\textbf{U}^{-1}(\bu)\right) + 
\left(\frac{1}{2} - \alpha\right) \left[ \omega^{\rho \sigma} \, \partial_\rho \partial_\sigma U^\mu \right]\left( \textbf{U}^{-1}(\bu) \right) 
\nonumber\\
& 
\quad
+ \left[g^{\rho i} \, \partial_\rho U^\mu \right]\left( \textbf{U}^{-1}(\bu) \right) \, \eta_i \, .
\end{align}
Therefore, as for one-dimensional systems, the rules of differential calculus hold in the Stratonovich scheme, $\alpha = 1/2$. In such a discretization, Eq.~\eqref{eq:change_var} shows that under a change of coordinates (that is, a reparametrization of the variables describing the system), $f^\mu$ and $g^{\mu i}$ transform as contravariant vectors do in Riemannian geometry regarding their $\mu$ index. Accordingly, $\omega^{\mu\nu}$ transforms as a rank-2 contravariant tensor. We will therefore borrow some of the language of Riemannian geometry to efficiently study transformations under a reparametrization of coordinates (see \textit{e.g.}~\cite{weinberg_gravitation_1972}).

Let us start by exploring how the Fokker--Planck description of the stochastic process transforms under  
reparametrization of coordinates. The Fokker--Planck equation associated to 
Eq.~\eqref{eq:multidim_langevin} in the Stratonovich discretization is 
\begin{align}
\partial_t P = - \partial_\mu \Big( f^\mu + \frac{1}{2}\sum_i g^{\nu i} \, \partial_\nu g^{\mu i} \Big) P + \frac{1}{2}\partial_\mu\partial_\nu \left(\omega^{\mu\nu}P\right) .
\end{align}
It is clear that $P(\bx,t)$ is not invariant under a change of coordinates. Indeed, the simplest scalar invariant object one can construct is the infinitesimal probability of finding the system in a box of size $\dd^d \bx$ around $\bx$ at time $t$,  that is written as
$\dd^d \bx \, P(\bx,t)$
(more formally, the Jacobians of the changes of variable in the volume element and in the probability density compensate).
 In what follows, we construct a scalar invariant probability density along the footsteps of~\cite{graham_covariant_1977}. This construction renders the connection to Riemannian geometry more explicit and it will serve as our starting point for constructing covariant path-integral representations of stochastic differential equations. Throughout this we work we assume that $\omega^{\mu\nu}(\bx)$ is invertible. This requires, in particular, that $n \geq d$ (where $n$ is the dimension of the noise space). The formulation of the path integral in cases in which the matrix $\omega^{\mu\nu}(\bx)$ is singular, for instance for an inertial Brownian particle in an external potential, was investigated in~\cite{grabert1980fluctuations}. Following standard conventions, we denote by
$\omega_{\mu\nu}(\bx)$ the inverse of $\omega^{\mu\nu}(\bx)$ ($\omega^{\mu\rho}\omega_{\rho\nu}=\delta^\mu_{\nu}$). The matrix $\omega_{\mu\nu}(\bx)$ is symmetric, positive-definite and transforms as a rank-2 covariant tensor under a change of coordinates. It can thus be promoted as the metric tensor of a $d$-dimensional Riemann space. Denoting by $\omega(\bx)$ the determinant of $\omega_{\mu\nu}(\bx)$, 
\begin{equation}
\omega(\bx) = \det \omega_{\mu\nu}(\bx)
\end{equation}
we construct the invariant volume element 
\begin{align}
\dd^d \bx \sqrt{\omega(\bx)}  
\end{align}
and we introduce
\begin{align}
K(\bx, t) = \frac{P(\bx, t)}{\sqrt{\omega(\bx)}}  \, ,
\end{align}
which is invariant under a change of coordinates, as follows from
\begin{equation}
 \dd^d \bx\,  P_{\bx}(\bx,t)
=
 \dd^d \bu\,  P_{\bu}(\bu,t)
\,,
\end{equation}
where the index refers to the process.
 It can be shown~\cite{graham_covariant_1977} that $K(\bx, t)$ evolves according to the manifestly covariant equation
\begin{align} \label{eq:covFP}
\partial_t K = - \nabla_\mu \left( h^\mu K \right) + \frac{1}{2}\nabla_\mu \nabla_\nu \left( \omega^{\mu\nu} K \right) ,
\end{align}
where $\nabla_\mu$ is the covariant derivative associated with the metric $\omega_{\mu\nu}$. While $\nabla_\mu=\partial_\mu$ when acting on a scalar, when applied to a contravariant vector $A^\nu$, one has
\begin{align}
\nabla_\mu A^\nu = \partial_\mu A^\nu + \Gamma^{\nu}_{\mu\rho}A^{{\rho}} \, ,
\end{align}
and when applied to a contravariant rank-2 tensor $T^{\rho\sigma}$:
\begin{align}
\nabla_\mu T^{\rho\sigma} = \partial_\mu T^{\rho\sigma} + \Gamma^{\rho}_{\mu\nu}T^{\nu\sigma} + \Gamma^{\sigma}_{\mu\nu}T^{\rho\nu} \, ,
\end{align}
where $\Gamma^{\mu}_{\rho\sigma}$ is the corresponding Christoffel symbol,
\begin{align}
\Gamma^{\mu}_{\rho\sigma} = \frac{1}{2}\omega^{\mu \nu}\left(\partial_\rho \omega_{\nu\sigma} + \partial_\sigma \omega_{\nu\rho} - \partial_\nu \omega_{\rho\sigma}\right) .
\end{align}
In addition, the vector $h^\mu$ appearing in Eq.~\eqref{eq:covFP} is defined by
\begin{align} \label{eq:defh}
h^\mu = f^\mu - \frac{1}{2} \, \Gamma^\rho_{\nu \rho}\omega^{\nu\mu} - \frac{1}{2} \partial_\nu g^{\nu i} g^{\mu j} \, \delta_{ij}  \, ,
\end{align}
which can be shown to transform as a contravariant vector under a change of coordinates.
(In dimension one, $h^\mu\mapsto f$).
We also introduce, as it will prove useful in many parts of this work, the Ricci scalar curvature associated to the $\omega_{\mu\nu}$ metric
\begin{align}\label{eq:Ricci}
R = \omega^{\mu\nu}\left(\partial_\eta \Gamma^{\eta}_{\mu\nu} - \partial_\mu \Gamma^{\eta}_{\eta \nu} + \Gamma^{\eta}_{\mu\nu}\Gamma^{\rho}_{\eta \rho} - \Gamma^{\eta}_{\mu\rho}\Gamma^{\rho}_{\eta \nu}\right) .
\end{align}
In this language, the stochastic dynamics with additive noise are associated to a flat space and a null Ricci curvature. Therefore, quite unlike the one-dimensional case, a multidimensional stochastic process with a nonzero Ricci curvature \textit{cannot} be mapped onto one with an additive form by a change of variables.

\subsubsection{Higher-order discretization schemes}
\label{sec:high-order-discr}

Nevertheless, 
similarly to the one-dimensional case, it is possible to generalize the non-linear $\alpha,\beta$ 
discretization scheme of Eq.~\eqref{eq:higher_order} to higher space dimensions. 
This is done by introducing a three index-quantity ${\B}^\mu_{\rho\sigma}(\mathbf{x}_k)$ in the definition of a \textit{quadratic} discretization rule
\begin{align}
\bar{x}_k^\mu = x_k^\mu + \alpha \Delta x_k^\mu + {\B}_{\rho\sigma}^\mu(\mathbf{x}_k) \Delta x_k^\rho \Delta x_k^\sigma \, .
\label{eq:defxbark_ddim}
\end{align}
The quantity  ${\B}^\mu_{\rho\sigma}$  is clearly symmetric with respect to the exchange of covariant indices $\rho, \sigma$.
The non-linear term in~\eqref{eq:defxbark_ddim} boils down to $\beta(x_k) \,\Delta x^2$ in the one-dimensional case of Eq.~\eqref{eq:higher_order}.
Such discretization schemes, that go one order in $\Delta x$ beyond the usual $\alpha$-discretization, 
will play an essential role in the discretized construction of the path-integral trajectory probability presented in Sec.~\ref{sec:explicit-covariant} and Sec.~\ref{sec:covLangevin}.
Although quadratic terms such as $\beta(x_k) \Delta x_k^2$ in one dimension or ${\B}_{\rho\sigma}^\mu(x_k) \Delta x_k^\rho \Delta x_k^\sigma $ in arbitrary dimensions play no role
in the continuous-time limit of the Langevin equation, it will become clear that a time-discretized interpretation of the path-integral action actually depends explicitly on them.
The stage is now set for investigating these additional subtleties that path integrals conceal.

\section{Path-integral representation of stochastic processes: conventional construction}\label{sec:pathint}

This section reviews the standard construction of path-integral representations for the transition probability of stochastic differential equations with Gaussian white noise. The simpler example of a one-dimensional process with additive noise is treated first. In line with the points raised in Sec.~\ref{sec:integration} for stochastic integrals, we show that one can resort to different discretizations to construct the path integral. We also recall the connection between the subtleties in the discretization of stochastic 
path integrals and the so-called operator ordering problem in quantum mechanics. We then turn to the more involved case of multi-dimensional processes with multiplicative noise for which we introduce the notion of covariant path-integral representation and construct explicitly the path integral in the linear $\alpha$-discretization. At the end of this section, we show that the rules of differential calculus, while known to hold at the level of Stratonovich-discretized stochastic differential equations, are not adequate for changing variables at the level of $\alpha$-discretized continuous-time path integrals, including the Stratonovich case $\alpha = 1/2$.
Therefore, an improved construction is needed and this will be the theme of sections~\ref{sec:ItocalculusPI} to~\ref{sec:covLangevin}. 

\subsection{The one-dimensional additive case}
\label{subsec:pi_additive}

We start by implementing our program on the example of a one-dimensional stochastic process with additive noise
\begin{align}
\frac{\dd x}{\dd t} = f + \eta \, .
\label{eq:langevin_additive}
\end{align}
We recall our notation so as to make the presentation in this section as self-contained as possible. 
We divide the interval $[t_0,\tf]$ into $N$ slices and we introduce the intermediate times $t_k = t_0 + k \Delta t$ with $k=0,\ldots,N$, $\Delta t = (\tf - t_0)/N$ and $t_N = \tf$. 
Clearly, the initial value is $x_0=x(t_0)$ and the final one $x_N=x(t_N) = x(\tf)$. 
The discrete-time companion process with time step $\Delta t$ associated to Eq.~\eqref{eq:langevin_additive}
is
\begin{align} \label{eq:additive_discrete}
\Delta x_k = f(x_k) \Delta t + \Delta \eta_k \, .
\end{align}

\subsubsection{The propagator in the Itō convention}

Let $\mathbb{P}(\xf, \tf | x_0, t_0)$ be the propagator associated to 
the Langevin Eq.~\eqref{eq:langevin_additive},
\textit{i.e.}~the probability to be at $\xf$ at time $\tf$ given that the motion starts from $x_0$ at time $t_0$, 
and $\mathbb{P}_{\Delta t}(x_{k+1}, t_{k+1} | x_k,t_k)$ the
one-step infinitesimal propagator associated to Eq.~\eqref{eq:additive_discrete}.
Relying on the fact that the processes described here are Markovian, we use the Chapman--Kolmogorov equation over the intermediate time windows $[t_k,t_{k+1}]$ to state 
\begin{align}
\mathbb{P}(\xf, \tf | x_0,t_0) = \lim_{N \rightarrow + \infty} \int \prod_{k=1}^{N-1} \dd x_k\prod_{k=0}^{N-1} \mathbb{P}_{\Delta t}(x_{k+1},t_{k+1} | x_{k},t_{k})~.
\label{eq:time-sliced-P}
\end{align}
In Eq.~\eqref{eq:additive_discrete}, the $\Delta \eta_k$ are independent and identically distributed random variables with a normal probability density
\begin{align}
P \left(\Delta \eta_k \right) =  \frac{1}{\sqrt{2 \pi \Delta t}} \, 
\ee^{\text{$\displaystyle-\frac{\Delta \eta_k^2}{ 2 \Delta t}$}} \, ,
\end{align}
so that the one-step propagator reads
\begin{align}\label{eq:1dItoinfprop}
\mathbb{P}_{\Delta t}(x_{k+1},t_{k+1} | x_{k},t_{k}) = \frac{1}{\sqrt{2 \pi \Delta t}}
\exp\left[-\frac{1}{2 \Delta t} \big(\Delta x_k - f(x_k) \Delta t\big)^2 \right] .
\end{align}
Such an expression corresponds to the Itō discretization of the argument of the exponential.
This result leads us to write the finite-time propagator as 
\begin{align}
\mathbb{P}(\xf,\tf | x_0,t_0) & = \lim_{N\rightarrow +\infty} 
\frac{1}{\sqrt{2 \pi \Delta t}}
\int \prod^{N-1}_{k=1} \left(\frac{\dd x_k}{\sqrt{2 \pi \Delta t}}\right) \exp\left[- \frac{\Delta t}{2} \sum_{k=0}^{N-1} \left(\frac{\Delta x_k}{\Delta t} - f(x_{k})\right)^2\right]
\nonumber \\
& 
= \int_{x(t_0)=x_0}^{x(\tf)=\xf} \mathcal{D}x \, \ee^{-\mathcal{S}[x(t)]} ~. 
\label{eq:pi_additive_ito}
\end{align}
The meaning of the formal continuous-time path integral is inferred from the limiting  discrete-time form:
the path measure is expressed as
\begin{align} \label{eq:path_measure_additive}
\mathcal{D}x = \lim_{N\rightarrow +\infty} 
\frac{1}{\sqrt{2 \pi \Delta t}} 
\prod^{N-1}_{k=1} \left(\frac{\dd x_k}{\sqrt{2 \pi \Delta t}}\right) ,
\end{align}
and the continuous-time action $\mathcal{S}[x(t)]$ as
\begin{align} \label{eq:action_additive}
\mathcal{S}[x(t)] \stackrel{0}{=} \,\, \frac{1}{2}\int_{t_0}^{\tf} \dd t \, \left( \frac{\dd x}{\dd t} - f(x) \right)^2 
\, .
\end{align}
The superscript $0$ above the equal sign is meant to specify the underlying time discretization, which is the Itō one as read from~(\ref{eq:pi_additive_ito}). 
The boundary conditions appearing in the path-integral sign in Eq.~\eqref{eq:pi_additive_ito} indicate that the sum is performed on continuous paths constrained to satisfy them.

\subsubsection{Changing discretization}\label{sec:changediscrPI}

Exactly in the same way as any integral observable can be expressed using different discretizations, 
and can thus correspond to visually different continuous-time expressions (see \textit{e.g.}~Eq.~\eqref{eq:integral_change_var}), we wish to rewrite the path-integral action  associated with the process in Eq.~\eqref{eq:langevin_additive} not only as the limit of an Itō discretized sum but as the limit of an $\alpha$-discretized one with $\alpha \in [0,1]$. This will concretely illustrate that the underlying discretization scheme must be prescribed to be able  
to work with path integrals in an unambiguous way (see~\cite{hertz2016path} for a review of the handling of path integrals in different discretizations). 
We start from the expression of the propagator in Eq.~\eqref{eq:pi_additive_ito} where we rewrite the action~(\ref{eq:action_additive}) by isolating the kinetic term $\int_{t_0}^{\tf}\dd t \, \dot{x}^2$ so as to obtain
\begin{align}
& \mathbb{P}(\xf,\tf | x_0,t_0)  = \left\langle \exp\left(-\frac{1}{2}\int_{t_0}^{\tf} \dd t \; f(x)^2 + 
\int_{\substack{
   \!\!\!\!\! t_0 \\
   \scriptstyle{\alpha=0}}
  }^{\tf} \dd t \; \frac{\dd x}{\dd t} \, f(x)\right) \right\rangle ,
\label{eq:Paveexp}
\end{align}
where the average is taken with respect to the free Brownian motion constrained to satisfy the boundary conditions $x(t_0)=x_0$ and $x(\tf)=\xf$. The subscript $\alpha = 0$ in the integral sign means that it is an Itō one. %
Within the expectation value, we can use Eq.~\eqref{eq:integral_change_var} (with $g=1$ since the process is additive) to
change discretization and
 replace the Itō integral by an $\alpha$-discretized one\footnote{
Importantly, when writing the time-discretized version of the integrals~\cite{Cugliandolo-Lecomte17a}, we see that the difference between the left- and right-hand-side of Eq.~\eqref{eq:chdiscrcontint} is of order $\sqrt{\Delta t}$,
validating the use of Eq.~(\ref{eq:chdiscrcontint}) in the exponential of~\eqref{eq:Paveexp}.
}:
\begin{equation}
\int_{\substack{
   \!\!\! t_0 \\
   \alpha=0}
  }^{\tf} \dd t \; \frac{\dd x}{\dd t} \, f(x) = 
  \int_{\substack{
   \, t_0 \\
   \scriptstyle{\alpha}}
  }^{\tf}  \dd t \; \frac{\dd x}{\dd t} \, f(x) - \alpha \int_{t_0}^{\tf} \dd t \; f'(x) \,.
\label{eq:chdiscrcontint}
\end{equation}
Therefore, the action in Eq.~\eqref{eq:action_additive} can be rewritten as
\begin{align} \label{eq:action_additive_alpha}
\mathcal{S}[x(t)] \stackrel{\alpha}{=} \,\, \int_{t_0}^{\tf} \dd t \, \left[\frac{1}{2}\left( \frac{\dd x}{\dd t} - f(x) \right)^2 + \alpha f'(x)\right] ,
\end{align}
where the superscript $\alpha$ in the continuous expression stands for $\alpha$-discretized integral.
Compared to the Itō-discretized action~(\ref{eq:action_additive}), we have to include a contribution $\alpha f'(x)$ in a generic 
$\alpha$-discretization.
This illustrates the known fact that~\cite{langouche1978functional,Langouche79,janssen_renormalized_1992}, \textit{even for an additive Langevin equation} (the writing of which 
is independent of the discretization convention), the path-integral action depends explicitly on the discretization scheme chosen to write it.

\subsubsection{Path integrals and operator ordering}

For the sake of completeness, we present another derivation of the path-integral representation of the transition probability $\mathbb{P}(\xf, \tf | x_0, t_0)$. This approach, 
which uses the Fokker--Planck equation as its starting point,
emphasizes the equivalence between the issue of the discretization of path integrals and the operator ordering problem in quantum mechanics, as extensively discussed in~\cite{chaichian_stochastic_2001, Tirapegui82}.

Using the standard momentum operator $\hat{p}=-i\frac{\dd}{\dd x}$ of quantum mechanics, we write the Fokker--Planck equation associated to Eq.~\eqref{eq:langevin_additive} as
\begin{align}\label{eq:FP-op}
\partial_t P = - \hat{\text{H}}_{\rm FP} P \qquad\qquad
\mbox{with}
\qquad\qquad
\hat{\text{H}}_{\rm FP} = \frac{1}{2}\hat{p}^2 + i \hat{p} \hat{f}
\; ,
\end{align}
which is naturally expressed in terms of a normal ordering of the non-commuting operators $\hat{p}$ and $\hat{f}$. Alternatively, the Fokker-Planck operator $\hat{\text{H}}_{\rm FP}$ can be written by using an $\alpha$-ordering of the operators $\hat{p}$ and $\hat{f}$ in which the products $\hat{p} \hat{f}$ and $\hat{f} \hat{p}$ only appear through the combination $\alpha \hat{f}\hat{p} + (1-\alpha)\hat{p}\hat{f}$. This leads to,
\begin{equation}\label{eq:ordering_alpha}
\hat{\text{H}}_{\rm FP} = \frac{\hat{p}^2}{2} + i \left( (1-\alpha)\hat{p}\hat{f} + \alpha \hat{f}\hat{p}\right) + \alpha \hat{f}' \,.
\end{equation}
The well-known Weyl ordering corresponds to the symmetric $\alpha=1/2$ case. When substituting the non-commuting operators $\hat{p}$ and $\hat{f}$ by commuting c-numbers, it is clear from Eq.~\eqref{eq:ordering_alpha} that different orderings lead to different c-Hamiltonian. In the following, we show that the c-Hamiltonian inferred from the $\alpha$-ordering of the Fokker-Planck operator $\hat{\text{H}}_{\rm FP}$ naturally appears in the $\alpha$-discretized action of the path integral representation of Eq.~\eqref{eq:FP-op}. Indeed, introducing the usual position eigenstates $\ket{x}$, 
the infinitesimal propagator can be expressed as
\begin{align}
\mathbb{P}_{\Delta t}(x + \Delta x,t+\Delta t | x,t) 
& = \bra{x + \Delta x }\text{e}^{-\Delta t\left(\hat{p}^2/2 + i \hat{p}\hat{f}\right)} \ket{x} 
\nonumber \\
& \simeq \bra{x + \Delta x } \text{e}^{-\Delta t \,\hat{p}^2/2}\text{e}^{-i \Delta t \hat{p}\hat{f}} \ket{x} ~,
\end{align}
where the last line is obtained using the Baker--Campbell--Hausdorff formula to first order, which is enough to correctly capture the continuous-time limit.
We recall (and this is proven in Sec.~\ref{sec:covLangevin}) that for any function $G$
\begin{align} \label{eq:commuteop}
\left[ \exp\left(f(x)\Delta t \frac{\dd}{\dd x}\right)G\right] (x) = G \left[\exp\left(f(x)\Delta t \frac{\dd}{\dd x}\right) \right] (x) \, ,
\end{align}
which in the present case translates into
\begin{align}
\exp\left(-i \hat{p} \hat{f} \Delta t\right)\ket{x} = \ket{\exp\left(f(x)\Delta t \frac{\dd}{\dd x}\right)x} = \ket{x + f(x)\Delta t + O(\Delta t^2)} \, .
\end{align}
We therefore obtain
\begin{align}
\mathbb{P}_{\Delta t}(x + \Delta x,t+\Delta t | x,t) & \simeq \bra{x + \Delta x} \text{e}^{-\Delta t \, \hat{p}^2/2} \ket{x + f(x)\Delta t +...} 
\nonumber \\
& \simeq \frac{1}{\sqrt{2 \pi \Delta t}}\exp\left[-\frac{\Delta t}{2}\left(\frac{\Delta x}{\Delta t} - f(x)\right)^2\right] \, ,
\end{align}
where the last expression comes from the Fourier expansion of position eigenstates $\ket{x}$ along the momentum eigenstates $\ket{p}$. This expresses that a $\hat{p}\hat{x}$ or normal ordering of the Hamiltonian naturally leads to an Itō discretized path integral (see Eq.~(\ref{eq:1dItoinfprop})). 

We could alternatively choose to order the operators differently by following an $\alpha$ ordering, 
\begin{align}
& 
\mathbb{P}_{\Delta t}(x + \Delta x,t+\Delta t | x,t) 
\nonumber\\
& 
\qquad
=
 \bra{x + \Delta x }\exp\left(-\frac{\hat{p}^2}{2} \Delta t - i \hat{p}\hat{f}\Delta t\right) \ket{x}  
\nonumber \\
& 
\qquad
=
\bra{x + \Delta x }\exp\left[-\frac{\hat{p}^2}{2} \Delta t - i \left( (1-\alpha)\hat{p}\hat{f} + \alpha \hat{f}\hat{p}\right) \Delta t - \alpha \hat{f}'\Delta t\right] \ket{x} .
\nonumber 
\end{align}
At this stage we can use the approximation
\begin{align}
& 
\mathbb{P}_{\Delta t}(x + \Delta x,t+\Delta t | x,t) 
\nonumber\\
& 
\qquad
\simeq
\ee^{-\alpha f'(x)\Delta t}
 \bra{x + \Delta x }
\ee^{-i \alpha \Delta t \hat{f}\hat{p}}
\; 
\ee^{-\frac{\hat{p}^2}{2} \Delta t} 
\; 
 \ee^{-i (1 - \alpha) \Delta t \hat{p}\hat{f}}
  \ket{x} 
\nonumber \\
& 
\qquad
=
\ee^{-\alpha f'(x)\Delta t}
\bra{x + \Delta x - \alpha \Delta t f(x+\Delta x) +... } \ee^{-\frac{\hat{p}^2}{2} \Delta t}   \ket{x + (1-\alpha) \Delta t f(x) + ...} 
\nonumber \\
& 
\qquad
=
 \frac{1}{\sqrt{2 \pi \Delta t}}\exp\left[-\alpha f'(x)\Delta t -\frac{\Delta t}{2}\left(\frac{\Delta x}{\Delta t}-f(x+\alpha \Delta x)\right)^2\right] ~.
\label{eq:infinit_prop}
\end{align}
Thus, the $\alpha$ ordering of the product $\hat{p}\hat{f}$ naturally leads to an $\alpha$-discretized path integral, see Eq.~(\ref{eq:action_additive}). In particular, the well known Weyl ordering $\alpha=1/2$ yields a Stratonovich-discretized path integral.

\subsection{Multidimensional processes with multiplicative noise}

We now turn to the more general multidimensional processes with multiplicative noise
\begin{align} \label{eq:pi_langevin_multi}
\frac{\dd x^\mu}{\dd t} \stackrel{\frac 12}{=} f^\mu(\bx) + g^{\mu i}(\bx) \eta_i \, .
\end{align}
Without loss of generality, Eq.~\eqref{eq:pi_langevin_multi} is understood as Stratonovich-discretized. Furthermore, the matrix with elements $g^{\mu i}(\bx)$ is hereafter assumed to be invertible, with inverse $g_{i \mu}(\bx)$. This imposes the necessary condition $d = n$. In the same vein as the presentation above, we introduce the propagator $\mathbb{P}(\bxf,\tf|\bx_0,t_0)$ of Eq.~\eqref{eq:pi_langevin_multi} and its discrete-time companion process defined with a time step of duration $\Delta t = (\tf - t_0)/N$, $N \in \mathbb{N}$. Accordingly, we introduce the scalar invariant propagators of the original process
\begin{align}
\label{eq:KPomega}
\K(\bxf,\tf|\bx_0,t_0) = \frac{\mathbb{P}(\bxf,\tf|\bx_0,t_0)}{\sqrt{\omega(\bxf)}} \, ,
\end{align}
and of the infinitesimal discrete process
\begin{align}
\K_{\Delta t}(\bx_{k+1},t_{k+1}|\bx_k,t_k) = \frac{\mathbb{P}_{\Delta t}(\bx_{k+1},t_{k+1}|\bx_k,t_k)}{\sqrt{\omega(\bx_{k+1})}}
\, .
\end{align}
Explicitly, the advantage of using such scalar invariant propagators is that, when changing variables as $\bu(t)=\textbf{U}(\bx(t))$ 
\begin{equation}
  \label{eq:jacchvarddim}
  \mathbb P_{\bx}(\bxf,\tf|\bx_0,t_0)
=
  \left|\det \frac{\partial U^\mu}{\partial x^\nu}(\bxf)\right|
  \mathbb P_{\bu}(\buf,\tf|\bu_0,t_0)
\end{equation}
(we indicated with a subscript the variable of the probability density),
the Jacobian prefactor of the change of probability is \textit{exactly} compensated, for $\K$, by the transformation of the prefactor $1/\sqrt{\omega(\bxf)}$ in Eq.~(\ref{eq:KPomega}).\footnote{
Indeed, denoting $G^{\mu i}$ and $\sqrt{\Omega}$ the noise amplitude and the volume measure of the process $\bu(t)$, we have $G^{\mu i}=\frac {\partial U^\mu}{\partial x^\nu} g^{\nu i}$
and thus $\sqrt{\omega(\bxf)} = \big|\det \frac{\partial U^\mu}{\partial x^\nu}(\bxf)\big| \sqrt{\Omega(\buf)}$.
}

We then write a time-sliced expression for the propagator $\mathbb{P}(\bxf,\tf|\bx_0,t_0)$, just as 
in Eq.~(\ref{eq:time-sliced-P}) but now extended to a vectorial process $\bx$. This  implies
\begin{align}
\K(\bxf,\tf|\bx_0,t_0) & 
= \lim_{N \to + \infty} \int \prod_{k=1}^{N-1}\left\{ \dd \bx_k \sqrt{\omega(\bx_k)}\right\} \prod_{k=0}^{N-1} \left\{\frac{\mathbb{P}_{\Delta t}(\bx_{k+1},t_{k+1}|\bx_{k},t_{k})}{\sqrt{\omega(\bx_{k+1})}}\right\}
\nonumber\\
& 
= \lim_{N \to + \infty} \int \prod_{k=1}^{N-1}\left\{ \dd \bx_k \sqrt{\omega(\bx_k)}\right\} \prod_{k=0}^{N-1} \K_{\Delta t}(\bx_{k+1},t_{k+1}|\bx_{k},t_{k}) 
\nonumber \\ 
\label{eq:KPIcont}
& 
= \int_{\bx(t_0)=\bx_0}^{\bx(\tf)=\bxf} \mathcal{D}\bx \, \,  \ee^{-\mathcal{S}[\bx(t)]}~. 
%
\end{align}
The meaning of the formal continuous-time path integral~\eqref{eq:KPIcont} is inferred from the limiting  of the corresponding discrete-time expressions of the path measure
\begin{align} \label{eq:path_measure}
\mathcal{D}\bx = \lim_{N \to + \infty} 
\left(\frac{1}{\sqrt{2\pi\Delta t}}\right)^{\!\!{d}}\;
 \prod_{k=1}^{N-1}\left\{ \frac{\dd \bx_k \sqrt{\omega(\bx_k)}}{\sqrt{2\pi\Delta t}^{\, d}}\right\}  ,
\end{align}
and of the action
\begin{align} \label{eq:action}
\ee^{-\mathcal{S}[\bx(t)]} = \lim_{N \to + \infty} \prod_{k=0}^{N-1} \left(2\pi\Delta t\right)^{\frac{d}{2}} \, \K_{\Delta t}(\bx_{k+1},t_{k+1}|\bx_{k},t_{k}) \, .
\end{align}

Note that different definitions of $\mathcal{D}\bx$ and $\mathcal{S}[\bx(t)]$ could give the same result in the $\Delta t \to 0$ limit, only the product $ \mathcal{D}\bx  \,  \ee^{-\mathcal{S}[\bx(t)]}$  
being prescribed.\footnote{%
The prefactor $\left(2\pi\Delta t\right)^{\frac{d}{2}}$ in~(\ref{eq:action}) allows one to eliminate a constant pre-exponential factor in $\K_{\Delta t}$.
} It is thus not uncommon  to find different discretizations of the path measure in the literature (see \textit{e.g.}~\cite{Langouche81, omote1977point}). The precise definition of $\mathcal{D}\bx$ in Eq.~\eqref{eq:path_measure}, with the function $\omega$ being evaluated at~$\bx_k$, is however the only one providing a scalar invariant path measure $\mathcal{D}\bx$ (up to a global constant factor). 
The convention~(\ref{eq:path_measure}) for $\mathcal{D}\bx$, which we adopt in this section, is very convenient since it allows one to focus solely on the transformation of the action when changing variables. 
In the previous section where the noise was additive, the discretization of the path measure $\mathcal{D}x$ in Eq.~\eqref{eq:path_measure_additive} was not an issue. 

Different discretizations can be used to represent the same action functional $\mathcal{S}[\bx(t)]$, as was illustrated by our study of the one-dimensional additive case in Sec.~\ref{subsec:pi_additive}. For a given discretization $\mathfrak{d}$, we write the action as the $\mathfrak{d}$-discretized integral of a Lagrangian 
\begin{align}
\mathcal{S}[\bx(t)] \stackrel{\mathfrak{d}}{=} \int_{t_0}^{\tf}\dd t \,\, \mathcal{L}_{\mathfrak{d}}[\textbf{x}(t),\dot{\textbf{x}}(t)] \, ,
\label{eq:SdiscLdisc}
\end{align}
where the $\mathfrak{d}$ subscript in $\mathcal{L}_{\mathfrak{d}}$ accounts for the fact that the functional form of the Lagrangian depends on the underlying discretization. 
We stress that the value of the action in~\eqref{eq:SdiscLdisc} is the \textit{same} for every discretization scheme~$\mathfrak d$, but that the continuous-time expression of $\mathcal{L}_{\mathfrak{d}}$
as a function of $\bx$ and $\dot\bx$
 will in general depend on the chosen scheme~$\mathfrak d$.

\subsection{Covariant path-integral representation of stochastic processes}
\label{sec:covar-path-integr}

In this section we define a central concept, that of a covariant path-integral representation for multidimensional stochastic differential equations with multiplicative noise.

 Let $\mathbb{P}_\bx (\bxf,\tf|\bx_0,t_0)$ 
be the transition probability density of being at $\bxf$ at  $\tf$ while being at $\textbf{x}_0$ at $t_0$. We consider time-discretized path-integral representations of this probability within a given discretization scheme $\mathfrak{d}$, written as%
\footnote{
\,In Eq.~\eqref{eq:PxPIgenforcov}, the convention we use is to have a prefactor $\sqrt{\omega(\bxf)}$ that encodes the fact that the propagator $\K$ of Eq.~(\ref{eq:KPomega}) transforms like a scalar,
and that both the measure $\mathcal{D}\bx$ and the action possess this same property, see Eqs.~(\ref{eq:DxDu}) and~(\ref{eq:scalar_lagrangian}).
Equivalently to~(\ref{eq:PxPIgenforcov}), one can also write that the \textit{probability} that $\bxf$ belongs to a domain $\mathcal X$ at time $\tf$ is
\begin{equation*}
\text{Prob}(\bxf\in\mathcal X, \tf| \bx_0, t_0)
\stackrel{\mathfrak{d}}{=} 
\int_{\textbf{x}(t_0) = \textbf{x}_0}^{\textbf{x}(\tf) \in \mathcal X } \mathcal{D}\textbf{x} ~ \exp\left\{-\int_{t_0}^{\tf} \dd t ~ \mathcal{L}_{\mathfrak{d}}^{\bx}[\textbf{x}(t),\dot{\textbf{x}}(t)] \right\} .
\end{equation*}
}
\begin{align}
\label{eq:PxPIgenforcov}
\mathbb{P}_\bx(\bxf,\tf|\bx_0,t_0)
\stackrel{\mathfrak{d}}{=} 
\sqrt{\omega(\bxf)}
\int_{\textbf{x}(t_0) = \textbf{x}_0}^{\textbf{x}(\tf) = \bxf} \mathcal{D}\textbf{x} ~ \exp\left\{-\int_{t_0}^{\tf} \dd t ~ \mathcal{L}_{\mathfrak{d}}^{\bx}[\textbf{x}(t),\dot{\textbf{x}}(t)] \right\} 
\end{align}
where the superscript $\bx$ stresses that this Lagrangian is associated to the variable $\bx$. We are free to reparametrize the phase space and define new variables $\textbf{u}$ through $\textbf{u} = \textbf{U}(\textbf{x})$ where $\textbf{U}$ is a 
smooth invertible transformation between two open subsets of  $\mathbb{R}^d$
 Using this new parametrization, the transition probability can be constructed accordingly:
\begin{align}
\mathbb{P}_\bu(\buf, \tf| \bu_0, t_0)
\stackrel{\mathfrak{d}}{=}
\sqrt{\Omega(\buf)}
 \int_{\textbf{u}(t_0) = {\bf U}(\textbf{x}_0) = \textbf{u}_0}^{\textbf{u}(\tf) =  {\bf U}(\bxf)= \buf} \mathcal{D}\textbf{u} ~ 
\exp\left\{ -\int_{t_0}^{\tf} \dd t ~ \mathcal{L}_{\mathfrak{d}}^{{\bf u}}[\textbf{u}(t),\dot{\textbf{u}}(t)]\right\} .
\end{align}

A path-integral representation is said to be \textit{covariant} if the following two conditions are fulfilled. Firstly, we must have
\begin{align}
\label{eq:DxDu}
\mathcal{D}\textbf{x} = \mathcal{D}\textbf{u} ~,
\end{align}
which expresses that the measure is a scalar invariant under changes of coordinates, consistently with our construction. Secondly, we require that
\begin{align}
\mathcal{L}^{\bx}_{\mathfrak{d}}[\textbf{x},\dot{\textbf{x}}] 
= 
\mathcal{L}^{\bu}_{\mathfrak{d}}[\textbf{u}, \dot{\textbf{u}}]
=
\mathcal{L}^{\bu}_{\mathfrak{d}}[\textbf{U}(\textbf{x}),(\partial \textbf{U}/\partial \textbf{x}) \cdot \dot{\textbf{x}}] ~,
\label{eq:scalar_lagrangian}
\end{align}
which means that one is free to use the standard chain rule directly at the level of the continuous-time Lagrangian. Stated otherwise, a discretization $\mathfrak{d}$ of the path integral is covariant if and only if the associated Lagrangian $\mathcal{L}^{\bx}_{\mathfrak{d}}[\textbf{x},\dot{\textbf{x}}]$ is manifestly covariant. 

In the following section we will compute $\mathcal{L}^{\bx}_{\alpha}[\textbf{x},\dot{\textbf{x}}]$, the continuous-time $\alpha$-discretized Lagrangian for all $\alpha \in [0,1]$. A rapid visual inspection will then show us that \textit{none} 
of these linear $\alpha$-discretizations, not even the Stratonovich one (which is nevertheless adapted to the use of the chain rule at the level of stochastic differential equations), are covariant: changing variables by applying blindly the chain rule at the level of the continuous-time Lagrangian is not an option in these $\alpha$-discretizations. We will then show in Secs.~\ref{sec:explicit-covariant} and~\ref{sec:covLangevin} that well-chosen 
higher-order discretizations can cure this problem --~which constitutes the main new result in this article.

\subsection{The $\alpha$-discretized path integral}\label{sec:pi_alpha_discrete}

\subsubsection{Statement of the result}
\label{sec:statement-resultPIalpha}

In this section, we construct the $\alpha$-discretized path-integral representation of the transition probability of Eq.~\eqref{eq:pi_langevin_multi}. Let us start by computing, in the limit $\Delta t \rightarrow 0 $ and up to $O(\Delta t)$ terms, the one-step  scalar propagator $\K_{\Delta t}(\bx_{k+1},t_{k+1}|\bx_{k},t_{k})$ associated to the Stratonovich-discretized equation:
\begin{align}\label{eq:pi_langevin_strato_discrete}
\Delta x^\mu_k = f^\mu(\bx_k)\Delta t + g^{\mu i}\left(\bx_k + \frac{\Delta \bx_k}{2}\right)\Delta \eta_{i,k} \, .
\end{align}
The propagator is straightforwardly given by
\begin{align} \label{eq:inf_scalar_prop}
\left(2\pi\Delta t\right)^{\frac{d}{2}} \K_{\Delta t}(\bx_{k+1},t_{k+1}|\bx_{k},t_{k}) = \left(2\pi\Delta t\right)^{\frac{d}{2}} \left|\det\left( \frac{\partial \Delta \eta_{i, k}}{\partial \Delta x_k^\sigma}\right)\right| \frac{ P[\Delta \boldsymbol{\eta}_{k}]}{\sqrt{\omega(\bx_{k+1})}} \, ,
\end{align}
with $P[\Delta \boldsymbol{\eta}_{k}]$ the probability distribution of the noise increments
\begin{align}
P[\Delta \boldsymbol{\eta}_{k}] = \left(\frac{1}{2\pi \Delta t}\right)^{\frac{d}{2}} \exp\left(- \frac{\Delta \eta_{i,k} \Delta \eta_{j,k} \delta^{ij}}{2 \Delta t}\right) ,
\end{align}
and where in~(\ref{eq:inf_scalar_prop}) $\Delta \eta_{i,k}$ is understood as expressed in terms of the $x^\mu_{k'}$'s, as directly obtained from~(\ref{eq:pi_langevin_strato_discrete}).
The point at which we choose to evaluate the functions appearing in $\K_{\Delta t}(\bx_{k+1},t_{k+1}|\bx_{k},t_{k})$ 
sets the discretization scheme of the action $\mathcal{S}[\bx(t)]$. As we will show next, we can infer from Eq.~\eqref{eq:inf_scalar_prop} the expression of the $\alpha$-discretized Lagrangian for any $\alpha \in [0,1]$. 
Using the notations introduced in Sec.~\ref{sec:multidim}, it  reads
\begin{align}
\label{eq:Lalpha}
\!\!\! \mathcal{L}^{\bx}_{\alpha}[\textbf{x},\dot{\textbf{x}}] 
 &
  = 
   \frac{1}{2}\left[\omega_{\mu\nu}\left(\frac{\dd x^\mu}{\dd t} - h^\mu\right)\left(\frac{\dd x^\nu}{\dd t} - h^\nu\right) + \left(1-2\alpha\right)\frac{\dd x^\mu}{\dd t}\left(\omega_{\mu\nu}\omega^{\rho\sigma}\Gamma_{\rho\sigma}^\nu + 2 \Gamma^\alpha_{\mu\alpha}\right)  \right. 
\nonumber\\ 
& \;\;
\left. + 2 \alpha \nabla_\mu h^\mu - (1-2\alpha)\omega_{\mu\nu}\omega^{\rho\sigma}\Gamma^{\nu}_{\rho\sigma}h^\mu + \left(\alpha - \frac{1}{2}\right)^2\omega_{\mu\nu}\omega^{\rho\sigma}\omega^{\alpha\beta}\Gamma_{\rho\sigma}^\mu \Gamma_{\alpha\beta}^\nu  \right. 
\nonumber\\
 &  \;\;
 - \alpha\left(1-\alpha\right)R + \alpha\left(1-\alpha\right)\omega^{\mu\nu}\Gamma_{\beta\mu}^\alpha \Gamma_{\alpha\nu}^\beta + \left(1-3\alpha(1-\alpha)\right)\omega^{\mu\nu}\partial_\nu \Gamma^\alpha_{\mu\alpha} \Bigg] \, ,
\end{align}
with $h^\mu$ defined in Eq.~(\ref{eq:defh}).
For the sake of completeness, we explicitly write here the  Itō-discretized Lagrangian,
\begin{align}
\label{eq:Lito}
\mathcal{L}^{\bx}_{0}[\textbf{x},\dot{\textbf{x}}] 
= & \, \frac{1}{2}\left[\omega_{\mu\nu}\left(\frac{\dd x^\mu}{\dd t} - h^\mu\right)\left(\frac{\dd x^\nu}{\dd t} - h^\nu\right) + \frac{\dd x^\mu}{\dd t}\left(\omega_{\mu\nu}\omega^{\rho\sigma}\Gamma_{\rho\sigma}^\nu + 2 \Gamma^\alpha_{\mu\alpha}\right) \right. 
\nonumber \\
& \left. 
\qquad 
- \omega_{\mu\nu}\omega^{\rho\sigma}\Gamma^{\nu}_{\rho\sigma}h^\mu 
+ \frac{1}{4}\omega_{\mu\nu}\omega^{\rho\sigma}\omega^{\alpha\beta}\Gamma_{\rho\sigma}^\mu \Gamma_{\alpha\beta}^\nu 
+ \omega^{\mu\nu}\partial_\nu \Gamma^\alpha_{\mu\alpha} \right] \,,
\end{align}
and the Stratonovich-discretized one,
\begin{align} \label{eq:Lstrato}
\mathcal{L}^{\bx}_{1/2}[\textbf{x},\dot{\textbf{x}}] = 
& \,  \frac{1}{2}\left[\omega_{\mu\nu}\left(\frac{\dd x^\mu}{\dd t} - h^\mu\right)\left(\frac{\dd x^\nu}{\dd t} - h^\nu\right) + \nabla_\mu h^\mu - \frac{1}{4}R + \frac{1}{4}\omega^{\mu\nu}\Gamma_{\beta\mu}^\alpha \Gamma_{\alpha\nu}^\beta 
\right.
\nonumber\\
& 
\left.
\qquad 
+ \frac{1}{4}\omega^{\mu\nu}\partial_\nu \Gamma^\alpha_{\mu\alpha} \right] \,.
\end{align}
Before diving into the details of the computation of the Lagrangian in Eq.~\eqref{eq:Lalpha}, note that there is no $\alpha$ such that its expression is manifestly covariant. 
This confirms our earlier claim that the standard chain rule of differential calculus cannot  be blindly used to change variables at the level of an $\alpha$-discretized continuous-time action, 
even if it is Stratonovich-discretized\footnote{
Note that, as will be discussed at the end of Sec.~\ref{subsec:ito_plus12}, choosing a different convention for the discretization of the path-integral measure,
one can arrive at a different expression of the Lagrangian in which applying the stochastic chain rule in the Itō and in the Hänggi--Klimontovich case leads to a correct computation,
as recently shown in Ref.~\cite{ding2021timeslicing}.
}. 

To make contact with existing results in one dimension,
note that in~(\ref{eq:Lalpha}) the covariant derivative $\nabla_\mu $ acts on the \textit{vector} $h^\mu$ (the covariant force defined in Eq.~\eqref{eq:defh}),
so that, for $d=1$, $\nabla_\mu h^\mu \mapsto f'-f {g'}/{g}$  (which, as expected, is equal to $F'-F {G'}/{G}$ for parameters $F\circ U = U' f$ and $G\circ U =U' g$ of the process $u(t)=U(x(t))$).
One obtains:
\begin{align}
\mathcal{L}^{x}_{\alpha}[{x},\dot{{x}}] 
&=
\frac 12
\frac {1}{g(x)^2}
\bigg(\frac{\dd x}{\dd t} -f_{(\alpha)\sns}(x) -(1-2\alpha) g(x) g'(x)\bigg)^2
  \label{eq:Lalpha1D}
\\
&
\qquad
+
\alpha f'_{(\alpha)\sns}(x)
-
(1-\alpha)\Big[f_{(\alpha)\sns}(x) -\alpha g(x) g'(x)\Big]
-
\frac 12 (1-\alpha)^2 g(x)g'(x)
\nonumber
\end{align}
%
where the result is expressed in terms of the force $f_{(\alpha)\sns}(x)=f(x)+(\frac 12-\alpha) g(x) g'(x)$ of the $\alpha$-discretized Langevin equation
equivalent to the Stratonovich-discretized equation~(\ref{eq:pi_langevin_multi}) (in one dimension).

\subsubsection{Derivation of the $\alpha$-discretized Lagrangian}

For a matter of convenience, in order to obtain a path-integral representation of Eq.~\eqref{eq:pi_langevin_multi} where all functions in the infinitesimal propagator are evaluated at $\bx_k + \alpha \Delta \bx_k$, we start by expressing this process by the equivalent $\alpha$-discretized stochastic differential equation:
\begin{equation} \label{eq:equivalpha}
\frac{\dd x^\mu}{\dd t} \stackrel{\alpha}{=} f_{(\alpha)}^\mu + g^{\mu i} \eta_i  \,,
\end{equation}
with
\begin{equation} \label{eq:driftalpha}
f_{(\alpha)}^\mu = f^\mu + \left(\frac{1}{2}-\alpha\right) g^{\nu i}\partial_\nu g^{\mu j} \delta_{ij} \,.
\end{equation}
In the discrete-time companion process of Eq.~\eqref{eq:equivalpha}, we express the noise increments in terms of the displacement $\Delta \bx$ by
\begin{align}
\Delta \eta_i = g_{i\nu}(\bx + \alpha \Delta \bx) \left(\Delta x^\nu - f_{(\alpha)}^\nu\sns(\bx + \alpha \Delta \bx)\,\Delta t\right),
\end{align} 
from where we can deduce the expression of the Jacobian of the change of variables when going from 
the noise to the position
\begin{align} \label{eq:jacobian}
& \left|\det\left(\frac{\partial\Delta \eta_i}{\partial \Delta x^\beta}\right)\right| 
= 
\nonumber\\
&
\qquad\quad
              \sqrt{\omega(\bx + \alpha \Delta \bx)}\left|\det \left(\delta^\nu_\beta - \alpha \partial_\beta f^\nu_{(\alpha)} \Delta t + \alpha g^{\nu j}\partial_\beta g_{j \rho} \left(\Delta x^\rho - f^\rho_{(\alpha)} \Delta t \right)\right)\right| \, .
\end{align}
\\
In Eq.~\eqref{eq:jacobian}, and until the end of this section, 
 functions are evaluated at $\bx + \alpha \Delta \bx$ unless explicitly stated otherwise. 
Following Eq.~\eqref{eq:inf_scalar_prop}, we first express $\omega(\bx + \alpha \Delta \bx)$ in terms of $\omega(\bx + \Delta \bx)$ up to terms 
$O(\Delta t)$
\begin{align}
\omega(\bx + \alpha \Delta \bx) & = \omega(\bx + \Delta \bx)\exp\left(\ln \omega(\bx + \alpha \Delta \bx) 
- \ln \omega(\bx + \Delta \bx) \right) 
\nonumber\\ 
& 
= \omega(\bx + \Delta \bx)\exp\left(-(1-\alpha)\Delta x^\mu \partial_\mu \ln \omega - \frac{(1-\alpha)^2}{2}\Delta x^\mu \Delta x^\nu \partial_\mu \partial_\nu \ln \omega\right) 
\nonumber \\ 
& 
\doteq  \omega(\bx + \Delta \bx)\exp\left(-2(1-\alpha)\Delta x^\mu \Gamma^\alpha_{\mu\alpha} - (1-\alpha)^2 \omega^{\mu\nu} \partial_\nu \Gamma^\alpha_{\mu\alpha} \Delta t\right) ,
\end{align}
where the last line was obtained after using the substitution relation Eq.~\eqref{eq:subs_multi_first} and the identity 
for the derivative of the determinant of the metric
\begin{equation}
\partial_\mu \ln \omega = \omega^{\alpha\beta}\partial_\mu \omega_{\alpha\beta} = 2 \Gamma^{\alpha}_{\mu\alpha} \, .
\end{equation}
Using now the formula valid for any matrix ${\mathbb H}$,
\begin{equation}
\det\left(1 + \epsilon\, {\mathbb H} \right) = 
\exp\left(\epsilon \, \mbox{Tr}\:{\mathbb H} - \frac{\epsilon^2}{2} \mbox{Tr}\: {\mathbb H}^2\right) + O\big(\epsilon^3\big) \, ,
\end{equation}
we can express the remaining determinant in Eq.~\eqref{eq:jacobian} as
\begin{equation}
\begin{split}
& \left|\det \left(\delta^\nu_\beta - \alpha \partial_\beta f^\nu_{(\alpha)} \Delta t + \alpha g^{\nu j}\partial_\beta g_{j \rho} \left(\Delta x^\rho - f^\rho_{(\alpha)} \Delta t \right)\right)\right| 
\\ & 
= \exp\left[ - \alpha \partial_\nu f_{(\alpha)}^\nu \Delta t + \alpha g^{\nu j} \partial_\nu g_{j\rho}\left(\Delta x^\rho - f_{(\alpha)}^\rho \Delta t \right) - \frac{\alpha^2}{2}\partial_\beta g^{\nu j} \partial_\nu g^{\beta i} \delta_{ij}\Delta t\right] .
\end{split}
\end{equation}
Grouping all terms and taking the continuum limit, this eventually allows us to express the $\alpha$-discretized Lagrangian as
\begin{align}
 \label{eq:alphaLagBis}
\!\!\!
\mathcal{L}^{\bx}_{\alpha}[\textbf{x},\dot{\textbf{x}}] & = \frac{\omega_{\mu\nu}}{2}\left(\frac{\dd x^\mu}{\dd t} - f^\mu_{(\alpha)}\right)\left(\frac{\dd x^\nu}{\dd t} - f^\nu_{(\alpha)}\right) - \frac{\dd x^\mu}{\dd t}\left(\alpha g^{\nu j}\partial_\nu g_{j \mu} - (1-\alpha)\Gamma_{\mu\alpha}^\alpha\right) 
\nonumber\\ 
& \;\;
+ \alpha\left(\partial_\nu f^\nu_{(\alpha)} + g^{\nu j}\partial_\nu g_{j\rho} f^\rho_{(\alpha)}\right) + \frac{(1-\alpha)^2}{2}\omega^{\mu\nu}\partial_\nu \Gamma_{\mu\alpha}^{\alpha} + \frac{\alpha^2}{2}\partial_\beta g^{\nu j}\partial_\nu g^{\beta i}\delta_{ij}  
\, .
\end{align}
In order to go from Eq.~\eqref{eq:alphaLagBis} to Eq.~\eqref{eq:Lalpha}, we substitute $f^\mu_{(\alpha)}$ in Eq.~\eqref{eq:alphaLagBis} by its expression as a  function of $h^\mu$ inferred from Eq.~\eqref{eq:driftalpha} and Eq.~\eqref{eq:defh},
\begin{align}
\label{eq:fin_alpha}
f^\mu_{(\alpha)} 
& = h^\mu + \frac{1}{2}\Gamma^\rho_{\nu\rho} \omega^{\mu\nu} + \frac{1}{2}\partial_\nu g^{\nu i} g^{\mu j} \delta_{ij} + \left(\frac{1}{2}-\alpha\right)g^{\nu i}\partial_\nu g^{\mu j}\delta_{ij} \nonumber \\ 
& = h^\mu - \frac{1}{2}\Gamma_{\alpha\beta}^\mu \omega^{\alpha\beta} - \alpha g^{\nu i}\partial_\nu g^{\mu j} \delta_{ij} \, .
\end{align}
The algebra is then tedious but straightforward to get to Eq.~\eqref{eq:Lalpha}.

\subsection{A free particle in the two-dimensional plane}
\label{subsec:free-particle}

Before delving in further mathematics, we illustrate the findings of this section with a simple example inspired from Edwards and Gulyaev's paper~\cite{edwards_path_1964}. We take a close look at a two-dimensional Brownian motion, when changing from Cartesian to polar coordinates, and we shed light on where the difficulties lie in the path-integral description of 
this specific example. We start from 
the equations of motion:
\begin{equation} \label{eq:2d_brownian_cartesian}
\frac{\dd x}{\dd t}=  (2D)^{1/2} \, \eta_x \, ,\qquad\qquad \frac{\dd y}{\dd t}= (2D)^{1/2} \, \eta_y \, ,
\end{equation}
where $\eta_x$ and $\eta_y$ are independent Gaussian white noises with correlations 
\begin{equation}
\langle\eta_x(t)\eta_x(t')\rangle=\langle\eta_y(t)\eta_y(t')\rangle= \delta(t-t')
\end{equation}
and $(2D)^{1/2}$ is a constant with dimensions of $[x]/[t]^{1/2}$, that is, $D$ is a diffusion coefficient.
The measure over trajectories, parametrized with Cartesian coordinates,  simply reads
\begin{equation} \label{eq:measure_2d_brownian_cartesian}
{\mathcal D}x {\mathcal D} y \; \ee^{\displaystyle{-\frac{1}{4D}\int_{t_0}^{\tf}\dd t \, \left(\dot{x}^2+\dot{y}^2\right)}} \, .
\end{equation}
Because the process in Eq.~\eqref{eq:2d_brownian_cartesian} has additive noise and vanishing drift, 
in writing Eq.~\eqref{eq:measure_2d_brownian_cartesian} there are no discretization issues to explicitly care about at
this level.

Following the work of Edwards and Gulyaev~\cite{edwards_path_1964} who 
pinpointed the difficulties that could arise in the functional formulation, 
we study this same problem using polar coordinates $r$ and $\phi$ with $x=r\cos\phi$ and $y=r\sin\phi$.
In the notation used in the general presentation, we have 
\begin{align}
& 
\bu = (r,\phi) = (U^1(\bx), \, U^2(\bx)) = \big((x^2+y^2)^{1/2}, \ \arctan(y/x)\big) 
\; . 
\end{align}
One can readily check that the coordinate transformation can 
be carried out at the level of the equations of motion. The resulting evolution of the radial and angular coordinates, 
$r$ and $\phi$, is governed by Langevin equations with multiplicative noise, which are sensitive to the discretization 
scheme.
The equation for the higher-dimensional change of variables, Eq.~(\ref{eq:change_var}), simplifies considerably thanks to $f^\mu=0$, $g^{\mu i } =(2D)^{1/2} \, \delta^{\mu i}$, and $\omega^{\mu\nu} =2D \, \delta^{\mu\nu}$. In the 
$\alpha$ scheme one has
\begin{align}
&
\frac{\dd u^\mu}{\dd t} \stackrel{\alpha}{=} 
\left(\frac{1}{2}-\alpha \right) \, 2D \, \delta^{\rho\sigma} \partial_\rho\partial_\sigma U^\mu \left( {\bf U}^{-1}(\bu) \right) 
+  (2D)^{1/2} \, \delta^{\rho i}\partial_\rho U^{\mu} \left( {\bf U}^{-1}(\bu)\right) \eta_i
\end{align}
and once the components are made explicit 
\begin{equation}
\label{eq:Langevin-polar}
\begin{aligned}
&
\frac{\dd r}{\dd t}  \stackrel{\alpha}{=} \left(\frac{1}{2}-\alpha \right) \frac{2D}{r} + (2D)^{1/2} \left[ \cos\phi \, \eta_x+\sin\phi \, \eta_y \right]
\; , 
\\
& 
\frac{\dd\phi}{\dd t} \stackrel{\alpha}{=} \frac{(2D)^{1/2}}{r}  \left[ - \sin \phi \, \eta_x + \cos \phi \, \eta_y   \right]
\; .
\end{aligned}
\end{equation}
 The transformation is reversible and one can go back to the simpler Cartesian evolution
 by applying the change of variables backwards.

From Eqs.~(\ref{eq:Langevin-polar}),
one reads $G^{\mu i}$, with the index $\mu$ labeling the polar coordinates $(r,\phi)$ and the Latin index $i$
labeling the Cartesian coordinates $(x,y)$. Therefore, from $G^{\mu i}$ one derives the  
metric, Eq.~(\ref{eq:def-omega}), in polar coordinates $\Omega^{\mu\nu}$
\begin{equation}
G^{\mu i}(r, \phi)  = (2D)^{\frac{1}{2}}
\begin{pmatrix}
\cos \phi & \sin \phi
\vspace{0.2cm}
\\
-\dfrac{\sin\phi}{r} & \dfrac{\cos\phi}{r}
\end{pmatrix}
\quad\Rightarrow\quad
\Omega^{\mu\nu}(r, \phi) = 
2D
\begin{pmatrix}
\; 1 & 0 \; 
\vspace{0.2cm}
\\
\; 0 & \dfrac{1}{r^2} \;
\end{pmatrix} 
\label{eq:sfdkjsdkjfsd}
\end{equation}
which turns out to be Euclidean. Note that $G^{ri}$ and $G^{\phi i}$ have different dimensions, 
as well as $\Omega^{r\nu}$ and $\Omega^{\phi\nu}$. The determinant of $\Omega_{\mu\nu}$ is $\Omega = [r/(2D)]^2$. The corresponding Ricci scalar curvature vanishes and the only nonzero Christoffel symbols are
\begin{align}
\hat \Gamma^{r}_{\phi\phi} = - r \, , 
\qquad\qquad \hat \Gamma^{\phi}_{r \phi} = \Gamma^{\phi}_{\phi r} = \frac{1}{r} \, .
\end{align}
The vectorial drift $h^\mu$, defined in Eq.~(\ref{eq:defh}), 
vanishes in the polar coordinate system since it does in the Cartesian coordinate system. One 
can also check this statement by doing the explicit calculation.

Next, we discuss how to reexpress the path probability~(\ref{eq:measure_2d_brownian_cartesian}) in terms of the polar coordinates $r$ and $\phi$.
Considering the Stratonovich scheme, if one (naively) assumes that the standard chain rule of calculus applies,
one obtains the following expression for the probability measure of the $r,\phi$ paths
\begin{equation}\label{eq:actionpolairenaive}
{\mathcal D}r  {\mathcal D} \phi \; \ee^{\displaystyle{-\frac{1}{4D}\int_{t_0}^{\tf}\dd t \, \left(\dot{r}^2+r^2\dot{\phi}^2\right)}} \, ,
\end{equation}
with ${\mathcal D}r  {\mathcal D} \phi $ the covariant volume element given from Eq.~\eqref{eq:path_measure}:
\begin{align}
{\mathcal D}r  {\mathcal D} \phi = \lim_{N \to + \infty} \frac{1}{2\pi \Delta t} \prod_{k=1}^{N-1}\left\{ r_k \frac{ \dd r_k \dd \phi_k}{4 \pi D \Delta t}\right\}  .
\end{align} 

However, the continuous-time expression~(\ref{eq:actionpolairenaive}) differs from the ones that can be derived from the Langevin dynamics~\eqref{eq:Langevin-polar} in polar coordinates,  using either the Itō discretization --~which may not come as a surprise~-- or the Stratonovich discretization. The latter is perhaps more surprising given that, at the level of the Langevin equations, one can readily see that Eqs.~(\ref{eq:Langevin-polar}) with $\alpha=1/2$ are directly derived from the Cartesian-coordinate Langevin equations~(\ref{eq:2d_brownian_cartesian}) using the standard chain rule of differential calculus. 

Indeed, in this particular problem, we have already identified all the ingredients of the generic expressions of Sec.~\ref{sec:statement-resultPIalpha}.  
Thus, from Eq.~\eqref{eq:Lito}, the Itō Lagrangian in polar coordinates 
reads
\begin{align}
\label{eq:Litopolar}
\mathcal{L}_0\big[r, \dot{r}, \phi, \dot{\phi}\big] = \frac{1}{2} \, \left[\frac{1}{2D} \left( \dot{r}^2 + r^2 \dot{\phi}^2 \right) + \frac{\dot{r}}{r} -  \frac{3 (2D)}{4 r^2}\right]  ,
\end{align}
and, from Eq.~(\ref{eq:Lstrato}),  the Stratonovich one 
\begin{align}
\label{eq:Lstratopolar}
\mathcal{L}_{1/2}\big[r, \dot{r}, \phi, \dot{\phi}\big] = \frac{1}{2}\left[ \frac{1}{2D} \left( \dot{r}^2 + r^2 \dot{\phi}^2 \right) 
- \frac{D}{ r^2}\right]  .
\end{align}
None of the above matches the expression in Eq.~\eqref{eq:actionpolairenaive} which was derived from the 
Cartesian functional measure after a (naive) change of variables to polar coordinates. 
This concretely illustrates that neither the Itō scheme nor the Stratonovich one (nor any $\alpha$ scheme) are covariant discretization schemes of path-integral actions, in the sense that they are not amenable to a blind use of the standard chain rule,

In the pioneering work of Edwards and Gulyaev~\cite{edwards_path_1964}, it was proposed to add a term $-\frac{1}{24D} r^2 \dot \phi^4 \dd t^2 $ in the naive Lagrangian\footnote{
See Sec.~4 of Ref.~\cite{edwards_path_1964} (in which $D=\frac 14$), where due to typos, one reads $-\frac{1}{12}r^4 \dot \phi^4 \dd t$ instead of the correct expression $-\frac 16 r^2 \dot \phi^4 \dd t^2$.
} 
of Eq.~(\ref{eq:actionpolairenaive}), so as to correctly represent the path probability. 
Such an extra term does not present a well-defined continuous-time expression, but we understand from the substitution rule~(\ref{eq:quarticsubstit}) that $\Delta\phi^4 \doteq  12 D^2 \Delta t^2/r^4$; hence,
the extra term proposed by Edwards and Gulyaev effectively corresponds to the contribution $-D/(2r^2)$ that is present in the Stratonovich Lagrangian~\eqref{eq:Lstratopolar}.

In Sec.~\ref{sec:DWG} and paragraphs~\ref{sec:disccovPI}-\ref{sec:higher-order-disc-point-B},
we present different and complementary constructions  of path-integral covariant  discretization schemes, that allow one to use the standard chain rule and thus to
describe the 2D Brownian motion with a Lagrangian $\frac{1}{4D}\big(\dot{r}^2+r^2\dot{\phi}^2\big)$ that possesses a proper meaning.
We also discuss a different discretization scheme in Sec.~\ref{subsec:ito_plus12}, where the Lagrangian and the normalization prefactor take a different form,
allowing one to apply the Itō modified chain rule, in the spirit of Ref.~\cite{ding2021timeslicing}.

We also remark on this example (and this is true in general, as we discuss later) that if one is only interested in the small-noise asymptotics $D\to 0$, the questions we are discussing are irrelevant at dominant order (since the main contribution to the Lagrangians~(\ref{eq:Litopolar})-(\ref{eq:Lstratopolar}) is of order $1/D$ and the remaining one, sensitive to the discretization scheme, is of order $D$).

\section{Extensions of Itō's lemma for path-integral calculus} \label{sec:ItocalculusPI}

As we have extensively discussed in Sec.~\ref{sec:pathint}, the chain rule does not operate at the level of a continuous-time $\alpha$-discretized Lagrangian.  
Indeed, for a generic non-linear transformation $\bu(t) = \textbf{U}\left(\bx(t)\right)$, with $\textbf{U}^{-1}=\textbf{X}$,
\begin{equation}
\mathcal{L}_{\alpha}^{\bu}[\textbf{u},\dot{\textbf{u}}] \neq \mathcal{L}_{\alpha}^{\bx}\left[\textbf{X}(\bu),\frac{\partial X^\mu}{\partial u^\nu}\dot{u}^\nu\right].
\end{equation}
 In this section, we thoroughly investigate the transformation properties of the Lagrangian under a non-linear change of variables and compute the corrections that need to be taken into account when applying blindly the modified chain rule  
\begin{equation} \label{eq:ito_naive}
\frac{\dd u^\mu}{\dd t} = \partial_\rho U^\mu \frac{\dd x^\rho}{\dd t} + \left(\frac{1}{2}-\alpha\right)\omega^{\rho\sigma}\partial_\rho\partial_\sigma U^\mu 
\end{equation}
within the $\alpha$-discretized continuous-time Lagrangian. Thus, we shall construct an extension of Itō's lemma for path-integral calculus, which is the main result of this section.
Our derivation is presented in Sec.~\ref{subsec:ito_plus1} and our findings summarized in Sec.~\ref{subsec:ito_plus12}.
 Concretely, in the rest of this section, we focus on the $\alpha$-discretized infinitesimal propagator and on its transformation under the change of variable $\bx(t) = \textbf{X}(\bu(t))$.

\subsection{Transformation of variables at the path-integral level}\label{subsec:ito_plus1}

Our starting point is the $\alpha$-discretized infinitesimal propagator for a diffusive process $\bx(t)$. Up to corrections of order $O(\Delta t^{3/2})$ in the exponential, it takes the general form
\begin{align}\label{eq:infi_prop_general0}
\mathbb{P}(\bx+\Delta\bx, t+\Delta t | \bx, t) = \frac{\sqrt{\omega(\bx + \gamma \Delta \bx)}}{(2 \pi \Delta t)^{d/2}}\exp\left(-\frac{1}{2}\omega_{\mu\nu}(\bx + \alpha \Delta \bx)\frac{\Delta x^\mu \Delta x^\nu}{\Delta t}\right) \nonumber \\ \exp\left(A^{(0)}(\bx) \Delta t + A_{\mu}^{(1)}(\bx + \alpha \Delta \bx)\Delta x^\mu\right).
\end{align}
In Sec.~\ref{sec:pathint}, we provided the expression of the $\alpha$-discretized infinitesimal propagator in the specific case $\gamma = 1$ which leads to a scalar invariant definition of the path measure $\mathcal{D}\bx$ (see Eq.~\eqref{eq:path_measure}). Most of the results presented in this section being insensitive to $\gamma$, we however choose to keep it unspecified in the following. We now introduce a non-linear change of variables $\bx(t) = \textbf{X}(\bu(t))$ and investigate the way in which 
the different terms in the exponential of Eq.~\eqref{eq:infi_prop_general0} transform, and which form they take in terms of $\bu$, 
while keeping all contributions up to order $O(\Delta t)$.

\subsubsection{Elementary transformation rules}

For this purpose, and recalling that $\Delta \bx = O(\Delta t^{1/2})$, we need to 
\begin{enumerate}
\item[(\textit{i})] 
express
$\Delta x^\alpha$ as a function of $\Delta u^\mu$ up to order $O\big(\Delta t^{3/2}\big)$ and
\item[(\textit{ii})] get the transformation law for a generic function $\varphi\left(\bx + \alpha \Delta \bx\right)$ in terms of $\varphi\left(\textbf{X}(\bu + \alpha \Delta \bu)\right)$ up to order $O(\Delta t)$. 
\end{enumerate}
 
For item (\textit{i}) we obtain\footnote{
For a function $\mathbf{X}(\bu)$ of the variable $\bu$, we denote by $\partial_{u^\mu}\sns X^\alpha$ the partial derivative with respect to $u^\mu$,
while we keep $\partial_\mu U^\alpha$ for the derivative with respect to $x^\mu$ of a function $\mathbf{U}(\bx)$. 
The notations $\frac{\partial X^\alpha}{\partial u^\mu}$ and $\frac{\partial U^\alpha}{\partial x^\mu}$ will also be used in some cases for readability purposes.
%
}
\begin{align} 
\label{eq:transfodx}
\Delta x^\alpha & =  X^\alpha(\bu + \Delta \bu) - X^\alpha(\bu)
\nonumber\\
& = 
 \partial_{u^\mu} X^\alpha(\bu + \alpha \Delta \bu) \Delta u^\mu + \frac{1-2\alpha}{2}\partial_{u^\mu}\partial_{u^\nu}
 X^{\alpha}(\bu + \alpha \Delta \bu)\Delta u^\mu \Delta u^\nu 
 \nonumber\\ 
 & \;\;\;
 + \frac{1-3\alpha(1-\alpha)}{6}\partial_{u^\mu}\partial_{u^\nu}\partial_{u^\rho} X^{\alpha}(\bu)\Delta u^\mu \Delta u^\nu\Delta u^\rho + O\left(\Delta u^4\right) .
\end{align}
The first two terms in the right-hand side of Eq.~\eqref{eq:transfodx}, from which one can infer Itō's lemma, are sufficient to express $\Delta x^\alpha$ up to order $O(\Delta t)$, which is the desired precision when working at the level of stochastic differential equations. The  next higher-order term is however necessary to study transformation laws at the path-integral level. 

Next we turn to (\textit{ii})
\begin{align}
\varphi\left(\bx + \alpha \Delta \bx\right) & = \varphi\left(\bf X(\bu + \alpha \Delta \bu - \alpha \Delta \bu) + \alpha \Delta \bx \right) 
\nonumber\\ 
& =  \varphi\left(\bf X(\bu + \alpha \Delta \bu)\right) + \frac{\alpha(1-\alpha)}{2}\partial_{u^\rho}\partial_{u^\sigma} X^{\nu}(\bu)\sms \partial_\nu \varphi({\bf X}(\bu))\, \Delta u^\rho \Delta u^\sigma 
\nonumber\\
& \;\;\; 
+ O(\Delta t^{3/2}) \, , 
 \label{eq:transfofunction}
\end{align}
where we used~(\ref{eq:transfodx}) to replace $\Delta \bx$ and we evaluated the factors in the second 
term at ${\bf X}(\bu)$ since we already have the desired second order in $\Delta \bu$.
Note that the correcting term in Eq.~\eqref{eq:transfofunction} is of order $O(\Delta t)$ and was negligible when studying transformation properties of $\alpha$-discretized stochastic differential equations. 

\subsubsection{Itō's lemma for zeroth and first-order terms in $\Delta \bx$}

Using these last two equations, together with the second-order substitution rule $\Delta u^\mu \Delta u^\nu \doteq  \Omega^{\mu\nu}(\bu)\Delta t$ where
\begin{equation}
\Omega^{\rho\sigma}(\bu) = \omega^{\mu\nu}(\textbf{X}(\bu)) \, \partial_\mu U^{\rho}(\textbf{X}(\bu)) \,\partial_\nu U^{\sigma}(\textbf{X}(\bu)) \, ,
\end{equation}
we see that blindly applying the discretized version of Itō's lemma, \textit{i.e.}~replacing 
\begin{align} 
\label{eq:ito_lemma_discrete}
\Delta x^\alpha & \rightarrow \partial_{u^\mu} X^\alpha(\bu + \alpha \Delta \bu) \Delta u^\mu + \frac{1-2\alpha}{2}\partial_{u^\mu\sns}\partial_{u^\nu}
 X^{\alpha}(\bu + \alpha \Delta \bu)\Omega^{\mu\nu}\Delta t \, ,
\end{align}
yields the correct result for the transformation rule of the zeroth order term in $\Delta \bf x$, $A^{(0)}(\bx)\Delta t$,  and the 
first-order one, $A^{(1)}_\mu(\bx+\alpha \Delta\bx)\Delta x^\mu$. Indeed,
\begin{equation}
A^{(0)}(\bx) \Delta t = A^{(0)}(\textbf{X}(\bu)) \Delta t 
\end{equation}
and
\begin{align}
A^{(1)}_\mu(\bx+\alpha \Delta\bx)\Delta x^\mu \doteq A^{(1)}_\mu(\textbf{X}(\bu + \alpha \Delta \bu))\left[\partial_{u^\mu} X^\alpha(\bu + \alpha \Delta \bu) \Delta u^\mu \right. \nonumber \\ \left. + \frac{1-2\alpha}{2}\partial_{u^\mu\sns}\partial_{u^\nu}
 X^{\alpha}(\bu + \alpha \Delta \bu)\Omega^{\mu\nu}\Delta t\right] + O\big(\Delta t^{3/2}\big) \,.
\end{align}
The next term is the tricky one, as we will show below.

\subsubsection{Extension of Itō's lemma for the quadratic term}

Collecting all terms in the transformation of the quadratic term \textit{{a priori}} requires to take into account systematic corrections to Itō's lemma. We denote by MCR the results obtained by blindly applying the modified chain rule inferred from Itō's lemma in Eq.~\eqref{eq:ito_lemma_discrete},
\textit{i.e.}
\begin{align}
\text{MCR} & = -\frac{1}{2\Delta t} \, \omega_{\mu\nu}\big(\textbf{X}(\bu + \alpha \Delta \bu )\big)
\nonumber\\
& 
\qquad\quad
\left[\partial_{u^\rho} X^\mu(\bu + \alpha \Delta \bu) \Delta u^\rho + \frac{1-2\alpha}{2}\partial_{u^\rho\sns}\partial_{u^\gamma}
 X^{\mu}(\bu + \alpha \Delta \bu)\Omega^{\rho\gamma}\Delta t \right]
 \nonumber \\
&
\qquad\quad
 \left[\partial_{u^\rho} X^\nu(\bu + \alpha \Delta \bu) \Delta u^\rho + \frac{1-2\alpha}{2}\partial_{u^\rho\sns}\partial_{u^\gamma}
 X^{\nu}(\bu + \alpha \Delta \bu)\Omega^{\rho\gamma}\Delta t\right],
\end{align}
and we carefully collect deviations from it in the transformation of the quadratic term. After using the second and fourth order substitution rules we obtain
\begin{align}
& 
-\frac{1}{2}\omega_{\mu\nu}(\bx + \alpha \Delta \bx)\frac{\Delta x^\mu \Delta x^\nu}{\Delta t} 
\doteq 
\nonumber\\
& 
\qquad\qquad
\text{MCR} - \frac{1-2\alpha}{2}\omega_{\mu\nu}\left[\frac{\partial X^\mu}{\partial u^\alpha}\frac{\partial^2 X^\nu}{\partial u^\beta \partial u^\gamma}\left(\frac{\Delta u^\alpha \Delta u^\beta \Delta u^\gamma}{\Delta t} - \Omega^{\beta \gamma}\Delta u^\alpha\right)\right] \nonumber \\ 
& 
\qquad\qquad
- \frac{1-3\alpha(1-\alpha)}{2}\omega_{\mu\nu}\frac{\partial X^\mu}{\partial u^\alpha}\frac{\partial^3 X^\nu}{\partial u^\rho \partial u^\beta \partial u^\gamma}\Omega^{\alpha\rho}\Omega^{\beta\gamma}\Delta t 
\nonumber\\
&
\qquad\qquad
- \frac{(1-2\alpha)^2}{4}\omega_{\mu\nu}\frac{\partial^2 X^\mu}{\partial u^\alpha \partial u^\beta}\frac{\partial^2 X^\nu}{\partial u^\rho \partial u^\sigma}\Omega^{\alpha\rho}\Omega^{\beta\sigma}\Delta t 
\nonumber \\
&
\qquad\qquad
- \frac{\alpha(1-\alpha)}{4}\frac{\partial \omega_{\mu\nu}}{\partial x^\eta}\frac{\partial^2 X^\eta}{\partial u^\alpha \partial u^\beta}\frac{\partial X^\mu}{\partial u^\rho}\frac{\partial X^\nu}{\partial u^\sigma}\left(\Omega^{\alpha\beta}\Omega^{\rho\sigma}+2\Omega^{\alpha\rho}\Omega^{\beta\sigma}\right)\Delta t 
\nonumber \\
&
\qquad\qquad
+ O\big(\Delta t^{3/2}\big)\,,
\label{eq:subsitution_kinetic_1}
\end{align}
where all functions in the right-hand-side are evaluated at $\bu + \alpha \Delta \bu$. Interestingly, this procedure naturally leads to the appearance of a third-order term $\Delta u^\alpha \Delta u^\beta \Delta u^\gamma$ in the infinitesimal propagator. It can nevertheless be rewritten in terms of zeroth and first-order terms as shown in~\cite{Cugliandolo-Lecomte17a} in one dimension and as we prove below in the general $d \geq 1$ case. This substitution rule has a weaker meaning than the second and fourth order ones as it does not correspond to $L^2$ convergence but rather to a weaker convergence in distribution.

\subsubsection{Third-order substitution rule}
\label{subsubsec:third-order}

Let us start by considering an infinitesimal propagator with a cubic term of the form
\begin{align}\label{eq:infi_prop_general}
\mathbb{P}(\bx+\Delta\bx, t+\Delta t | \bx, t) = \frac{\sqrt{\omega(\bx + \gamma \Delta \bx)}}{(2 \pi \Delta t)^{d/2}}\exp\left(-\frac{1}{2}\omega_{\mu\nu}(\bx + \alpha \Delta x)\frac{\Delta x^\mu \Delta x^\nu}{\Delta t}\right) \nonumber \\ \exp\left(A^{(0)}(\bx) \Delta t + A_{\mu}^{(1)}(\bx + \alpha \Delta \bx)\Delta x^\mu + A_{\mu\nu\rho}^{(3)}(\bx + \alpha \Delta \bx)\frac{\Delta x^\mu \Delta x^\nu \Delta x^\rho}{\Delta t}\right).
\end{align}
Without loss of generality,
we assume in the following that $A_{\mu\nu\rho}^{(3)}$ is a fully symmetric tensor. To replace the cubic term, we look for $\tilde{A}^{(0)}$ and $\tilde{A}_\mu^{(1)}$ such that the new infinitesimal propagator 
\begin{align}\label{eq:infi_prop_generaltilde}
\tilde{\mathbb{P}}(\bx+\Delta\bx, t+\Delta t | \bx, t) 
= 
\frac{\sqrt{\omega(\bx + \gamma \Delta \bx)}}{(2 \pi \Delta t)^{d/2}}\exp\left(-\frac{1}{2}\omega_{\mu\nu}(\bx + \alpha \Delta \bx)
\frac{\Delta x^\mu \Delta x^\nu}{\Delta t}\right) \nonumber \\ \exp\left(\tilde{A}^{(0)}(\bx) \Delta t + \tilde{A}_{\mu}^{(1)}(\bx + \alpha \Delta \bx)\Delta x^\mu \right),
\end{align}
generates the same stochastic process. This requires that $\tilde{\mathbb{P}}$ is well normalized and that both distributions have, up to corrections of order $O(\Delta t^{3/2})$, the same first and second moments. 
Both second moments coincide in a simple way as 
\begin{equation}
\left\langle \Delta x^\mu \Delta x^\nu \right\rangle = \left\langle \Delta x^\mu \Delta x^\nu \right\rangle_{\sim} = \omega^{\mu\nu}(\bx) \, \Delta t + O\big(\Delta t^{3/2}\big) \,.
\end{equation}
where $\langle \dots \rangle$ is the average using ${\mathbb P}$ and  $\langle \dots \rangle_\sim$ is the average using $\tilde{\mathbb{P}}$. Note that in the following, we omit the discretization point in places where it is irrelevant. For any function $f$ we next introduce the notation
\begin{equation}
\overbar{f(\Delta \bx)} = \int \dd \Delta \bx  \; \frac{\sqrt{\omega(\bx + \gamma \Delta \bx)}}{(2 \pi \Delta t)^{d/2}}\exp\left(-\frac{1}{2}\omega_{\mu\nu}(\bx + \alpha \Delta x)\frac{\Delta x^\mu \Delta x^\nu}{\Delta t}\right)f(\Delta \bx) \,, 
\end{equation}
so that the first moments are obtained as
\begin{align}
\left\langle \Delta x^\mu \right\rangle &= \overbar{\Delta x^\mu} + \left(A^{(1)}_\nu + 3 A^{(3)}_{\nu\rho\sigma}\omega^{\rho\sigma}\right)\omega^{\mu\nu}\Delta t + O\big(\Delta t^{3/2}\big) 
\\
\left\langle \Delta x^\mu \right\rangle_{\sim} &= \overbar{\Delta x^\mu} + \tilde{A}^{(1)}_\nu \omega^{\mu\nu}\Delta t + O\big(\Delta t^{3/2}\big) \,.
\end{align}
The equality of the last two right-hand-sides imposes
\begin{equation}\label{eq:subslin}
\tilde{A}^{(1)}_\mu = A^{(1)}_\mu + 3 A^{(3)}_{\mu\rho\sigma}\omega^{\rho\sigma} \,.
\end{equation}
The normalization of the propagator $\mathbb{P}$ yields
\begin{align}
1 = 
& 
\; \overbar{1} + A^{(0)} \Delta t + \overbar{A^{(1)}_\mu(\bx + \alpha \Delta \bx)\Delta x^\mu} + \overbar{A^{(3)}_{\mu\nu\rho}(\bx + \alpha \Delta \bx)\frac{\Delta x^\mu\Delta x^\nu \Delta x^\rho}{\Delta t}} 
\nonumber\\
& 
+ \frac{1}{2}A^{(1)}_\mu A^{(1)}_\nu \omega^{\mu\nu}\Delta t 
+ \frac{1}{2}A^{(3)}_{\mu\nu\rho}A^{(3)}_{\alpha\beta\sigma}\left(9\omega^{\mu\nu}\omega^{\alpha\beta}\omega^{\rho\sigma} + 6 \omega^{\mu\alpha}\omega^{\nu\beta}\omega^{\rho\sigma}\right)\Delta t 
\nonumber\\
& 
+ 3 A^{(1)}_\mu A^{(3)}_{\nu\alpha\beta} \omega^{\mu\nu} \omega^{\alpha\beta} \Delta t + O\big(\Delta t^{3/2}\big)\,,
\end{align}
while the one of $\tilde{\mathbb{P}}$ reads accordingly
\begin{align}
1 = \overbar{1} + \tilde{A}^{(0)} \Delta t + \overbar{\tilde{A}^{(1)}_\mu(\bx + \alpha \Delta \bx)\Delta x^\mu} + \frac{1}{2}\tilde{A}^{(1)}_\mu \tilde{A}^{(1)}_\nu \omega^{\mu\nu}\Delta t + O\big(\Delta t^{3/2}\big) \,.
\end{align}
We furthermore note that
\begin{equation}
\overbar{A^{(1)}_\mu(\bx + \alpha \Delta \bx)\Delta x^\mu} = A^{(1)} \overbar{\Delta x^\mu} + \alpha \partial_{\nu} A^{(1)}_\mu \omega^{\mu\nu}\Delta t + O\big(\Delta t^{3/2}\big)\,,
\end{equation}
and
\begin{align}\label{eq:overbar3}
\overbar{A^{(3)}_{\mu\nu\rho}(\bx + \alpha \Delta \bx)\frac{\Delta x^\mu\Delta x^\nu \Delta x^\rho}{\Delta t}} 
= 
& \;
A^{(3)}\overbar{\frac{\Delta x^\mu\Delta x^\nu \Delta x^\rho}{\Delta t}} 
\nonumber\\
& 
+ 3 \alpha \partial_\sigma A^{(3)}_{\mu\nu\rho}\omega^{\mu\nu}\omega^{\rho\sigma}\Delta t + O\big(\Delta t^{3/2}\big)\,.
\end{align}
Altogether Eqs.~\eqref{eq:subslin}-\eqref{eq:overbar3} lead to 
\begin{align}\label{eq:a0}
\tilde{A}^{(0)} = 
&
A^{(0)} + 3 A^{(3)}_{\mu\nu\rho}A^{(3)}_{\alpha\beta\sigma} \omega^{\mu\alpha}\omega^{\nu\beta}\omega^{\rho\sigma} + A^{(3)}_{\mu\nu\rho}\overbar{\frac{\Delta x^\mu\Delta x^\nu \Delta x^\rho}{\Delta t^2}} 
\nonumber\\
& 
- 3 \omega^{\rho\sigma}A^{(3)}_{\mu\rho\sigma}\frac{\overbar{\Delta x^\mu}}{\Delta t} - 3 \alpha A^{(3)}_{\mu\rho\sigma}\omega^{\mu\nu}\partial_\nu \omega^{\rho\sigma} \,.
\end{align}
We lastly evaluate $\overbar{\Delta x^\mu}$ and $\overbar{\Delta x^\mu \Delta x^\nu \Delta x^\rho}$. We first have, 
keeping all terms up to $O(1)$
\begin{align}
\frac{\overbar{\Delta x^\mu}}{\Delta t} 
= 
& \int \dd \Delta \bx \;  \frac{\Delta x^\mu}{\Delta t}  \; \frac{\sqrt{\omega(\bx + \gamma \Delta \bx)}}{(2 \pi \Delta t)^{d/2}} 
\nonumber\\
& \qquad\qquad
\exp\left(-\frac{1}{2}\omega_{\alpha\beta}(\bx + \alpha \Delta x)\frac{\Delta x^\alpha \Delta x^\beta}{\Delta t}\right)
\nonumber \\ 
= 
& \int \dd \Delta \bx  \; \frac{\Delta x^\mu}{\Delta t}  \; \frac{\sqrt{\omega(\bx)}} {(2 \pi \Delta t)^{d/2}}\frac{\sqrt{\omega(\bx + \gamma \Delta \bx)}}{\sqrt{\omega(\bx)}} 
\nonumber\\
& \qquad\qquad
\exp\left(-\frac{1}{2}\omega_{\alpha\beta}(\bx)\frac{\Delta x^\alpha \Delta x^\beta}{\Delta t} 
- \frac{\alpha}{2}\partial_\gamma\omega_{\alpha\beta} (\bx) \frac{\Delta x^\alpha \Delta x^\beta \Delta x^\gamma}{\Delta t} \right)
\nonumber \\ 
= 
& \int \dd \Delta \bx  \frac{\sqrt{\omega(\bx)}} {(2 \pi \Delta t)^{d/2}}
\exp\left(-\frac{1}{2}\omega_{\alpha\beta}(\bx)\frac{\Delta x^\alpha \Delta x^\beta}{\Delta t}\right)
\nonumber\\
& \qquad\qquad
\left(\frac{\gamma}{2} \partial_\gamma \ln \omega(\bx) \frac{\Delta x^\mu \Delta x^\gamma}{\Delta t} 
- \frac{\alpha}{2} \partial_\gamma\omega_{\alpha\beta}(\bx) 
\frac{\Delta x^\mu \Delta x^\alpha \Delta x^\beta \Delta x^\gamma}{\Delta t}\right) 
\nonumber \\ 
= & \; \; \frac{\gamma}{2} \omega^{\mu\gamma} \partial_\gamma \ln \omega - \frac{\alpha}{2}\partial_\gamma \omega_{\alpha\beta}\left(\omega^{\alpha \beta}\omega^{\mu\gamma} + 2 \omega^{\alpha\mu}\omega^{\beta \gamma}\right).
\end{align}
Accordingly, when contracting against any fully symmetric tensor $T_{\mu\nu\gamma}$, we have
\begin{align}
& T_{\mu\nu\gamma}\frac{\overbar{\Delta x^\mu \Delta x^\nu \Delta x^\gamma}}{\Delta t^2} =
T_{\mu\nu\gamma}\left[\frac{3\gamma}{2}\omega^{\mu\nu}\partial_\rho\omega^{\gamma\rho} 
\right.
\nonumber\\
&
\qquad\qquad\;\;
\left.
- \frac{\alpha}{2}\partial_\rho \omega_{\alpha\beta} \left(3 \omega^{\mu\nu}\omega^{\alpha\beta}\omega^{\gamma\rho} + 6 \omega^{\mu\nu}\omega^{\alpha\gamma}\omega^{\beta\rho} + 6 \omega^{\mu\rho}\omega^{\nu\alpha}\omega^{\gamma\beta}\right)\right]
\,.
\end{align}
Consequently, the condition in Eq.~\eqref{eq:a0} reduces  to
\begin{equation}
\tilde{A}^{(0)} = A^{(0)} + 3 A^{(3)}_{\mu\nu\rho}A^{(3)}_{\alpha\beta\sigma} \omega^{\mu\alpha}\omega^{\nu\beta}\omega^{\rho\sigma} \,.
\end{equation}
We are therefore free to proceed with the replacement 
\begin{equation}
A^{(3)}_{\mu\nu\rho}\frac{\Delta x^\mu \Delta x^\nu \Delta x^\rho}{\Delta t} \doteq 3 A^{(3)}_{\mu\nu\rho} \omega^{\mu\nu}\Delta x^{\rho} + 3 A^{(3)}_{\mu\nu\rho}A^{(3)}_{\alpha\beta\sigma} \omega^{\mu\alpha}\omega^{\nu\beta}\omega^{\rho\sigma}\Delta t \,,
\label{eq:3rdordersubstanyd}
\end{equation}
independently of the discretization parameters $\alpha$ and $\gamma$. We stress that the substitution rule~\eqref{eq:3rdordersubstanyd} is only valid in the exponential of the propagator.\footnote{
Note that in prefactor of the exponential of the propagator, the cubic substitution rule is independent of its prefactor and takes the general form
$
\frac{\Delta x^\mu \Delta x^\nu \Delta x^\rho}{\Delta t} 
\doteq
 \omega^{\mu\nu}\Delta x^{\rho} +
 \omega^{\mu\rho}\Delta x^{\nu} +
 \omega^{\nu\rho}\Delta x^{\mu} 
$
(in the spirit of Refs.~\cite{mclaughlin_path_1971,gervais_point_1976}).
The rule~(\ref{eq:3rdordersubstanyd}) can also be formally inferred by expanding the exponential of its l.h.s., using the above substitution rule (valid in prefactor) and re-exponentiating;
but the complete justification of~Eq.~\eqref{eq:3rdordersubstanyd} is the one provided in the present section.
}
In addition we note that, importantly, its right-hand-side depends quadratically on the coefficient of $\Delta x^\mu \Delta x^\nu \Delta x^\rho$
through its last term (which is subdominant in $\Delta t$ but of relevant order in the propagator).
We note that in $d=1$  this last term simplifies to $3 (A^{(3)})^2 g^6 \Delta t$ which is compatible 
with Eq.~(74) in~\cite{Cugliandolo-Lecomte17a}\footnote{%
In Ref.~\cite{Cugliandolo-Lecomte17a} Eq.~(74) 
has a typo and should read:
$
A^{(3)} \Delta x^3\Delta t^{-1}
\doteq
3 A^{(3)}  2D g(x)^2 \Delta x 
+
3\big[A^{[3)}  2 D g(x)^2 \Delta x\big]^2
$.
}
for $D=\frac 12$.

We are now in a position to proceed with the substitution of the cubic term appearing in Eq.~\eqref{eq:subsitution_kinetic_1}. We first note that
\begin{equation}
\omega_{\mu\nu} = \frac{\partial U^\alpha}{\partial x^\mu} \frac{\partial U^\beta}{\partial x^\nu}\Omega_{\alpha\beta} \,,
\end{equation}
so that 
\begin{align}
& - \frac{1-2\alpha}{2}\omega_{\mu\nu}\frac{\partial X^\mu}{\partial u^\alpha}\frac{\partial^2 X^\nu}{\partial u^\beta \partial u^\gamma}\frac{\Delta u^\alpha \Delta u^\beta \Delta u^\gamma}{\Delta t} 
\nonumber \\ 
& 
\qquad\qquad 
=
- \frac{1-2\alpha}{2} \Omega_{\alpha\sigma} \frac{\partial U^\sigma}{\partial x^\nu}\frac{\partial^2 X^\nu}{\partial u^\beta \partial u^\gamma}\frac{\Delta u^\alpha \Delta u^\beta \Delta u^\gamma}{\Delta t} 
\nonumber \\  
& 
\qquad\qquad
=
- \frac{1-2\alpha}{6}\left(\Omega_{\alpha\sigma} \frac{\partial U^\sigma}{\partial x^\nu}\frac{\partial^2 X^\nu}{\partial u^\beta \partial u^\gamma} +  \Omega_{\beta\sigma} \frac{\partial U^\sigma}{\partial x^\nu}\frac{\partial^2 X^\nu}{\partial u^\alpha \partial u^\gamma} 
\right.
\nonumber\\
& 
\qquad\qquad\qquad\qquad\qquad
\left.
+  \Omega_{\gamma\sigma} \frac{\partial U^\sigma}{\partial x^\nu}\frac{\partial^2 X^\nu}{\partial u^\alpha \partial u^\beta}\right)\frac{\Delta u^\alpha \Delta u^\beta \Delta u^\gamma}{\Delta t} 
\nonumber \\ 
& 
\qquad\qquad
\doteq
- \frac{1-2\alpha}{2}\Delta u^\gamma \left(2 \frac{\partial U^\beta}{\partial x^\nu}\frac{\partial^2 X^\nu}{\partial u^\beta \partial u^\gamma} + \Omega^{\alpha\beta}\Omega_{\gamma\sigma}\frac{\partial U^\sigma}{\partial x^\nu}\frac{\partial^2 X^\nu}{\partial x^\alpha \partial x^\beta}\right) \nonumber \\ 
& 
\qquad\qquad
\;\;\;\;
+ \frac{(1-2\alpha)^2}{4}\left(\Omega^{\beta \eta}\Omega^{\gamma \phi} \Omega_{\sigma \psi} \frac{\partial U^\sigma}{\partial x^\nu}\frac{\partial U^\psi}{\partial x^\mu}\frac{\partial^2 X^\nu}{\partial u^\beta \partial u^\gamma}\frac{\partial^2 X^\mu}{\partial u^\eta \partial u^\phi} 
\right.
\nonumber\\
& 
\left.
\qquad\qquad\qquad\qquad\qquad\;\;\;\;
+ 2 \Omega^{\gamma \rho} \frac{\partial U^\sigma}{\partial x^\nu}\frac{\partial^2 X^\nu}{\partial u^\beta \partial u^\gamma}\frac{\partial U^\beta}{\partial x^\mu}\frac{\partial^2 X^\mu}{\partial u^\sigma \partial u^\rho}\right)\Delta t 
\end{align}
(where we kept a fully explicit notation for the derivatives since we have a coexistence of partial derivatives with respect to~$x^\mu$ and $u^\nu$).
The last line proceeds by applying Eq.~\eqref{eq:3rdordersubstanyd} with
\begin{align}
A^{(3)}_{\alpha\beta\gamma} 
= & \; 
- \frac{1-2\alpha}{6}\left(\Omega_{\alpha\sigma} \frac{\partial U^\sigma}{\partial x^\nu}\frac{\partial^2 X^\nu}{\partial u^\beta \partial u^\gamma} +  \Omega_{\beta\sigma} \frac{\partial U^\sigma}{\partial x^\nu}\frac{\partial^2 X^\nu}{\partial u^\alpha \partial u^\gamma}
\right.
\nonumber\\
&
\qquad\qquad\quad
\left.
+  \; \Omega_{\gamma\sigma} \frac{\partial U^\sigma}{\partial x^\nu}\frac{\partial^2 X^\nu}{\partial u^\alpha \partial u^\beta}\right),
\end{align}
which is written in a fully symmetric form as requested for Eq.~\eqref{eq:3rdordersubstanyd} to hold.
Inserting the above equation into Eq.~\eqref{eq:subsitution_kinetic_1}, we obtain\footnote{%
In dimension one, this becomes:
\begin{align*}
& 
 -\frac 12 g(x+\alpha \Delta x)^2 \frac{\Delta x^2}{\Delta t}
\doteq
\text{MCR}
-
(1-2 \alpha) \frac{X''(u)}{X'(u)} \Delta u
+
\frac{(1-2 \alpha )^2}{2} \frac{G(u)^2 X''(u)^2}{ X'(u)^2} \Delta t
\\
&
\qquad
\qquad
\qquad
-\frac{1-3 \alpha(1-\alpha )  }{2} \frac{G(u)^2 X'''(u)}{X'(u)} \Delta t
+
\frac{3 \alpha(1-\alpha ) }{2} \frac{ G'(u) X'(u)+G(u) X''(u)}{ X'(u)^2}  G(u) X''(u)\Delta t
\end{align*}
which corrects Eq.~(88) of Ref.~\cite{Cugliandolo-Lecomte17a} which had a calculation mistake.
}
\begin{align}
& 
-\frac{1}{2}\omega_{\mu\nu}(\bx + \alpha \Delta \bx)\frac{\Delta x^\mu \Delta x^\nu}{\Delta t} 
\doteq 
\;\; 
\text{MCR} -(1-2\alpha)\Delta u^\gamma \frac{\partial U^\beta}{\partial x^\nu}\frac{\partial^2 X^\nu}{\partial u^\beta \partial u^\gamma} 
\nonumber \\ 
& 
\qquad\qquad
+ \frac{(1-2\alpha)^2}{2}\Omega^{\gamma \rho} \frac{\partial U^\sigma}{\partial x^\nu}\frac{\partial^2 X^\nu}{\partial u^\beta \partial u^\gamma}\frac{\partial U^\beta}{\partial x^\mu}\frac{\partial^2 X^\mu}{\partial u^\sigma \partial u^\rho}\Delta t 
\nonumber \\ 
& 
\qquad\qquad
- \frac{1-3\alpha(1-\alpha)}{2}\omega_{\mu\nu}\frac{\partial X^\mu}{\partial u^\alpha}\frac{\partial^3 X^\nu}{\partial u^\rho \partial u^\beta \partial u^\gamma}\Omega^{\alpha\rho}\Omega^{\beta\gamma}\Delta t 
\nonumber \\
& 
\qquad\qquad
- \frac{\alpha(1-\alpha)}{4}\frac{\partial \omega_{\mu\nu}}{\partial x^\eta}\frac{\partial^2 X^\eta}{\partial u^\alpha \partial u^\beta}\frac{\partial X^\mu}{\partial u^\rho}\frac{\partial X^\nu}{\partial u^\sigma}\left(\Omega^{\alpha\beta}\Omega^{\rho\sigma}+2\Omega^{\alpha\rho}\Omega^{\beta\sigma}\right)\Delta t 
\nonumber \\
& 
\qquad\qquad
+ O\big(\Delta t^{3/2}\big)\,.
\label{eq:subsitution_kinetic_2}
\end{align}
The rest of the derivation consists in  noting that
\begin{equation}
\frac{\partial U^\beta}{\partial x^\nu}\frac{\partial^2 X^\nu}{\partial u^\beta \partial u^\gamma} = \frac{\partial}{\partial u^\gamma}\ln \det \left(\frac{\partial X^\mu}{\partial u^\nu}\right),
\end{equation}
so that the term ${O}(\Delta t^{1/2})$ in Eq.~(\ref{eq:subsitution_kinetic_2}) takes the form of a total derivative. Within the path integral, it will therefore generate a boundary term. Recall however that in Eq.~\eqref{eq:subsitution_kinetic_2} all functions are $\alpha$-discretized so we first need to rewrite it in the Stratonovich discretization to be able to use the normal chain rule. We obtain 
\begin{align}
& 
-(1-2\alpha)\Delta u^\gamma \left.\frac{\partial}{\partial u^\gamma}
\left(\ln \det \left(\frac{\partial X^\mu}{\partial u^\nu}\right)\right)\right|_{\bu + \alpha \Delta \bu}
 \doteq 
 \nonumber\\
 & 
 \qquad\qquad \qquad\qquad
  -(1-2\alpha)\Delta u^\gamma \left.\frac{\partial}{\partial u^\gamma}\left(\ln \det \left(\frac{\partial X^\mu}{\partial u^\nu}\right)\right)\right|_{\bu + \Delta \bu/2} 
\nonumber \\  
&
\qquad\qquad \qquad\qquad
+ \frac{(1-2\alpha)^2}{2}\Omega^{\gamma\rho} \frac{\partial^2}{\partial u^\gamma \partial u^\rho}\ln \det \left(\frac{\partial X^\mu}{\partial u^\nu}\right)\Delta t \,.
\end{align}
Once integrated over time between $t_0$ and $\tf$, the boundary term will bring in a contribution to the normalization prefactor equal to
\begin{equation}
\left(\frac{\det \left.\partial X^\mu/\partial u^\nu\right|_{\buf}}{\det \left.\partial X^\mu/\partial u^\nu\right|_{\bu_0}}\right)^{-(1-2\alpha)} \;.
\label{eq:boundary_term_ext}
\end{equation}

Overall, we therefore obtain the generalization of Itō's lemma for the transformation of the $\alpha$-discretized kinetic term of the path-integral Lagrangian
\begin{align}\label{eq:ito_extended_final}
& 
-\frac{1}{2}\omega_{\mu\nu}(\bx + \alpha \Delta \bx)\frac{\Delta x^\mu \Delta x^\nu}{\Delta t} 
\doteq 
\text{MCR} + \text{BT} 
\nonumber\\
& 
\qquad\quad
 - \frac{\alpha(1-\alpha)}{2}\left[\Omega^{\alpha\beta}\frac{\partial U^\rho}{\partial x^\nu}\frac{\partial^3 X^\nu}{\partial u^\alpha \partial u^\beta \partial u^\rho} \right. 
\nonumber \\ 
& 
\qquad\qquad \qquad\qquad \quad
\left.
+ \frac{1}{2}\frac{\partial \omega_{\mu\nu}}{\partial x^\eta}\frac{\partial^2 X^\eta}{\partial u^\alpha \partial u^\beta}\frac{\partial X^\mu}{\partial u^\rho}\frac{\partial X^\nu}{\partial u^\sigma}\left(\Omega^{\alpha\beta}\Omega^{\rho\sigma}+2\Omega^{\alpha\rho}\Omega^{\beta\sigma}\right)\right] \Delta t \nonumber\\
&
\qquad\quad 
+ O\big(\Delta t^{3/2}\big) \,,
\end{align}
where BT accounts for the boundary terms.
Up to these boundary terms, corrections to the results inferred from a blind use of Itō's lemma take the form of a universal term (in the sense that it does not depend on the $\alpha$-discretization scheme used to construct the path integral) with an amplitude proportional to $\alpha(1-\alpha)$ that vanishes when $\alpha=0$ and $\alpha=1$.

\subsection{Transformation rules for $\alpha$-discretized path integrals}\label{subsec:ito_plus12}

We summarize here the above results to state how $\alpha$-discretized path integrals transform under changes of coordinates. 
We take as a starting point the $\alpha$-discretized scalar invariant propagator for a $d$-dimensional process $\bx(t)$
\begin{align}
\K_\bx(\bxf,\tf|\bx_0,t_0) & = \int_{\bx(t_0)=\bx_0}^{\bx(\tf)=\bxf} \mathcal{D}\bx \, \,  \ee^{-\mathcal{S}[\bx(t)]}\,,
\end{align}
with 
\begin{align}
\mathcal{D}\bx = \lim_{N \to + \infty} 
\left(\frac{1}{\sqrt{2\pi\Delta t}}\right)^{\!\!{d}}\;
 \prod_{k=1}^{N-1}\left\{ \frac{\dd \bx_k \sqrt{\omega(\bx_k)}}{\sqrt{2\pi\Delta t}^{\, d}}\right\}  ,
\end{align}
discretized so as to be invariant under a change of variables and the action $\mathcal{S}[\bx(t)]$ as an $\alpha$-discretized integral
\begin{equation}
\mathcal{S}[\bx(t)] \stackrel{\alpha}{=} \int_{t_0}^{\tf}\dd t \,\, \mathcal{L}^{\bx}_{\alpha}[\textbf{x}(t),\dot{\textbf{x}}(t)]\,.
\end{equation}
Equation~\eqref{eq:ito_extended_final} then states that for a transformation $\bu(t) = \textbf{U}(\bx(t))$, the propagator reads
\begin{align}
\K_\bu(\buf,\tf|\bu_0,t_0) & = \left(\frac{\det \left.\partial X^\mu/\partial u^\nu\right|_{\buf}}{\det \left.\partial X^\mu/\partial u^\nu\right|_{\bu_0}}\right)^{-(1-2\alpha)} \int_{\bu(t_0)=\bu_0}^{\bu(\tf)=\buf} \mathcal{D}\bu \, \,  \ee^{-\tilde{\mathcal{S}}[\bu(t)]}\,,
\end{align}
where the action writes as the $\alpha$-discretized integral of a Lagrangian $\tilde{\mathcal{L}}^{\bu}_{\alpha}$,
\begin{equation}
\tilde{\mathcal{S}}[\bu(t)] \stackrel{\alpha}{=} \int_{t_0}^{\tf}\dd t \,\, \tilde{\mathcal{L}}^{\bu}_{\alpha}[\textbf{u}(t),\dot{\textbf{u}}(t)]\,,
\end{equation}
which can be inferred from the original one $\mathcal{L}^{\bx}_{\alpha}[\textbf{x}(t),\dot{\textbf{x}}(t)]$ as
\begin{equation}\label{eq:transfo_lagrange}
\tilde{\mathcal{L}}^{\bu}_{\alpha}[\textbf{u},\dot{\textbf{u}}] 
= 
\mathcal{L}^{\bx}_{\alpha}\left[\textbf{X}(\bu),\partial_{u^\mu} X^\alpha \dot{u}^\mu + \frac{1-2\alpha}{2}\Omega^{\mu\nu}\partial_{u^\mu\sns}\partial_{u^\nu}
 X^{\alpha}\right] + \alpha(1-\alpha)\sms\delta\sns\mathcal{L}[\textbf{u}]\,.
\end{equation}
The first term in the right-hand side of Eq.~\eqref{eq:transfo_lagrange} is the one obtained by naively using the modified chain rule in the $\alpha$-discretized continuous-time Lagrangian $\mathcal{L}^{\bx}_{\alpha}[\bx,\dot{\bx}]$. The second one quantifies the deviations from the modified chain rule. In Eq.~\eqref{eq:ito_extended_final}, we have shown that
\begin{align}
\delta\sns\mathcal{L}[\textbf{u}] = &
\; \frac{1}{2}\left[\Omega^{\alpha\beta}\frac{\partial U^\rho}{\partial x^\nu}\frac{\partial^3 X^\nu}{\partial u^\alpha \partial u^\beta \partial u^\rho} 
\right.
\nonumber\\
& 
\left.
\;\; \;\;\;+ \; \frac{1}{2}\frac{\partial \omega_{\mu\nu}}{\partial x^\eta}\frac{\partial^2 X^\eta}{\partial u^\alpha \partial u^\beta}\frac{\partial X^\mu}{\partial u^\rho}\frac{\partial X^\nu}{\partial u^\sigma}\left(\Omega^{\alpha\beta}\Omega^{\rho\sigma}+2\Omega^{\alpha\rho}\Omega^{\beta\sigma}\right)\right] \,.
\end{align}
For $\alpha \neq 0,1$, blindly applying the modified chain rule of stochastic calculus in the continuous-time Lagrangian therefore leads to incorrect results --~and it is indeed not correct to use the usual chain rule for $\alpha=1/2$. Remarkably, if $\alpha = 0$ or $\alpha = 1$, the correcting term vanishes and, up to boundary terms, the modified chain rule can be directly applied at the level of continuous-time path integrals,
which is compatible with the results of~\cite{ding2021timeslicing}. 

An instructive illustration comes from the study of the free particle in the two-dimensional plane, see Sec.~\ref{subsec:free-particle}. In Cartesian coordinates, the action reads,
\begin{equation}\label{eq:cont_time_action_2d}
\mathcal{S}[x(t),y(t)] = \frac{1}{4D}\int_{t_0}^{\tf}\dd t \, \left(\dot{x}^2+\dot{y}^2\right).
\end{equation}
Furthermore, from Eq.~\eqref{eq:Langevin-polar}, we have at the level of Itō discretized Langevin equations,
\begin{equation}
 \begin{pmatrix}
\dot{x} \\
\dot{y}
\end{pmatrix} =  \begin{pmatrix}
\cos\phi & -r \sin\phi \\
\sin\phi & r\cos\phi
\end{pmatrix} \begin{pmatrix}
\dot{r} - \frac{D}{r} \\
\dot{\phi}
\end{pmatrix}\,.
\end{equation}
Hence, by blindly applying Itō's lemma to the continuous-time action Eq.~\eqref{eq:cont_time_action_2d}, and upon interpreting the result as an Itō integral, we would obtain
\begin{align}\label{eq:cont_time_action_2d_rphi}
\mathcal{S}[r(t),\phi(t)] & \stackrel{0}{=} \frac{1}{2}\int_{t_0}^{\tf}\dd t \left[\frac{1}{2D}\left(\dot{r}^2+ r^2 \dot{\phi}^2\right)-\frac{\dot{r}}{r}+\frac{(2D)}{4r^2}\right] \nonumber \\
& \stackrel{0}{=} \frac{1}{2}\int_{t_0}^{\tf}\dd t \left[\frac{1}{2D}\left(\dot{r}^2+ r^2 \dot{\phi}^2\right)+\frac{\dot{r}}{r}-\frac{3(2D)}{4r^2}\right] - \int_{t_0}^{\tf}\dd t \left[\frac{\dot{r}}{r} - \frac{D}{r^2}\right] \,.
\end{align}
Furthermore, the Itō integral of $\dot{r}/r$ can be rewritten as
\begin{align}
\int_{\substack{
   \!\!\! t_0 \\
   \scriptstyle{\alpha=0}}
  }^{\tf} 
\dd t \; \frac{\dot{r}}{r} 
& = 
\int_{\substack{
   \!\!\! \!\!\! \!\!\! t_0 \\
   \scriptstyle{\alpha=1/2}}
  }^{\tf} 
\dd t \; \frac{\dot{r}}{r} 
+ \int_{t_0}^{\tf}\dd t \; \frac{2D}{2r^2}  = \ln\left(\frac{r_{\text{f}}}{r_0}\right) + \int_{t_0}^{\tf}\dd t \; \frac{D}{r^2} \, ,
\end{align}
so that 
\begin{align}\label{eq:cont_time_action_2d_rphi_ln}
\mathcal{S}[r(t),\phi(t)] & \stackrel{0}{=}  \frac{1}{2}\int_{t_0}^{\tf}\dd t \left[\frac{1}{2D}\left(\dot{r}^2+ r^2 \dot{\phi}^2\right)+\frac{\dot{r}}{r}-\frac{3(2D)}{4r^2}\right] - \ln\left(\frac{r_{\text{f}}}{r_0}\right).
\end{align}
which indeed matches the correct result of Eq.~\eqref{eq:Litopolar} up to a boundary term. Additionally, in the Itō case $\alpha = 0$, this boundary term can be absorbed by a redefinition of the path measure as follows~\cite{ding2021timeslicing}. For the stochastic process,
\begin{equation}
\frac{\dd x^\mu}{\dd t} \stackrel{0}{=} f^\mu(\bx) + g^{\mu i}(\bx) \eta_i \, ,
\end{equation}
we can choose to write
\begin{align}
\K(\bxf,\tf|\bx_0,t_0) = \int_{\bx(t_0)=\bx_0}^{\bx(\tf)=\bxf} \tilde{\mathcal{D}}\bx \, \,  \ee^{-\mathcal{S}[\bx(t)]}
\end{align}
with the path measure now defined as 
\begin{align} \label{eq:path_measure_ito}
\tilde{\mathcal{D}}\bx = \lim_{N \to + \infty} \left(\frac{1}{\sqrt{2\pi\Delta t}}\right)^d \int \prod_{k=1}^{N-1}\left\{ \frac{\dd \bx_k \sqrt{\omega(\bx_{k-1})}}{\sqrt{2\pi\Delta t}^{\, d}}\right\}  ,
\end{align}
corresponding to $\gamma = 0$ in the language of Eq.~\eqref{eq:infi_prop_general}, and the action reading
\begin{align} \label{eq:Sforpath_measure_ito}
\mathcal{S}[\bx(t)] \stackrel{0}{=} \frac{1}{2}\int_{t_0}^{\tf} \dd t \; \omega_{\mu\nu}\left(\frac{\dd x^\mu}{\dd t} - f^\mu\right)\left(\frac{\dd x^\nu}{\dd t} - f^\nu\right) .
\end{align}
If such a convention is taken, the path measure is not invariant under changes of coordinate but transforms as  
\begin{align} \label{eq:transfoDxIto}
\mathcal{D}\bx = \mathcal{D}\bu \left(\frac{\det \left.\partial X^\mu/\partial u^\nu\right|_{\buf}}{\det \left.\partial X^\mu/\partial u^\nu\right|_{\bu_0}}\right) ,
\end{align}
which exactly compensates the boundary term in Eq.~\eqref{eq:boundary_term_ext}. This result can be illustrated on our introductory example about the kinetic energy ${\mathcal K}$ of a Brownian particle, see Sec.\ref{sec:brownian-particle}. The kinetic energy obeys the Itō discretized stochastic differential equation
\begin{align}
\label{eq:langevin_kineticI}
\frac{\dd K}{\dd t}\stackrel{0}{=}&-\frac{2\gamma}{m}{\mathcal K}+\frac{\gamma T}{m}+\sqrt{\frac{4\gamma T {\mathcal K}}{m}}\eta
\; . 
\end{align}
Regarding the corresponding path-integral formulation, the  kinetic energy scalar invariant propagator reads, in the Itō scheme, 
\begin{equation}\label{eq:PItrivialItoK}
{\mathbb K}({\Kf},\tf|K_0,0)=\int_{\mathcal K(0)=0}^{\mathcal K(\tf)=\Kf}{\mathscr D}{\mathcal K}\:
\exp
\left[-\frac{m}{8\gamma T}\int_0^{\tf}\dd t \; \frac{1}{ {\mathcal K}}\left(\frac{\dd {\mathcal K}}{\dd  t}+\frac{2\gamma}{m}{\mathcal K}-\frac{\gamma T}{m}\right)^2
\right],
\end{equation}
where
\begin{equation}
\frac{1}{\sqrt{2\pi\Delta t}}\prod_{k=1}^{\tf/\Delta t-1}
\left(\sqrt{\frac{m}{8\pi\gamma T \Delta t\, {\mathcal K}_{k-1}}}\right)\dd {\mathcal K}_k\to {\mathscr D} {\mathcal K} 
\end{equation}
is discretized according to Eq.~\eqref{eq:path_measure_ito} and
\begin{equation}
\Delta t\sum_{k=0}^{\tf/\Delta t-1} \! 
\frac{1}{{\mathcal K}_k}\left(\frac{{\mathcal K}_{k+1}-{\mathcal K}_{k}}{\Delta t}+\frac{2\gamma}{m} {\mathcal K}_{k}-\frac{\gamma T}{m}\right)^2\to \int_0^{\tf}\!\! \dd t \, \frac{1}{ {\mathcal K}}\left(\frac{\dd {\mathcal K}}{\dd  t}+\frac{2\gamma}{m}{\mathcal K}-\frac{\gamma T}{m}\right)^2
\!\! .
\end{equation}
If we now consider the velocity $v=\sqrt{2{\mathcal K}/m}$, the path-integral weight
\begin{equation}
-\frac{1}{4\gamma T}\int_0^{\tf}\dd t\left(m\frac{\dd v}{\dd t}+\gamma v\right)^2
\end{equation} 
of Eq.~\eqref{eq:PItrivialIto} can indeed be readily obtained from Eq.~\eqref{eq:PItrivialItoK} by a blind use of Itō's lemma,
\begin{equation}
\frac{\dd {\mathcal K}}{\dd t} = m v \frac{\dd v}{\dd t} + \frac{\gamma T}{m}\,.
\end{equation} 
Such manipulations would not be valid for $\alpha\neq 0,1$ and in particular in the Stratonovich discretization.

\subsection{An alternative discretization of the measure $\mathcal{D}\bx$}\label{subsec:new_measure}

For the sake of generality, we introduce an alternative discretization of the path measure which 
extends Eq.~\eqref{eq:path_measure_ito} to any $\alpha$ and is commonly found in the literature (see for instance~\cite{lau_state-dependent_2007,Aron_etal_2014,itami_universal_2017,Cugliandolo-Lecomte17a}). 
We also present here the new corresponding action and its transformation rules under changes of variables. The 
resulting expressions will be used in Sec.~\ref{sec:msrjd} to construct a Martin--Siggia--Rose--Janssen--De$\,\,$Dominicis (MSRJD) 
path integral with manifestly covariant path measure over the physical and response fields. 

We define the $\alpha$ measure as
\begin{equation}\label{eq:path_measure_new}
\mathcal{D}^{(\alpha)}\bx = \lim_{N \to + \infty} \left(\frac{1}{\sqrt{2\pi\Delta t}}\right)^d \int \prod_{k=1}^{N-1}\left\{ \frac{\dd \bx_k \sqrt{\omega(\bx_{k-1} + \alpha \Delta \bx_{k-1})}}{\sqrt{2\pi\Delta t}^{\, d}}\right\} .
\end{equation}
We note that the manifestly covariant path measure $\mathcal{D}\bx$ of Eq.~\eqref{eq:path_measure} corresponds to 
the case $\alpha=1$. 
The $\alpha$ measure can be re-expressed in terms of $\mathcal{D}\bx$, 
up to negligible corrections in the continuous-time limit, as
\begin{equation}
\mathcal{D}\bx = \mathcal{D}^{(\alpha)}\bx \, \exp\left(-\delta \mathcal{S}^{(\alpha)}[\bx(t)]\right) 
\end{equation}
where
\begin{equation}\label{eq:shift_action}
\delta \mathcal{S}^{(\alpha)}[\bx(t)] \stackrel{\alpha}{=} - (1-\alpha)\int_{t_0}^{\tf} \dd t \, \dot{x}^\mu \Gamma_{\mu\alpha}^{\alpha} - \frac{(1-\alpha)^2}{2}\int_{t_0}^{\tf} \dd t \, \omega^{\mu\nu} \partial_\nu\Gamma_{\mu\alpha}^{\alpha} \,.
\end{equation}
Therefore, the scalar invariant propagator can be written in different ways
\begin{align}
\K(\bxf,\tf|\bx_0,t_0) 
 = \int_{\bx(t_0)=\bx_0}^{\bx(\tf)=\bxf} \mathcal{D}\bx \, \,  \ee^{-\mathcal{S}[\bx(t)]}
= \int_{\bx(t_0)=\bx_0}^{\bx(\tf)=\bxf} \mathcal{D}^{(\alpha)}\bx \, \,  
\ee^{-\mathcal{S}^{(\alpha)}[\bx(t)]}
\;, 
\end{align}
with 
\begin{equation}
\mathcal{S}^{(\alpha)}[\bx(t)] \equiv \mathcal{S}[\bx(t)]+\delta \mathcal{S}^{(\alpha)}[\bx(t)]\:.
\end{equation}
We stress that, because the definition of the measure depends explicitly  on $\alpha$, we have $\mathcal{S}^{(\alpha)}[\bx(t)] - \mathcal{S}^{(\alpha')}[\bx(t)] \neq 0$ in general. The new action can be written 
as the $\alpha$-discretized integral of a new Lagrangian $\tilde{\mathcal{L}}_{(\alpha)}^\bx\sns[\dot{\bx},\bx]$ which, from Eqs.~\eqref{eq:Lalpha} and~\eqref{eq:shift_action}, is given by
\begin{align}
\label{eq:Lalpha_new}
\!\!\! \tilde{\mathcal{L}}_{(\alpha)}^\bx[\dot{\bx},\bx]
 &
  = 
   \frac{1}{2}\left[\omega_{\mu\nu}\left(\frac{\dd x^\mu}{\dd t} - h^\mu\right)\left(\frac{\dd x^\nu}{\dd t} - h^\nu\right) + \frac{\dd x^\mu}{\dd t}\left(\left(1-2\alpha\right)\omega_{\mu\nu}\omega^{\rho\sigma}\Gamma_{\rho\sigma}^\nu - 2\alpha \Gamma^\alpha_{\mu\alpha}\right)  \right. 
\nonumber\\ 
& \;\;
\left. + 2 \alpha \nabla_\mu h^\mu - (1-2\alpha)\omega_{\mu\nu}\omega^{\rho\sigma}\Gamma^{\nu}_{\rho\sigma}h^\mu + \left(\alpha - \frac{1}{2}\right)^2\omega_{\mu\nu}\omega^{\rho\sigma}\omega^{\alpha\beta}\Gamma_{\rho\sigma}^\mu \Gamma_{\alpha\beta}^\nu  \right. 
\nonumber\\
 &  \;\;
 - \alpha\left(1-\alpha\right)R + \alpha\left(1-\alpha\right)\omega^{\mu\nu}\Gamma_{\beta\mu}^\alpha \Gamma_{\alpha\nu}^\beta + \alpha(2\alpha - 1)\omega^{\mu\nu}\partial_\nu \Gamma^\alpha_{\mu\alpha} \Bigg] \, .
\end{align}

Under a non-linear change of variables, the transformation properties of the new Lagrangian $\tilde{\mathcal{L}}_{(\alpha)}^\bx\sns[\dot{\bx},\bx]$ are altered with respect to Eq.~\eqref{eq:transfo_lagrange} as the non-invariance of the redefined path measure also needs to be taken into account. In order to investigate the transformation properties of the measure $\mathcal{D}^{(\alpha)}\bx$ we start by noticing that
\begin{align}
\delta \mathcal{S}^{(\alpha)}[\bx(t)] & \stackrel{\frac{1}{2}}{=} - \frac{(1-\alpha)}{2}\int_{t_0}^{\tf} \dd t \, \dot{x}^\mu \partial_\mu \ln \omega - \frac{\alpha(1-\alpha)}{2}\int_{t_0}^{\tf} \dd t \, \omega^{\mu\nu} \partial_\nu\Gamma_{\mu\alpha}^{\alpha} \, 
\nonumber \\ & = - \frac{(1-\alpha)}{2}\ln \left(\frac{\omega(\bxf)}{\omega(\bx_0)}\right) - \frac{\alpha(1-\alpha)}{2}\int_{t_0}^{\tf} \dd t \, \omega^{\mu\nu} \partial_\nu\Gamma_{\mu\alpha}^{\alpha} \,.
\end{align}
Under a non-linear change of variables $\bu(t) = \textbf{U}(\bx(t))$, and upon introducing 
\begin{equation}
\mathcal{D}^{(\alpha)}\bu = \lim_{N \to + \infty} \left(\frac{1}{\sqrt{2\pi\Delta t}}\right)^d \int \prod_{k=1}^{N-1}\left\{ \frac{\dd \bu_k \sqrt{\Omega(\bu_{k-1} + \alpha \Delta \bu_{k-1})}}{\sqrt{2\pi\Delta t}^{\, d}}\right\} ,
\end{equation}
we thus obtain
\begin{align}\label{eq:transform_measure}
\dfrac{\mathcal{D}^{(\alpha)}\bu }{ \mathcal{D}^{(\alpha)}\bx }
= 
&
\left(\frac{\det \left.\partial X^\mu/\partial u^\nu\right|_{\buf}}{\det \left.\partial X^\mu/\partial u^\nu\right|_{\bu_0}}\right)^{-(1-\alpha)} 
\nonumber\\
& \qquad
\exp\left[-\frac{\alpha(1-\alpha)}{2}\int_{t_0}^{\tf} \dd t \, \left( \Omega^{\mu\nu} \frac{\partial \hat\Gamma^{\alpha}_{\nu\alpha}}{\partial u^\mu} - \omega^{\mu\nu} \frac{\partial \Gamma^{\alpha}_{\nu\alpha}}{\partial x^\mu}\right)\right]
\end{align}
where $\hat\Gamma^{\alpha}_{\nu\beta}$ is the Christoffel symbol associated to the metric $\Omega_{\mu\nu}$. Lastly, using the transformation properties of the Christoffel symbol, \textit{i.e.}
\begin{equation}
\hat\Gamma^{\alpha}_{\nu\beta} = \frac{\partial U^\alpha}{\partial x^\rho}\frac{\partial X^\sigma}{\partial u^\nu}\frac{\partial X^\lambda}{\partial u^\beta}\Gamma^{\rho}_{\sigma\lambda} + \frac{\partial^2 X^\rho}{\partial u^\nu\partial u^\beta}\frac{\partial U^\alpha}{\partial x^\rho} \, ,
\end{equation}
we deduce that
\begin{align}
\Omega^{\mu\nu} \frac{\partial \hat\Gamma^{\alpha}_{\nu\alpha}}{\partial u^\mu} - \omega^{\mu\nu} \frac{\partial \Gamma^{\alpha}_{\nu\alpha}}{\partial x^\mu} 
& = 
\frac{1}{2}\Omega^{\alpha\beta}\omega^{\rho\sigma}\frac{\partial \omega_{\rho\sigma}}{\partial x^\mu}\frac{\partial^2 x^\mu}{\partial u^\alpha \partial u^\beta} 
\nonumber\\
& \;\; + \omega^{\alpha\beta}\frac{\partial U^\nu}{\partial x^\beta}\frac{\partial X^\mu}{\partial u^\phi \partial u^\nu}\frac{\partial U^\phi}{\partial x^\alpha \partial x^\mu} 
 + \Omega^{\alpha\beta} \frac{\partial^3 X^\mu}{\partial u^\alpha \partial u^\beta \partial u^\nu}\frac{\partial U^\nu}{\partial x^\mu} \,,
\end{align}
which we use to state the transformation properties of the Lagrangian as follows. We take as a starting point the $\alpha$-discretized scalar invariant propagator for a $d$-dimensional process $\bx(t)$ with the new discretization of the path measure in Eq.~\eqref{eq:path_measure_new},
\begin{align}
\K_\bx(\bxf,\tf|\bx_0,t_0) & = \int_{\bx(t_0)=\bx_0}^{\bx(\tf)=\bxf} \mathcal{D}^{(\alpha)}\bx \, \,  \ee^{-\mathcal{S}^{(\alpha)}[\bx(t)]}\,,
\end{align}
where the action $\mathcal{S}^{(\alpha)}[\bx(t)]$ is the $\alpha$-discretized integral
\begin{equation}
\mathcal{S}^{(\alpha)}[\bx(t)] \stackrel{\alpha}{=} \int_{t_0}^{\tf}\dd t \,\, \tilde{\mathcal{L}}_{(\alpha)}^\bx[\dot{\bx},\bx]\,.
\end{equation}
Equation~\eqref{eq:ito_extended_final} together with Eq.~\eqref{eq:transform_measure} then states that for a transformation $\bu(t) = \textbf{U}(\bx(t))$, the propagator reads
\begin{align}\label{eq:new_BT}
\K_\bu(\buf,\tf|\bu_0,t_0) & = \left(\frac{\det \left.\partial X^\mu/\partial u^\nu\right|_{\buf}}{\det \left.\partial X^\mu/\partial u^\nu\right|_{\bu_0}}\right)^{\alpha} \int_{\bu(t_0)=\bu_0}^{\bu(\tf)=\buf} \mathcal{D}^{(\alpha)}\bu \, \,  \ee^{-\tilde{\mathcal{S}}^{(\alpha)}[\bu(t)]}\,,
\end{align}
where the action is the $\alpha$-discretized integral of a Lagrangian $\tilde{\mathcal{L}}^{\bu}_{(\alpha)}$,
\begin{equation}
\tilde{\mathcal{S}}^{(\alpha)}[\bu(t)] \stackrel{\alpha}{=} \int_{t_0}^{\tf}\dd t \,\, \tilde{\mathcal{L}}^{\bu}_{(\alpha)}[\textbf{u}(t),\dot{\textbf{u}}(t)]\,,
\end{equation}
which can be inferred from the original one $\tilde{\mathcal{L}}_{(\alpha)}^\bx[\dot{\bx},\bx]$ as
\begin{equation}\label{eq:transfo_lagrange_new}
\tilde{\mathcal{L}}^{\bu}_{(\alpha)}[\textbf{u},\dot{\textbf{u}}]
 =
 \tilde{\mathcal{L}}^{\bx}_{(\alpha)}
 \!\! \left[\textbf{X}(\bu),\partial_{u^\mu\sns} X^\alpha \sms \dot{u}^\mu +
 \frac{1-2\alpha}{2}\Omega^{\mu\nu}\partial_{u^\mu\sns}\partial_{u^\nu\sns} X^{\alpha}\right]
 + \alpha(1-\alpha)\sms\delta\sns\mathcal{L}[\textbf{u}]\,.
\end{equation}
The first term in the right-hand side of Eq.~\eqref{eq:transfo_lagrange} is the one obtained by naively using the modified chain rule in the $\alpha$-discretized continuous-time Lagrangian $\mathcal{L}^{\bx}_{\alpha}[\bx,\dot{\bx}]$. The second one quantifies the deviations from the modified chain rule and now writes
\begin{align}
\delta\sns\mathcal{L}[\textbf{u}] = \frac{1}{2}\frac{\partial \omega_{\mu\nu}}{\partial x^\eta}\frac{\partial^2 X^\eta}{\partial u^\alpha \partial u^\beta}\frac{\partial X^\mu}{\partial u^\rho}\frac{\partial X^\nu}{\partial u^\sigma}\Omega^{\alpha\rho}\Omega^{\beta\sigma}- \frac{1}{2}\omega^{\alpha\beta}\frac{\partial U^\nu}{\partial x^\beta}\frac{\partial^2 X^\mu}{\partial u^\phi \partial u^\nu}\frac{\partial^2 U^\phi}{\partial x^\alpha \partial x^\mu} \,.
\end{align}
This additional contribution comes both from the transformation properties of the quadratic term given in Eq.~\eqref{eq:ito_extended_final} which,
importantly enough, are 
independent of the discretization of the path measure, and that of the path measure itself as shown in Eq.~\eqref{eq:transform_measure}. In the Itō discretization, we recover the fact that the modified chain rule can be blindly used directly at the level of the continuous-time action without generating any additional boundary term. In the following sections we explain how to restore covariance in the action (for the standard chain rule) in a general manner, either by DeWitt's continuous-time path-integral representation, or by building explicit higher-order discretization schemes in a time-sliced approach.

\section{Covariant path-integral representation à la DeWitt}\label{sec:DWG}

In Sec.~\ref{sec:pi_alpha_discrete}, we discussed the non covariance of the formal continuous-time Lagrangian associated to the $\alpha$-discretized path-integral representation of multidimensional stochastic equations. Designing constructions of path integrals that would be manifestly covariant in continuous time thus became a challenge for the theoretical physics community, especially for those attempting to apply Feynman's path-integral methods to quantum mechanics on curved spaces. 

In a seminal paper~\cite{dewitt_dynamical_1957}, later complemented by the work of~\cite{mclaughlin_path_1971}, DeWitt came up with a first answer to this question. In DeWitt's construction, which follows the spirit of path integrals as introduced by Feynman in quantum mechanics~\cite{feynman_space-time_1948}, the infinitesimal propagator is given, up to a proportionality constant, as the exponential of the manifestly covariant action of a classical massive particle in curved space evaluated at the infinitesimal classical path with appropriate boundary conditions. In~\cite{dewitt_dynamical_1957}, it is then shown that such a path integral indeed propagates the solution of a Schrödinger equation with Hamiltonian given, up to an additional potential term proportional to the Ricci curvature of the embedding space, by that of a classical particle quantized in such a way that it remains covariant at the quantum level. 

DeWitt's approach for defining the path integral was later transposed to the context of diffusion processes, similar to the ones we focus on in the present work, by Graham~\cite{graham_path_1977} and later Graham and Deininghaus~\cite{deininghaus_nonlinear_1979} who used a similar construction for the definition of path integrals for multidimensional diffusion processes. 
Early approaches defining a covariant path-integral in dimension one include those of Stratonovich~\cite{Strato-original_1962,stratonovich1971probability} and Horsthemke--Bach~\cite{horsthemke_onsager-machlup_1975}
(who resort to a change of variable to an additive process in order to define the path-integral measure).
While in this section we mostly follow the lines of the derivation by 
Langouche \textit{{et al.}}~\cite{langouche1980short,Tirapegui82}, we will comment on the difference with Graham's approach~\cite{graham_covariant_1977} in the end. 

Following DeWitt, we start by assuming that the process in Eq.~\eqref{eq:pi_langevin_strato_discrete} can be formally described by the following (scalar invariant) path-integral propagator\footnote{%
We recall that the relation between the scalar invariant propagator $\K$ and the actual propagator $\mathbb P$ of the process is
$\K(\bxf,\tf|\bx_0,t_0) = {\mathbb{P}(\bxf,\tf|\bx_0,t_0)}/{\sqrt{\omega(\bx)}}$, 
see Eq.~(\ref{eq:KPomega}).
}
\begin{align} \label{eq:pathdewitt}
%
\K(\bxf,\tf|\bx_0,t_0) = \int_{\bx(t_0)=\bx_0}^{\bx(\tf)=\bxf} \mathcal{D}\bx ~ 
\exp\left[-\int_{t_0}^{\tf} \dd  t \left( \frac{1}{2} \omega_{\mu\nu} \dot{x}^\mu\dot{x}^\nu + a_\mu \dot{x}^\mu + b\right) \right] \, ,
\end{align}
where the path measure is defined in Eq.~\eqref{eq:path_measure} 
and the infinitesimal scalar invariant propagator reads
\begin{align} 
\label{eq:propdewitt}
& 
\K_{\Delta t}^\dW(\bx+\Delta \bx,t+\Delta t | \bx,t) 
= 
\nonumber\\
&
\qquad\qquad\qquad
\frac 1{\big( 2\pi \Delta t\big)\!^{\frac d 2}}
\exp \left[-\int_t^{t+\Delta t}
 \!
 \dd  s \left( \frac{1}{2}\omega_{\mu\nu} \dot{x}_{\cl}^\mu\dot{x}_{\cl}^\nu + a_\mu \dot{x}_{\cl}^\mu + b\right)  \right]
\! .\!
\end{align}
\\
In Eq.~\eqref{eq:propdewitt}, $\bx_{\cl}(s)$ is the infinitesimal minimal-action path such that $\bx_{\cl}(t) = \bx$ and $\bx_{\cl}(t+\Delta t) = \bx+\Delta \bx$. The vector field $a_\mu(\bx)$ and the function $b(\bx)$ are then set by requiring agreement between the formula in  Eq.~\eqref{eq:propdewitt} and the already known Stratonovich-discretized infinitesimal propagator inferred from Eq.~\eqref{eq:Lstrato}. In the following, we evaluate the integral in the exponential of Eq.~(\ref{eq:propdewitt}) and write the result in a Stratonovich-discretized form amenable to immediate comparison. 
\begin{figure}[t!]
~\\[6mm]
\centerline{
\includegraphics[width=.9\columnwidth]{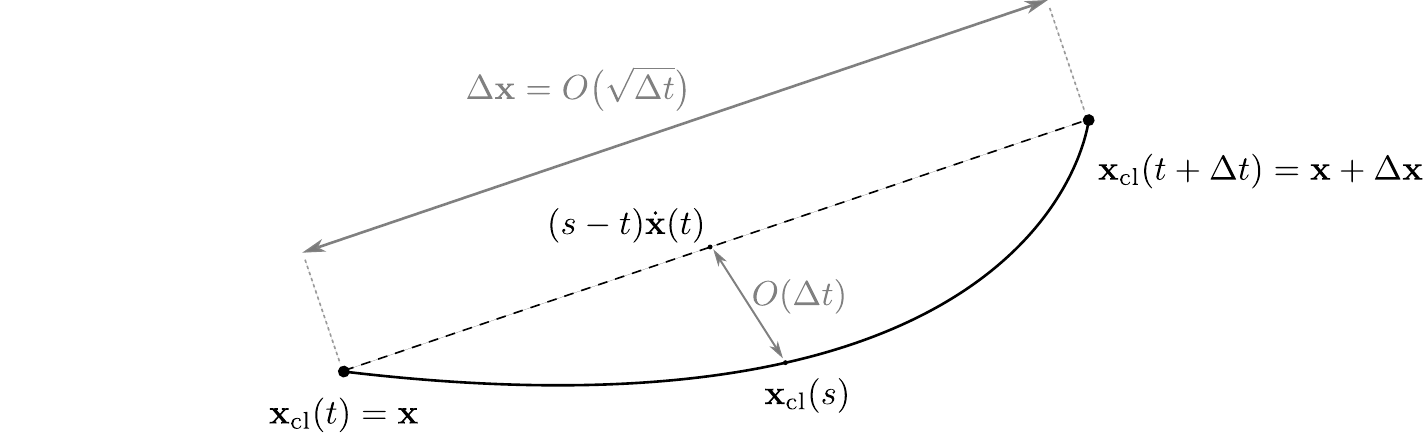}
}
~\\[-5mm]
\caption{Classical trajectory $x_{\text{cl}}(s)$ between the two infinitesimally separated points $\bx$ and $\bx+\Delta x$,
which becomes dominant in the $\Delta t$ limit. It verifies the Euler--Lagrange equation~(\ref{eq:classical}).
}
\label{fig:schemaxcl}
\end{figure}%
We write the classical trajectory (see Fig.~\ref{fig:schemaxcl}) as 
\begin{equation}
x^{\mu}_{\cl}(s) = x^\mu + \delta x^\mu(s)
\end{equation} 
and, due to the boundary conditions,
$\delta x^\mu(s) \sim O(\sqrt{\Delta t})$ and $\dot{x}_{\cl}^\mu(s) \sim O(\Delta t^{-1/2})$. The minimal-action path satisfies the Euler--Lagrange equations (see \textit{e.g.}~Ref.~\cite{adler_introduction_1975}):
\begin{align} \label{eq:classical}
\ddot{x}_{\cl}^\nu + \Gamma^\nu_{\alpha \beta}\dot{x}_{\cl}^\alpha \dot{x}_{\cl}^\beta = \omega^{\nu\rho}\partial_\rho b + \omega^{\nu\rho}(\partial_\rho a_\mu - \partial_\mu a_\rho)\dot{x}_{\cl}^\mu \, .
\end{align} 
Hence, we have $\ddot{x}_{\cl}^\mu(s) \sim O(\Delta t^{-1})$ and recursively the $n^{\text{th}}$ time derivative scales as $x^{\mu,(n)}_{\cl}(s) \sim O(\Delta t^{-n/2})$. Note that for $a_\mu = 0$ and $b = 0$, Eq.~\eqref{eq:classical} is just a geodesic equation. From the above discussed scalings we have 
\begin{equation}
\int_{t}^{t+\Delta t}  \dd s \, b(\bx_{\cl}(s)) = b(\bx) \Delta t + O(\Delta t^{3/2}) 
\end{equation}
and furthermore
\begin{align}
& \int_{t}^{t + \Delta t}  \dd s \; a_\mu(\bx_{\cl}( s)) \, \dot{x}^\mu( s) 
\nonumber\\
& 
\qquad\qquad
 = \int_{t}^{t + \Delta t}  \dd s \left[ a_\mu(\bx) + \delta x^\nu \, \partial_\nu a_\mu(\bx) \right] \,  \dot{x}^\mu( s)  
+ O(\Delta t^{3/2}) 
\end{align}
which becomes
\begin{align}
& \int_{t}^{t + \Delta t}  \dd s \; a_\mu(\bx_{\cl}( s)) \, \dot{x}^\mu( s) 
\nonumber\\
 & 
 \qquad\qquad
 = a_\mu(\bx) \Delta x^\mu 
 + \partial_\nu a_\mu(\bx) \int_{t}^{t+\Delta t} \dd  s \;  \delta x^\nu(s) {\, \dot x^\mu}(s) 
  + O(\Delta t^{3/2}) \, .
 \label{eq:int_a_dW}
\end{align}
We now use the expansions
\begin{equation}\begin{split}\label{eq:scalingsclass}
\delta x^\mu( s) & = ( s - t) \dot{x}^\mu(t) + O(\Delta t) \, , \\
\dot{x}_{\cl}^\mu( s) & = \dot{x}_{\cl}^\mu(t) + O(\Delta t^0) \, ,
\end{split}
\end{equation}
to conclude that to the desired order
\begin{align}
\int_{t}^{t + \Delta t}  \dd  s \, \delta x^\nu( s) {\, \dot x^\mu}( s) = \frac{1}{2}\Delta x^\mu \Delta x^\nu + O(\Delta t^{3/2}) 
\, .
\end{align}
Therefore  the integral in Eq.~\eqref{eq:int_a_dW} can be expressed as
\begin{align}
\int_{t}^{t+\Delta t}  \dd s  \, a_\mu(\bx_{\cl}( s)) \dot{x}^\mu( s)= a_\mu(\bx + \Delta \bx/2) \Delta x^\mu + \mathcal{O}(\Delta t^{3/2}) \,.
\end{align}
Finally, we want to integrate the kinetic term
\begin{align}
\int_{t}^{t+\Delta t} \dd  s \; \omega_{\mu\nu}(\bx_{\cl}( s)) \, \dot{x}_{\cl}^\mu( s) \dot{x}_{\cl}^\nu( s) \, .
\end{align}
First, note that the integrand, that would be strictly conserved over time along a geodesic, remains conserved in the presence of the fields $a_\mu$ and $b$ to the desired order in $\Delta t$ since, 
from Eq.~(\ref{eq:classical}),
\begin{equation}
    \frac{\dd}{\dd  s}\left(\omega_{\mu\nu}\dot{x}_{\cl}^\mu\dot{x}_{\cl}^\nu\right) = 2 \dot{x}^\mu \partial_\mu b \, ,
\label{eq:dSqVelds}
\end{equation}
so that
\begin{equation}
    \omega_{\mu\nu}(\bx_{\cl}( s))\dot{x}_{\cl}^\mu( s)\dot{x}_{\cl}^\nu( s) = \omega_{\mu\nu}(\bx)\dot{x}_{\cl}^\mu(t)\dot{x}_{\cl}^\nu(t) + O(\sqrt{\Delta t}) \, .
\end{equation}
Thus,
\begin{align}
\int_{t}^{t+\Delta t}  \dd  s \, \omega_{\mu\nu}(\bx_{\cl}( s)) \dot{x}_{\cl}^\mu( s) \dot{x}_{\cl}^\nu( s) =  \omega_{\mu\nu}(\bx)\dot{x}^\mu(t)\dot{x}^\nu(t) \Delta t+  O(\Delta t^{3/2}) ~.
\end{align}
Next, we Taylor expand 
\begin{equation}
\Delta x^\mu = \dot{x}^\mu(t) \Delta t + \frac{1}{2} \ddot{x}^\mu(t) \Delta t^2 + \frac{1}{6} \dddot{x}^\mu(t)\Delta t^3 + ...
\end{equation}
and using Eqs.~\eqref{eq:classical}-\eqref{eq:scalingsclass} we obtain
\begin{align}
& 
\frac{1}{2}\int_{t}^{t+\Delta t} \dd  s  \; \omega_{\mu\nu}(\bx_{\cl}( s)) \dot{x}_{\cl}^\mu( s) \dot{x}_{\cl}^\nu( s)  
\nonumber\\
& 
\qquad
=  \frac{1}{2}\omega_{\mu\nu}(\bx)\frac{\Delta x^\mu \Delta x^\nu}{\Delta t} + \frac{1}{4 \Delta t} \partial_\alpha \omega_{\nu \beta}(\bx) \Delta x^\alpha \Delta x^\beta \Delta x^\nu 
\nonumber\\ & 
\qquad\quad
+ \frac{1}{\Delta t}\left[ \frac{1}{12}\partial_\mu\partial_\nu \omega_{\alpha \beta}(\bx) - \frac{1}{24}\Gamma^{\rho}_{\alpha\beta}(\bx) \Gamma^{\sigma}_{\mu\nu}(\bx)\omega_{\rho\sigma}(\bx)\right]
\Delta x^\alpha \Delta x^\beta \Delta x^\mu \Delta x^\nu 
\nonumber\\
&
\qquad\quad
+ O(\Delta t^{3/2}) \, ,
\end{align}
where all functions are evaluated at $\bx$. The details of the computation can be found in~\cite{cheng1972quantization}. Note that to the desired order in $\Delta t$, the result is independent of $a_\mu$ and $b$. This means that evaluating the integral in Eq.~\eqref{eq:propdewitt} along the minimal-action path or along the geodesic path with appropriate boundary conditions is completely equivalent. We can now expand all the functions around $\bx + \Delta \bx/2$ and use the substitution rules to compare this result  to the Stratonovich Lagrangian~(\ref{eq:Lstrato}) obtained in Sec.~\ref{sec:pathint}. Agreement between the two requires
\begin{align}
\label{eq:dWresa}
a_\mu & = - \omega_{\mu\nu} h^\nu \, , \\
\label{eq:dWresb}
b & = \frac{1}{2}\omega_{\mu\nu}h^\mu h^\nu + \frac{1}{2}\nabla_\mu h^\mu + \frac{1}{6}R \, .
\end{align}
Thus, the continuous-time action that describes the stochastic process in Eq.~\eqref{eq:pi_langevin_strato_discrete}, constructed following DeWitt's approach (\textit{i.e.}~using Eq.~\eqref{eq:propdewitt}) takes, unsurprisingly as it was built using covariant notions, 
a manifestly covariant form and reads
\begin{align}\label{eq:actionFeynman}
\mathcal{S}_{\dW}[\bx(t)] = \int_{t_0}^{\tf} \dd  t ~ 
\left[ 
\frac{1}{2}\omega_{\mu\nu}(\dot{x}^\mu - h^\mu)(\dot{x}^\nu - h^\nu) + \frac{1}{2}\nabla_\mu h^\mu + \frac{1}{6}R
\right] 
.
\end{align}
This way of constructing path integrals is therefore, in the continuous-time limit, compatible with the naive use of the chain rule for performing changes of variables. 

In his 1977 paper~\cite{graham_path_1977}, Graham derived the following manifestly covariant continuous-time action, 
\begin{align}\label{eq:actionGraham}
\mathcal{S}_{\rm G}[\bx(t)] = \int_{t_0}^{\tf} \dd  t~
\left[ 
\frac{1}{2}\omega_{\mu\nu}(\dot{x}^\mu - h^\mu)(\dot{x}^\nu - h^\nu) + \frac{1}{2}\nabla_\mu h^\mu + \frac{1}{12}R 
\right]
,
\end{align}
where the curvature contribution has a coefficient $1/12$ instead of the $1/6$ of Eq.~\eqref{eq:actionFeynman}. The reason for this is that in Graham's construction, as was shown in~\cite{deininghaus_nonlinear_1979}, the infinitesimal propagator actually takes the WKB form
\begin{align} \begin{split}\label{eq:graham}
& \K(\bx_1 = \bx+\Delta \bx,t+\Delta t | \bx_0= \bx,t) \\ &
= \left(\Delta t\right)^{d/2}\left(\omega(\bx + \Delta x)\omega(\bx)\right)^{-1/4} \sqrt{ \det{-\frac{\partial^2 \mathcal{S}_{\rm G}[\bx_{\cl}(t)]}{\partial x_0^\mu \partial x_1^\nu}}} \; \exp{-\mathcal{S}_{\rm G}[\bx_{\cl}(t)]} \, ,
\end{split}
\end{align}
with the so-called Van Vleck--Pauli--Morette determinant~\cite{morette1951definition} given by
\begin{align}
&   \left(\Delta t\right)^{d/2}\left(\omega(\bx + \Delta x)\omega(\bx)\right)^{-1/4} \sqrt{ \det{-\frac{\partial^2 \mathcal{S}_{\rm G}[\bx_{\cl}(t)]}{\partial x_0^\mu \partial x_1^\nu}}} 
  \nonumber\\
  & 
  \qquad\quad
  = \exp{-\frac{1}{12}R \Delta t + O(\Delta t^{3/2})}  \, .
\end{align}
The main difference between the two Lagrangians in Eqs.~\eqref{eq:actionFeynman} and~\eqref{eq:actionGraham} thus lies in whether the WKB-type prefactor is incorporated or not within the action itself, which results in the $R/6$ or $R/12$ contributions.
A contribution $R/8$ was obtained by Dekker~\cite{dekker_path_1980, dekker1981proof} following a different discretization procedure compatible with the use of Fourier analysis in Riemann normal coordinates.
As we explain in the next Section, 
in a time-discretized approach, one can use discretization schemes of the action that yield a Lagrangian with a generic contribution $-\frac 12 \lambda R$ for any $\lambda \in \mathbb R$.

\section{An explicitly covariant discretization scheme for path integrals}
\label{sec:explicit-covariant}

At odds with the DeWitt and Graham approaches, we now construct a covariant path integral using time-discretization schemes which 
involve explicit discretization points, in the spirit of the Itō or Stratonovich ones,
but including a quadratic term in $\Delta \bx$ that goes one order higher in powers of $\sqrt{\Delta t}$ than the
traditional linear schemes.
The construction we propose in this section is new (and, in one dimension, the results become equivalent to those obtained in Ref.~\cite{cugliandolo2019building}).
Early approaches in this spirit were provided by Dekker, who analyzed the discretization of the normalization prefactor in 1D~\cite{dekker_functional_1976}.
Other analyses involving other time discretization can be found in Refs.~\cite{Langouche79,arnold_symmetric_2000}.

\subsection{Higher-order discretization}
\label{sec:statement-result-explicit-covariant}

One way to proceed is to consider the mid-point discretized Lagrangian of Eq.~(\ref{eq:Lstrato})
\begin{align} 
\mathcal{L}^{\bx}_{1/2}[\textbf{x},\dot{\textbf{x}}] = 
& 
\,  \frac{1}{2}
\left[
\omega_{\mu\nu}\left(\frac{\dd x^\mu}{\dd t} - h^\mu\right)\left(\frac{\dd x^\nu}{\dd t} - h^\nu\right) + \nabla_\mu h^\mu - \frac{1}{4}R 
\right.
\nonumber\\
& \qquad 
\left.
+ \frac{1}{4}\omega^{\mu\nu}\Gamma_{\beta\mu}^\alpha \Gamma_{\alpha\nu}^\beta + \frac{1}{4}\omega^{\mu\nu}\partial_\nu \Gamma^\alpha_{\mu\alpha} \right] 
\label{eq:LxStrato2lines}
\end{align}
and to absorb the non-covariant terms by extending the Stratonovich discretization by one order in $\sqrt{\Delta t}$.
To do so, we introduce a tensor ${\B}^{\alpha}_{\rho \sigma}$, 
see Eq.~(\ref{eq:defxbark_ddim}),  so as to define the discretization point to be 
\begin{equation}
\bar x^\alpha
=
x^\alpha + \frac{1}{2}\Delta x^\alpha + {\B}^\alpha_{\rho\sigma}\Delta x^\rho \Delta x^\sigma
\; . 
\label{eq:covariant-M-discretization}
\end{equation}

This more refined scheme is relevant only in the $\omega_{\mu\nu}\Delta x^\mu \Delta x^\nu$ term of the time-discrete version of~(\ref{eq:LxStrato2lines}).
We rewrite this kinetic term in the form
\begin{align}
& 
\omega_{\mu\nu}\Big(x^\alpha + \frac{1}{2}\Delta x^\alpha\Big)\,\Delta x^\mu \Delta x^\nu 
\nonumber\\[1mm]
& 
\qquad
=  \omega_{\mu\nu}\Big(x^\alpha + \frac{1}{2}\Delta x^\alpha + {\B}^\alpha_{\rho\sigma}\Delta x^\rho \Delta x^\sigma -{\B}^\alpha_{\rho\sigma}\Delta x^\rho \Delta x^\sigma \Big)\,\Delta x^\mu \Delta x^\nu 
\nonumber\\[1mm] 
& 
\qquad
= \omega_{\mu\nu}\Big(x^\alpha + \frac{1}{2}\Delta x^\alpha + {\B}^\alpha_{\rho\sigma}\Delta x^\rho \Delta x^\sigma\Big)\,\Delta x^\mu \Delta x^\nu 
\nonumber\\
& 
\qquad\qquad\qquad
- {\B}^\alpha_{\rho\sigma}\partial_{\alpha}\omega_{\mu\nu}\,\Delta x^\rho \Delta x^\sigma \Delta x^\mu \Delta x^\nu 
+ O(\Delta t^3)
\end{align}
and we then use the substitution rule~(\ref{eq:quarticsubstit}) to derive
\begin{align}
& 
\omega_{\mu\nu}\Big(x^\alpha + \frac{1}{2}\Delta x^\alpha\Big)\,\Delta x^\mu \Delta x^\nu 
\nonumber\\[1mm]
& 
\qquad
\doteq  \omega_{\mu\nu}\Big(x^\alpha + \frac{1}{2}\Delta x^\alpha + {\B}^\alpha_{\rho\sigma}\Delta x^\rho \Delta x^\sigma\Big)\,\Delta x^\mu \Delta x^\nu  
\nonumber\\
& 
\qquad\qquad\qquad
 -  {\B}^\alpha_{\rho\sigma}\partial_{\alpha}\omega_{\mu\nu}
 \big(\omega^{\mu\nu} \omega^{\rho\sigma}  + \omega^{\mu\rho} \omega^{\nu\sigma}  + \omega^{\mu\sigma} \omega^{\nu\rho} \big)
 + O(\Delta t^3)
 \; . 
\end{align}
%
%
Here, using the symmetry of ${\B}^\alpha_{\rho\sigma}$ with respect to~$\rho\leftrightarrow\sigma$, one can
replace $\omega^{\mu\rho} \omega^{\nu\sigma}  + \omega^{\mu\sigma} \omega^{\nu\rho}$ by $2 \omega^{\mu\rho} \omega^{\nu\sigma}$. 
One then rewrites the last line using Christoffel symbols as
$
{\B}^\alpha_{\rho\sigma}\partial_{\alpha}\omega_{\mu\nu}\big(\omega^{\mu\nu}\omega^{\rho\sigma} + 2 \omega^{\mu\rho}\omega^{\sigma\nu}\big)  
= {\B}^\alpha_{\rho\sigma} 
\big(2 \omega^{\rho\sigma} \Gamma^\beta_{\alpha \beta} + 4 \omega^{\rho \mu} \Gamma^\sigma_{\alpha \mu}\big)
$,
to finally obtain
\begin{align}
\omega_{\mu\nu}&\Big(x^\alpha + \frac{1}{2}\Delta x^\alpha\Big)\Delta x^\mu \Delta x^\nu 
\nonumber
\\
&
\qquad
\doteq  \omega_{\mu\nu}\Big(x^\alpha + \frac{1}{2}\Delta x^\alpha + {\B}^\alpha_{\rho\sigma}\Delta x^\rho \Delta x^\sigma\Big)\Delta x^\mu \Delta x^\nu
\nonumber
\\
&\qquad\qquad\qquad
-{\B}^\alpha_{\rho\sigma}
\big(2 \omega^{\rho\sigma} \Gamma^\beta_{\alpha \beta} + 4 \omega^{\rho \mu} \Gamma^\sigma_{\alpha \mu}\big)
+ O(\Delta t^3)
\,.
\label{eq:lastlineomegaomega}
\end{align}
The goal is to absorb the non-covariant terms of the mid-point discretized Lagrangian (given by the last line of Eq.~(\ref{eq:LxStrato2lines}))
with the help of the terms generated by the change of discretization (given by the last line of~\eqref{eq:lastlineomegaomega}).
This is possible provided that ${\B}$ satisfies the condition
\begin{equation}
\label{eq:M}
{\B}^\alpha_{\rho\sigma} \left(2 \omega^{\rho\sigma} \Gamma^\beta_{\alpha \beta} + 4 \omega^{\rho \mu} \Gamma^\sigma_{\alpha \mu}\right) 
=
 \frac{1}{4}\omega^{\mu\nu}\left(\partial_\nu \Gamma^{\rho}_{\rho\mu} + \Gamma^\rho_{\mu\sigma}\Gamma^{\sigma}_{\rho\nu}\right) + \Big(\lambda - \frac{1}{4}\Big)R
\end{equation}
 (which can be solved in general, as discussed below).
The parameter $\lambda \in \mathbb{R}$ allows one to incorporate a portion of the Ricci curvature $R$ present in the 
Stratonovich-discretized Lagrangian~(\ref{eq:LxStrato2lines}) into the discretization scheme,
so that the associated continuous-time Lagrangian then reads
\begin{equation}
\label{eq:covStrato}
\mathcal{L}^{\bx}_{\B}[\textbf{x},\dot{\textbf{x}}] = \frac{1}{2} \left[\omega_{\mu\nu}\left(\frac{\dd x^\mu}{\dd t}-h^\mu\right)\left(\frac{\dd x^\nu}{\dd t}-h^\nu\right) + \nabla_\mu h^\mu - \lambda R \right]
.
\end{equation}
%
Here the index ${\B}$ indicates that every function of $\bx$ on the r.h.s.~is evaluated at 
the discretization point $\bar\bx$ with components
\begin{equation}
  \label{eq:defMdiscreti}
 \bar x^\alpha =   x^\alpha + \frac{1}{2}\Delta x^\alpha + {\B}^\alpha_{\rho\sigma}\Delta x^\rho \Delta x^\sigma 
 \, ,
\end{equation}
where we recall that ${\B}$ depends on $\bx$.
Thus, we have just shown that \textit{it is possible to obtain a covariant quadratic discretization scheme} (in the sense of Sec.~\ref{sec:covar-path-integr}) by requiring that the tensor ${\B}$ solves Eq.~(\ref{eq:M})
--~since this condition ensures that the Lagrangian, written as~\eqref{eq:covStrato}, indeed transforms  covariantly under a change of variables.
We stress that the coefficients ${\B}^\alpha_{\rho\sigma}$ defining the higher-order discretization scheme do not transform as the coordinates of a (1,2) tensor field: when changing variables from $\bx(t)$ to $\bu(t)=\bU(\bx(t))$, the new Lagrangian for the process $\bu(t)$ is discretized with a tensor ${\B}$ that solves Eq.~(\ref{eq:M}) with
the metric, Christoffel symbols and scalar curvature attached to the noise amplitude\footnote{
We recall that $G^{\mu i}=\frac {\partial U^\mu}{\partial x^\nu} g^{\nu i}$ is read as $G^{\mu i}(\bu)=\frac {\partial U^\mu}{\partial x^\nu}(\bU^{-1}(\bu))\, g^{\nu i}(\bU^{-1}(\bu))$, see \textit{e.g.}~Eq.~\eqref{eq:FGoffg} in one dimension.
}
$G^{\mu i}=\frac {\partial U^\mu}{\partial x^\nu} g^{\nu i}$ of the process $\bu(t)$.
 
\smallskip
We now discuss the satisfiability of the condition~(\ref{eq:M}) on ${\B}$.
On the one hand, ${\B}^\alpha_{\rho\sigma}$ is symmetric in $\rho\sigma$ and possesses $[d(d-1)/2+d] d$ degrees of freedom. 
On the other hand, Eq.~\eqref{eq:M} specifies only one constraint on ${\B}$, so that we expect a solution to exist in general.
For instance, in $d=1$ the equation 
fixes ${\B}$ completely. Actually, the equation simplifies considerably since 
indices become superfluous, though one needs to keep track of whether they were covariant or contravariant:
\begin{eqnarray}
&&
 \omega^{\mu\nu} \ \, \mapsto \ g^2
\; , 
\qquad \;\;
\omega_{\mu\nu} \mapsto g^{-2} 
\; , 
\nonumber
\\[4pt]
&& 
\Gamma^\rho_{\mu\nu} \ \ \mapsto  \ \
\Gamma = 
\frac{1}{2}   \, g^2 \,  \frac{\dd g^{-2}}{\dd x} = -g^{-1} \, \frac{\dd g}{\dd x} 
\; , 
\\[4pt]
&& 
\frac{\dd \Gamma^\rho_{\mu\nu}}{\dd x} 
\mapsto \
\frac{\dd \Gamma }{\dd x} =
g^{-2} \left(\frac{\dd g}{\dd x}\right)^2 - g^{-1} \, \frac{\dd^2 g}{\dd x^2}
\; , 
\nonumber
\end{eqnarray}
and $R=0$. Then Eq.~\eqref{eq:M} becomes
$6 {\B} \Gamma  = 
\left(\dd \Gamma/\dd x + \Gamma^2 \right)/4$
which, replacing the results above yields
\begin{eqnarray}
{\B}  
= 
- \frac{1}{12} \frac{g'}{g}  + \frac{1}{24} \frac{g''}{g'} 
\equiv \betag
\; , 
\label{eq:Mbeta1d}
\end{eqnarray}
with $g'=\dd g/\dd x$ and $g''=\dd^2 g/\dd x^2$. 
Here,
$\betag$ is the $g$-dependent value that the coefficient of the quadratic term in the discretization scheme for a one-dimensional process, see Eq.~\eqref{eq:higher_order}, should take for the covariance of the action to be ensured.
As expected, we  recover in Eq.~\eqref{eq:Mbeta1d} the result of~\cite{cugliandolo2019building} for the one dimensional 
covariant quadratic discretization scheme  --~yet following a completely different route.

For $d>1$ 
and any $\lambda \in \mathbb{R}$, we generically expect Eq.~\eqref{eq:M} to have multiple solutions (see Secs.~\ref{sec:two-dimens-diff} and~\ref{sec:diffusion-d-1} for examples). Equation~\eqref{eq:M} is singular at points where the first derivatives of the metric vanish but the second ones do not. While it is always possible to define, for any point, a coordinate system such that this is the case we expect  
these points to be isolated. Hence, there is always a discretization scheme of the form~(\ref{eq:defMdiscreti}) allowing to recover both Graham's or DeWitt's action in continuous time (or any other action of the form~(\ref{eq:covStrato}) with an arbitrary prefactor in 
front of the Ricci scalar curvature). 

\smallskip

Let us now assume that $\lambda = - 1/3$ so as to coincide with DeWitt's action~(\ref{eq:actionFeynman}). 
DeWitt's infinitesimal propagator reads (see Eqs.~(\ref{eq:propdewitt}) and~\eqref{eq:dWresa}-\eqref{eq:dWresb})
\begin{align}
& 
\K_{\Delta t}^\dW(\bx + \Delta \bx , t + \Delta t | \bx,t) =
(2\pi \Delta t)^{-d/2} 
\nonumber\\
&
\quad 
\times \exp \left[ 
- \int_t^{t+\Delta t}\!\!\!\! \dd  s \left( \frac{1}{2} \omega_{\mu\nu} \dot{x}^{\mu}\dot{x}^{\nu} - \omega_{\mu\nu} h^{\nu} \dot{x}^\mu +\frac{1}{2}\omega_{\mu\nu}h^{\mu}h^{\nu} + \frac{1}{2}\nabla_\mu h^\mu + \frac{1}{6}R\right) 
\right],
\end{align}
where the previous equation is evaluated along the trajectory minimizing the continuous-time action, \textit{i.e} along the infinitesimal geodesic going from $(\bx,t)$ to $(\bx + \Delta \bx, t + \Delta t)$ at the desired order in $\Delta t$. We thus obtain 
\begin{align}
& 
\K_{\Delta t}^\dW(\bx + \Delta \bx, t + \Delta t| \bx,t) = 
(2\pi \Delta t)^{-d/2} 
\nonumber\\
& 
\qquad
\times 
\exp \left[ - \frac{1}{2}\Delta t\left(- \omega_{\mu\nu} h^{\nu} \frac{\Delta x^\mu}{\Delta t} +\frac{1}{2}\omega_{\mu\nu}h^{\mu}h^{\nu} + \frac{1}{2}\nabla_\mu h^\mu + \frac{1}{6}R\right)
\right.
\nonumber\\
&
\qquad\qquad\quad\;\;
\left. 
-\frac{1}{2}\int_t^{t+\Delta t} \dd  s \, \omega_{\mu\nu} \dot{x}^{\mu}\dot{x}^{\nu} 
\right]
\end{align}
%
%
%
where the first terms we have integrated out are midpoint-evaluated. In the higher-order discretization 
scheme~(\ref{eq:defMdiscreti}), we therefore have
\begin{align}
&
\omega_{\mu\nu}\Big(x^\alpha + \frac{1}{2}\Delta x^\alpha + {\B}^\alpha_{\rho\sigma}\Delta x^\rho \Delta x^\sigma\Big)\Delta x^\mu \Delta x^\nu = 
\nonumber\\
& 
\qquad\qquad\qquad\qquad\qquad\qquad
\Delta t \int_{t}^{t+\Delta t} \!\!\!\! \dd  s \, \omega_{\mu\nu} \dot{x}^{\mu}\dot{x}^{\nu} + O(\Delta t^{5/2})
\,,
\end{align}
provided ${\B}$ solves the condition~(\ref{eq:M}).
Along a geodesic, the squared velocity $\omega_{\mu\nu}\dot{x}^\mu\dot{x}^\nu$ 
being constant to the desired order (see Eq.~(\ref{eq:dSqVelds})), we obtain
\begin{equation}
\omega_{\mu\nu}\Big(x^\alpha + \frac{1}{2}\Delta x^\alpha + {\B}^\alpha_{\rho\sigma}\Delta x^\rho \Delta x^\sigma\Big)\frac{\Delta x^\mu}{\Delta t} \frac{\Delta x^\nu}{\Delta t}
 =
 \omega_{\mu\nu}(\bx) \dot{x}^\mu \dot{x}^\nu + O(\Delta t^{1/2}).
\label{eq:discvsdW}
\end{equation}
The right-hand side represents the squared velocity along the geodesic going from $(\bx,t)$ to $(\bx + \Delta \bx, t + \Delta t)$. The left-hand side is the norm of the squared velocity along a trajectory of constant speed vector $\Delta x^\mu / \Delta t$ (\textit{i.e.}~a straight line in the sense of $\mathbb{R}^d$) evaluated at the point $x^\alpha + \frac{1}{2}\Delta x^\alpha + {\B}^\alpha_{\rho\sigma}\Delta x^\rho \Delta x^\sigma$.
This proves that for $\lambda=-\frac 13$, the quadratic discretization scheme we have introduced
ensures that $\omega_{\mu\nu} \frac{\Delta x^\mu}{\Delta t} \frac{\Delta x^\nu}{\Delta t}$
is equal to $\omega_{\mu\nu}\dot{x}^\mu\dot{x}^\nu$ evaluated along the considered infinitesimal geodesic, 
\textit{up to irrelevant terms of order} $O(\Delta t^{1/2})$.
This would not be true in general for other discretization schemes.
From this, we see that Eq.~\eqref{eq:discvsdW} provides the bridge between DeWitt's original description of path integrals and the time-discretized construction that we put forward. 

\smallskip

Finally, we note that the precise discretization scheme that we have derived does not play a role when one is only interested in the small-noise asymptotics 
(as we already noticed in the example of Sec.~\ref{subsec:free-particle}).
Indeed, if $D$ denotes the noise amplitude (say, $g^{\mu i}\sim \sqrt{D}$ in the Langevin equation~(\ref{eq:pi_langevin_multi})),
comparing the Stratonovich Lagrangian~(\ref{eq:LxStrato2lines}) and the covariant one~(\ref{eq:covStrato}), we remark that the non-covariant terms of~(\ref{eq:LxStrato2lines})
scale as $D$ while the dominant contribution scales as $1/D$ (since indeed $\omega^{\mu \nu}\sim D$, $\omega_{\mu\nu}\sim 1/D$ and $\Gamma^\mu_{\rho\sigma}\sim D^0$).

\subsection{Two-dimensional diffusion in polar coordinates}
\label{sec:two-dimens-diff}

We consider again here the free stochastic motion of a particle  in the two-dimensional plane, a problem which we 
already discussed in Sec.~\ref{subsec:free-particle}, and that we can now revisit with the help of the quadratic covariant discretization scheme.
The equations of motion are
\begin{equation}
\begin{split}
\dot{x}  = \eta_x \, , \qquad\qquad
\dot{y}  = \eta_y \, ,
\end{split}
\end{equation}
with $\eta_x$ and $\eta_y$ two independent zero-mean and Gaussian white noises with variance $\langle\eta_i(t)\eta_j(t')\rangle=2D
\delta_{ij} \delta(t-t')$. In these Cartesian $\bx$ coordinates, the continuous-time Lagrangian reads
\begin{equation}
\mathcal{L}^{\bf x}[x,y,\dot{x}, \dot{y}] = \frac{1}{4D}\left(\dot{x}^2 + \dot{y}^2\right)
\, .
\label{eq:sjghkhfps}
\end{equation}
We then transform variables to $\bf u$, the polar coordinates $r$ and $\phi$ defined by $x = r \cos\phi$ and $y = r \sin\phi$. In any covariant discretization scheme, the polar coordinates continuous-time Lagrangian must be 
\begin{equation}
\mathcal{L}^{\bf u}[r,\phi,\dot{r}, \dot{\phi}] = \frac{1}{4D}\left(\dot{r}^2 + r^2 \dot{\phi}^2\right)
\, .
\label{eq:sjgfdgdk}
\end{equation}
The Lagrangian shown in Eq.~\eqref{eq:covStrato} is of course compatible with the above expression. Indeed, the associated metric  in polar coordinates  is the one of flat space
\begin{equation}
\Omega_{\mu\nu} = (2D)^{-1}  \begin{pmatrix}
1 & 0 \\
0 & r^2
\end{pmatrix}
\end{equation}
that is $2D \, \Omega_{rr} = 1$ and $2D \, \Omega_{\phi\phi} = r^2$. The non-vanishing connections are
$\hat\Gamma^\phi_{r\phi} = \hat\Gamma^\phi_{\phi r}  = r^{-1}$ and $\hat \Gamma^r_{\phi\phi}=-r$,
which yield vanishing curvature, $R=0$,  and $h^\mu = 0$ by covariance. 

We would like to identify now  the corresponding appropriate quadratic discretization, the one 
determined by the choice of ${\B}^\alpha_{\rho\sigma}$. In polar coordinates, the constraint~\eqref{eq:M} that ${\B}^\alpha_{\rho\sigma}$ has to satisfy reads
\begin{equation}
2 \frac{{\B}^{r}_{rr}}{r} + 6 \frac{{\B}^r_{\phi \phi}}{r^3} = - \frac{1}{2 r^2}
\:.
\end{equation}
Among the many solutions to this equation, a simple one is 
%
\begin{equation}
  \label{eq:solMpolaire}
{\B}^r_{\phi\phi}=-\frac{r}{12}
\; , 
\end{equation}
and all other components of ${\B}$, including ${\B}^r_{rr}$, equal to zero. 

The consequence is interesting: 
to write a time-discretized path-integral representation of Brownian motion in the two-dimensional plane using 
polar coordinates, one can follow two alternative routes:
\begin{itemize}
\item[(\textit{i})] Take a Stratonovich-discretized Lagrangian with a ``spurious'' contribution $\propto D/r^2$ (see Eq.~(\ref{eq:Lstratopolar})).
It cannot be obtained by applying the standard chain rule to the Cartesian Lagrangian~(\ref{eq:sjghkhfps}) 
but instead by using the rather complicated modified Itō lemma, Eq.~\eqref{eq:subsitution_kinetic_2} for $\alpha=\frac 12$,
which contains extra terms compared to the blind application of the chain rule.

\item[
(\textit{ii})] Simply interpret the Lagrangian~(\ref{eq:sjgfdgdk}) in polar coordinates (obtained from~(\ref{eq:sjghkhfps}) by applying the standard chain rule) as
discretized with the \textit{quadratic} scheme~(\ref{eq:defMdiscreti}), where ${\B}$ is given by~(\ref{eq:solMpolaire}).
\end{itemize}
Both routes are equivalent, although the second one is simpler.
They also bear the same content as DeWitt's representation of Sec.~\ref{sec:DWG}; the advantage of route (\textit{ii}) is that, to actually compute the weight of a path,
one does not need to explicitly find the infinitesimal optimal path at every step --~the quadratic contribution ${\B}^\alpha_{\rho\sigma}\Delta x^\rho \Delta x^\sigma$ to the
discretization scheme~(\ref{eq:defMdiscreti}) being explicit in the increments.

\subsection{Diffusion on the $(d-1)$-dimensional unit sphere of $\mathbb R^d$}
\label{sec:diffusion-d-1}

We now consider the diffusion of a particle on the $(d-1)$-dimensional unit sphere. In $d$-dimensional Cartesian coordinates, the equation of motion is
\begin{equation}
\frac{\dd \mathbf{r}}{\dd t} 
\stackrel{\frac 12} 
=
 - \mathbf{r}\cdot\left(\mathbf{r}\cdot\mathbb{\beeta}\right) + 
r^2 \, \mathbf{\beeta}
\end{equation}
with $\beeta$ a $d$-dimensional zero-mean Gaussian white noise with uncorrelated components and 
$\langle \eta_i(t) \eta_j(t')\rangle = 2D \delta_{ij} \delta(t-t')$. 
The previous equation is written in the Stratonovich discretization and is thus norm-preserving. In Itō discretization, one has to add a drift term to ensure the proper 
normalization and the 
corresponding equation reads
\begin{equation}
\frac{\dd \mathbf{r}}{\dd t} 
\stackrel{0} 
=
-D(d-1)\mathbf{r}  -\mathbf{r}\left(\mathbf{r}\cdot\mathbb{\beeta}\right) + 
r^2 \mathbf{\beeta}
\end{equation}
from which one can easily deduce the associated Fokker--Planck equation 
\begin{equation}
\partial_t P(\mathbf{r},t) = 
D\,
\nabla^2_{\mathcal{S}_d}P(\mathbf{r},t)
\end{equation}
where $\nabla^2_{\mathcal{S}_d}$ is the Laplacian on the $(d-1)$-dimensional unit sphere. Focusing on $d = 3$, for which the transition probability was provided in~\cite{riseborough1982diffusion}, we 
parametrize the unit sphere by the two usual angles $\theta$ and $\phi$ of the spherical coordinates.  The 
time-dependent probability density is then $P (\theta,\phi,t)$ and its evolution is ruled by the equation
\begin{equation}
\partial_t P(\theta,\phi,t) = 
D
\left[\frac{1}{\sin{\theta}}\partial_\theta\left(\sin{\theta}\,\partial_\theta P(\theta,\phi,t)\right)+\frac{1}{\sin^2{\theta}}\partial_\phi^2P(\theta,\phi,t)\right]
. 
\end{equation}
We now introduce the probability density $\hat{P}(\theta,\phi,t)$ of the $\theta$ and $\phi$ variables defined by 
\begin{equation}
\dd x \dd y \dd z \, P(x,y,z,t) \, \delta\big(\sqrt{x^2 + y^2 + z^2}-1\big) = \dd \theta \dd \phi \, \hat{P}(\theta,\phi,t)
\end{equation}
yielding $\hat{P}(\theta,\phi,t) = \sin\theta \, P(\theta,\phi,t)$ and hence
\begin{align}
\partial_t \hat{P}(\theta,\phi,t) = 
&
\;\;
D\,
\partial_\theta \Big(-\frac{\cos{\theta}}{\sin{\theta}}\hat{P}(\theta,\phi,t)\Big) 
\nonumber\\
\;\;
&
+ 
D
\left[\partial_\theta^2 \hat{P}(\theta,\phi,t)+ \partial_\phi^2 \left(\frac{1}{\sin^2{\theta}}\hat{P}(\theta,\phi,t)\right)\right]
. 
\end{align}
We have thus rewritten our constrained three-dimensional problem as a well defined two-dimensional Langevin process on the angles $\theta$ and $\phi$: 
\begin{equation}
\dot{\theta}  = D\frac{\cos{\theta}}{\sin{\theta}} + (2D)^{1/2} \, \eta_\theta 
\; , \qquad\qquad
\dot{\phi}  = \frac{1}{\sin\theta}\, (2D)^{1/2} \, \eta_\phi
\; , 
\end{equation}
in which the discretization scheme is irrelevant (since the increments of $\theta$ are independent of $\eta_\phi$). 
The noises are Gaussian and uncorrelated  with $\langle \eta_\theta(t) \rangle =\langle \eta_\phi (t) \rangle =0$
and $\langle \eta_\theta(t) \eta_\theta(t')\rangle =\langle \eta_\phi(t) \eta_\phi(t') \rangle = \delta(t-t')$.
 The associated metric is proportional to the one of the two-dimensional unit sphere
\begin{equation}
\omega_{\mu\nu} =
\frac{1}{2D}
 \begin{pmatrix}
1 & 0 \\
0 & \sin^2\theta
\end{pmatrix}
,
\end{equation}
$2D \omega_{\theta\theta} = 1$ and $2D \omega_{\phi\phi} = \sin^2\theta$. The 
non-vanishing connections are  $\Gamma^\theta_{\phi\phi} = - \sin \theta \cos\theta$, 
$\Gamma^\phi_{\theta\phi} = \Gamma^\phi_{\phi\theta} = 1/\tan \, \theta$.
The curvature is  $R=4D$ (the scalar curvature is twice the Gaussian one on two-dimensional surfaces). 
Moreover, one can show that $h^\mu = 0$. This seems to reflect the fact that even though there is a drift in the equation of motion for $\theta$, the vectorial drift $h^\mu$ remembers the absence of physical drift in the original three-dimensional problem. 
In the above defined covariant discretization, the continuous-time Lagrangian is
\begin{equation}
\mathcal{L}[\theta,\phi,\dot{\theta}, \dot{\phi}] = \frac{1}{4D} \left(\dot{\theta}^2 + \sin^2\theta \,\dot{\phi}^2\right) - 2 \lambda D
\,.
\label{eq:fghdfghdfgd}
\end{equation}
For a generic value of $\lambda$, the associated discretization matrix satisfies, see Eq.~(\ref{eq:M}):
\begin{equation}
2 \, \sin(2\theta)  {\B}^{\theta}_{\theta\theta} + 
12 \, \frac{\cos\theta}{\sin\theta} {\B}^{\theta}_{\phi\phi} =
\left(4 \lambda-\frac{1}{2}\right) \, \sin^2(\theta)-1
\,.
\label{eq:eqforMspherical}
\end{equation}
Once again, there are many solutions to this equation. 
A simple one is given by
\begin{equation}
  \label{eq:solMspheric}
{\B}^\theta_{\theta\theta}=\left(\lambda - \frac 1 8 \right) \tan \theta
\; , 
\qquad\qquad
{\B}^\theta_{\phi\phi}=-\frac{1}{12} \tan \theta
\; , 
\end{equation}
and other components equal to zero. The spherical-coordinate Lagrangian~(\ref{eq:fghdfghdfgd}) for any  $\lambda$ can be read in a time-discretized form with the quadratic discretization scheme~(\ref{eq:defMdiscreti}), where the tensor ${\B}$ has non-zero components given by~Eq.~\eqref{eq:solMspheric}. A Stratonovich-discretized Lagrangian would instead include ``spurious'' terms (that one cannot derive by applying the standard chain rule to the Cartesian-coordinate action), similarly to what we explained
in the presentation of the polar coordinates bidimensional planar Brownian motion considered in the previous paragraph.

\section{Covariant Langevin equation in discrete time -- and an application to covariant path integrals }\label{sec:covLangevin}

In~\cite{cugliandolo2019building}, some of the present authors used a completely different route than the one presented in Sec.~\ref{sec:explicit-covariant}, based on a discretization of Langevin equations which is covariant in discrete time (for a finite time step), to build a covariant and explicit discretization scheme of the path integral valid for one-dimensional processes.
While, as shown in Eq.~\eqref{eq:Mbeta1d}, the approach of Sec.~\ref{sec:explicit-covariant} leads to identical results 
when the dimension is set to $1$, this is no longer true when $d>1$. 
We extend in this section the results of~\cite{cugliandolo2019building} to the multidimensional case and show that this approach generically leads to devising a covariant discretization scheme which is indeed different from the one of Sec.~\ref{sec:explicit-covariant}.

\subsection{Covariant Langevin equation in discrete time at finite $\Delta t$}\label{sec:covLangevin-sub}

Consider a discrete-time Stratonovich Langevin equation. As already discussed, this equation is covariant up to terms scaling as $O(\Delta t^{1/2})$. While this is enough to warrant the covariance of the Langevin equation 
in the limit $\Delta t \to 0$, we have seen that midpoint discretized infinitesimal propagators have a non-covariant formal continuous-time limit due to these $O(\Delta t^{1/2})$ corrections. In this section, we introduce an alternative discretization of Langevin equations that describes the same process as the Stratonovich-discretized one in the continuous-time limit and that has the remarkable property of being covariant in discrete time to all orders in $\Delta t$ in a way that will be made precise below. It extends the results previously obtained 
in~\cite{cugliandolo2019building} for one-dimensional systems to multidimensional ones. Lastly, we use this discretization as a guide to construct a new path-integral representation of the continuous-time process. 

We introduce the stochastic equation
\begin{equation}
\label{eq:Tdis}
\begin{split}
\Delta x^\alpha = \mathbb{T}_{\bff, \bg} \big( f^{\alpha} \Delta t + g^{\alpha i} \Delta \eta_i \big) \, ,
\end{split}
\end{equation}
with $\Delta \eta_i(t)$ the same zero-mean Gaussian noise used
throughout this work
and the \textit{linear} operator $\mathbb{T}_{\bff, \bg}$ defined by its action on a generic function $h (\bx) $:
\begin{equation}
\mathbb{T}_{\bff, \bg}\,  h = \frac{\exp{\left(f^\mu \Delta t + g^{\mu i}\Delta\eta_i\right)\partial_{\mu}}-1}{\left(f^\mu \Delta t + g^{\mu i}\Delta\eta_i\right)\partial_{\mu}} \, h 
\: , 
\label{eq:defTfg}
\end{equation}
where, in the fraction, the functions $f^\mu$ and $g^{\mu i}$ and the partial derivatives are understood as operators.\footnote{
For an operator $\mathbb O$, the fraction $\frac{\ee^{\mathbb O}-1}{\mathbb O}$ appearing \textit{e.g.}~in~(\ref{eq:defTfg}) is understood as a series $\sum_{k\geq 1}\frac{1}{n!}\mathbb O^{n-1}$.
}
The direct expansion of Eq.~\eqref{eq:Tdis} shows that this process and the Stratonovich-discretized one in Eq.~\eqref{eq:alphadiscrete} are equivalent in the limit $\Delta t \to 0$,
since indeed from~\eqref{eq:defTfg} one has $(\mathbb{T}_{\bff, \bg}\,  h)(\bx) = h\big(\bx+\frac 12 \Delta \bx \big) + O(\Delta t)$.

This discretization scheme, inspired by the field of calculus with Poisson point processes~\cite{falsone2018stochastic}, is transparent to the chain rule, \textit{non-perturbatively for any finite} $\Delta t$ (hence to all orders in small $\Delta t$), in the sense that the process $\bu(t) = \textbf{U}(\bx(t))$ evolves according to
\begin{equation}\label{eq:Tdischangevar}
\Delta u^\alpha = \mathbb{T}_{\bF, \bG} \big( F^{\alpha} \Delta t + G^{\alpha i} \Delta \eta_i \big)
\end{equation}
with the right-hand-side a function of $\bf u$ written in terms of
\begin{equation}
   \label{eq:notationchangevar}
        F^\alpha  =  f^\beta \, \partial_\beta U^\alpha \, ,
        \qquad\qquad
        G^{\alpha i}  =  g^{\beta i} \, \partial_\beta U^\alpha \, .
\end{equation}
This statement is proven in the next section. 

\subsubsection{Proving the covariance of the discretization scheme}

Let $\textbf{U}$, be an invertible transformation of the initial coordinates $x^\alpha$. 
We define the new process $\bu(t) = \textbf{U}(\bx(t))$. Let us  first assume  that the chain rule holds with the 
discretization rule~(\ref{eq:Tdis}) specified above. Then,
\begin{align}
\Delta u^\sigma 
& = 
\left[\frac{\exp{\partial_{\alpha} U^\mu \, f^{\alpha} \Delta t + \partial_{\alpha}  U^\mu \, g^{\alpha i} \Delta \eta_i}\partial_{u^\mu}-1}{\left(\partial_{\alpha}  U^\mu \, f^{\alpha} \Delta t + \partial_{\alpha}  U^\mu \, g^{\alpha i} \Delta \eta_i\right) \partial_{u^\mu}}\right]
\left(\partial_{\beta} U^\sigma f^{\beta}\Delta t + \partial_{\beta} U^\sigma g^{\beta j} \Delta \eta_j\right) 
\nonumber \\ 
& = \left[\frac{\exp{f^{\alpha} \Delta t + g^{\alpha i} \Delta \eta_i}\partial_{\alpha}-1}{\left(f^{\alpha} \Delta t + g^{\alpha i} \Delta \eta_i \right)\partial_{\alpha}}\right]
\left(\partial_{\beta} U^\sigma f^{\beta}\Delta t + \partial_{\beta} U^\sigma g^{\beta j} \Delta \eta_j\right) 
 \nonumber\\ 
& = \left[\exp{f^{\alpha} \Delta t + g^{\alpha i} \Delta \eta_i}\partial_{\alpha}-1\right]U^\sigma
\; , 
\label{eq:Deltausigmafinal}
\end{align}
Independently of the chain rule, we have
\begin{equation}
\Delta u^\sigma = U^{\sigma}(\bx + \Delta \bx) - U^{\sigma}(\bx) \, ,
\end{equation}
with, moreover,
\begin{equation}
\begin{split}
\Delta x^\mu & = \left[\frac{\exp{f^{\alpha} \Delta t +  g^{\alpha i} \Delta \eta_i}\partial_{\alpha}-1}{\left(f^{\alpha} \Delta t + g^{\alpha i} \Delta \eta_i\right) \partial_{\alpha}}\right]
\left(f^{\mu}\Delta t + g^{\mu j} \Delta \eta_j\right) 
 \\ & = 
\left[\frac{\exp{f^{\alpha} \Delta t + g^{\alpha i} \Delta \eta_i}\partial_{\alpha}-1}{\left(f^{\alpha} \Delta t + g^{\alpha i} \Delta \eta_i \right)\partial_{\alpha}}\right]\left(\frac{\partial x^\mu}{\partial x^\nu} f^{\nu}\Delta t + \frac{\partial x^\mu}{\partial x^\nu} g^{\nu j} \Delta \eta_j\right) 
\\ & = 
\left[\exp{f^{\alpha} \Delta t + g^{\alpha i} \Delta \eta_i}\partial_{\alpha}-1\right]x^\mu
\,. 
\end{split}
\end{equation}
Therefore, proving the validity of the chain rule in this discretization scheme amounts to proving the functional identity (for any function $\textbf V$), 
\begin{eqnarray}\label{eq:identity_cov}
&& 
V^\sigma \big[\left(\exp{f^{\alpha} \Delta t + g^{\alpha i} \Delta \eta_i}\partial_{x^\alpha}\right) \bx \big] = 
\left[\exp{f^{\alpha} \Delta t + g^{\alpha i} \Delta \eta_i}\partial_{x^\alpha}\right]V^\sigma(\bx) \,,
\qquad\quad
\end{eqnarray}
which generalizes Eq.~\eqref{eq:commuteop}. In order to prove the previous equation, we follow the route taken in the one-dimensional case~\cite{cugliandolo2019building} and we define
\begin{equation}
\Psi(a, \bx) = \exp{a \left(f^\mu \Delta t + g^{\mu i}\Delta \eta_i\right)\partial_{\mu}}
\left[V^{\sigma}({\boldsymbol \chi}(a,\bx)) \right] \, ,
\end{equation}
with 
\begin{equation}
\chi^\beta(a,{\mathbf x}) = \exp{(1-a)\left(f^{\mu} \Delta t + g^{\mu i} \Delta \eta_i\right)\partial_{\mu}}x^\beta \, .
\end{equation}
We can then prove that $\Psi'(a, {\mathbf x}) \equiv \dd \Psi(a, {\mathbf x})/ \dd a= 0$, a property that leads to
\begin{equation}
\Psi(0,{\mathbf x}) = \Psi(1,\bx) \, ,
\label{eq:Psi01}
\end{equation}
so that the functional equation established earlier is verified. Indeed, 
\begin{eqnarray*}
&& \Psi'(a,\bx) 
=  \sum_{m \geq 1} \frac{a^{m-1}}{(m-1)!}\left[\left(f^\mu \Delta t + g^{\mu j}\Delta \eta_j\right)\partial_{\mu}\right]^m \big[V^{\sigma}({\boldsymbol \chi}(a,\bx))\big]
\\ 
&& \;\;\;\;\;\;
+ \sum_{m \geq 0} \frac{a^{m}}{m!}\left[\left(f^\mu \Delta t + g^{\mu j}\Delta \eta_j\right)\partial_{\mu}\right]^m \left[\partial_\beta V^\sigma({\boldsymbol \chi}(a,\bx))\frac{\partial \chi^\beta}{\partial a}\right]
\\ 
&& 
\; = \sum_{m \geq 1} \frac{a^{m-1}}{(m-1)!}\left[\left(f^\mu \Delta t + g^{\mu j}\Delta \eta_j\right)\partial_{\mu}\right]^m \left[V^{\sigma}({\boldsymbol \chi}(a,\bx))\right] 
\\ 
&& 
\;\;\;\;\;\;
- \sum_{m \geq 0} \frac{a^{m}}{m!}\left[\left(f^\mu \Delta t + g^{\mu j}\Delta \eta_j\right)\partial_{\mu}\right]^m \left[\partial_\beta V^\sigma({\boldsymbol \chi}(a,\bx))\left(f^\nu \Delta t + g^{\nu i}\Delta \eta_i\right)\frac{\partial \chi^\beta}{\partial x^\nu}\right]
\qquad
\\ 
&& 
\;
= \sum_{m \geq 1} \frac{a^{m-1}}{(m-1)!}\left[\left(f^\mu \Delta t + g^{\mu j}\Delta \eta_j\right)\partial_{\mu}\right]^m \left[V^{\sigma}({\boldsymbol \chi}(a,\bx))\right] 
\\ 
&& 
\;\;\;\;\;\;
- \sum_{m \geq 0} \frac{a^{m}}{m!}\left[\left(f^\mu \Delta t + g^{\mu j}\Delta \eta_j\right)\partial_{\mu}\right]^m 
 \left[\left(f^\nu \Delta t + g^{\nu i}\Delta \eta_i\right)\partial_\nu \big\{V^\sigma({\boldsymbol \chi}(a,\bx))\big\}\right] 
\\ 
&& 
\;
= 0 \, ,
\end{eqnarray*}
which completes the proof.

\subsubsection{Implicit discretization scheme of the Langevin equation}
\label{subsubsec:implicit}

In order to use Eq.~\eqref{eq:Tdis} as a guide for building covariant path-integral representations, 
we rewrite the latter in an implicit form closer to Eq.~\eqref{eq:alphadiscrete}. We expand Eq.~\eqref{eq:Tdis} neglecting terms of order $O(\Delta t^{2})$ and obtain,
\begin{equation} \label{eq:Tdisimplicit}
\Delta x^\mu = f^{\mu}\Big(\bx+\frac{\Delta \bx}{2}\Big)\Delta t + \left[g^{\mu i}\Big(\bx+\frac{\Delta \bx}{2}\Big) + 
{\mathsf {\B}}^{\mu i}_{\alpha \beta} \Delta x^\alpha \Delta x^\beta\right] \Delta \eta_i + O(\Delta t^2) \, ,
\end{equation}
with
\begin{equation} \label{eq:MTdis}
    {\mathsf {\B}}^{\mu i}_{\alpha\beta} = \frac{1}{24}\left(\partial_\alpha\partial_\beta g^{\mu i} - 2 g_{\beta j}\partial_\alpha g^{\gamma j}\partial_\gamma g^{\mu i}\right).
\end{equation}
One can also write
\begin{equation}
{\mathsf {\B}}^{\mu i}_{\alpha \beta} = \partial_\nu g^{\mu i} {\B}^\nu_{\alpha\beta}
\,,
\label{eq:BgprimeB}
\end{equation}
where, in arbitrary dimension $d$, ${\B}^\nu_{\alpha\beta}$ is a priori different from the tensor introduced in Sec.~\ref{sec:explicit-covariant}.
In the specific case of dimension one, one obtains
 ${\mathsf {\B}}^{\mu i}_{\alpha \beta} \mapsto
g''/24 - (g')^2/(12 \, g) = g' \betag $, with $\betag$ the coefficient of the quadratic discretization scheme: $\overline x = x + \Delta x/2 + \betag  (\Delta x)^2$ of Eq.~\eqref{eq:higher_order}, with $\betag$ given by Eq.~(\ref{eq:Mbeta1d}).
This is the route initially taken in Ref.~\cite{cugliandolo2019building} to build a discretized covariant action in $d=1$.

In the following, we introduce the notations
\begin{equation}
    \bar{g}^{\mu i}\left(\bx, \Delta \bx\right) = g^{\mu i}\Big(\bx+\frac{\Delta \bx}{2}\Big) + {\mathsf {\B}}^{\mu i}_{\alpha \beta} \Delta x^\alpha \Delta x^\beta \, ,
\end{equation}
and
\begin{equation} \label{eq:barg}
    \bar{g}_{i\mu}(\bx, \Delta \bx) = {g}_{i\mu}\Big(\bx + \frac{\Delta x}{2}\Big) + T_{i \mu \alpha \beta}(\bx)\Delta x^\alpha \Delta x^\beta \, ,
\end{equation}
with
\begin{equation} \label{eq:Tdisinverse}
    T_{i \mu \alpha \beta} = - g_{j \mu}g_{i \nu} {\mathsf {\B}}^{\nu j}_{\alpha \beta} \, ,
\end{equation}
so that
\begin{equation}
    \bar{g}_{i\mu}(\bx, \Delta \bx)\, \bar{g}^{\mu j}(\bx, \Delta \bx) = \delta_i^j + O(\Delta t^{3/2}) \, .
\end{equation}
Inverting Eq.~\eqref{eq:Tdisimplicit} thus yields
\begin{equation} \label{eq:covKinx}
    \Delta \eta_i = \bar{g}_{i\mu}(\bx, \Delta \bx)\left[\Delta x^\mu - f^{\mu}\Big(\bx + \frac{\Delta\bx}{2}\Big)\Delta t\right] + O(\Delta t^2) \,.
\end{equation}
Furthermore, for any invertible and smooth transformation $\bu = \textbf{U}(\bx)$ of the original variables, Eq.~\eqref{eq:Tdischangevar} tells us that 
\begin{equation} \label{eq:covKinu}
        \Delta \eta_i = \bar{G}_{i\mu}(\bu, \Delta \bu)
        \left[\Delta u^\mu - F^{\mu}\Big(\bu + \frac{\Delta\bu}{2}\Big) \Delta t\right] + O(\Delta t^2) \, ,
\end{equation}
where the notations follow Eq.~\eqref{eq:notationchangevar}. All in all, this warrants that the accordingly discretized kinetic term transforms formally as a scalar under changes of variables up to corrections of order $O(\Delta t^{3/2})$. Indeed, 
by taking the squares of Eqs.~\eqref{eq:covKinx} and~\eqref{eq:covKinu} and equating the results
divided by $\Delta t^2$
\begin{eqnarray}
        & & \!\!\!\!\!\!\!\! \bar{G}_{i\mu}(\bu, \Delta \bu)\bar{G}_{j\nu}(\bu, \Delta \bu) \!
        \left[\frac{\Delta u^\mu}{\Delta t} - F^{\mu}\Big(\bu + \frac{\Delta\bu}{2}\Big)\right]
        \!
        \left[\frac{\Delta u^\nu}{\Delta t} - F^{\nu}\Big(\bu + \frac{\Delta\bu}{2}\Big)\right] 
        \!
        \delta^{ij}\Delta t 
        \quad
        \nonumber\\ 
        & & \!\! = \bar{g}_{i\mu}(\bx, \Delta \bx)\bar{g}_{j\nu}(\bx, \Delta \bx)
        \!
        \left[\frac{\Delta x^\mu}{\Delta t} - f^{\mu}\Big(\bx + \frac{\Delta\bx}{2}\Big)\right]
        \!
        \left[\frac{\Delta x^\nu}{\Delta t} - f^{\nu}\Big(\bx + \frac{\Delta\bx}{2}\Big)\right]
        \! 
        \delta^{ij}\Delta t 
        \quad
    \nonumber    \\ 
        & & \quad + O(\Delta t^{3/2}) \, .
        \label{eq:covKinPi}
\end{eqnarray}
The realization that in this implicit discretization scheme the kinetic contribution to the
 infinitesimal propagator transforms covariantly under the change of variables  
 is at the basis of our subsequent construction of manifestly covariant path integrals.

\subsection{A discretization scheme for covariant path integrals}
\label{sec:disccovPI}

In order to obtain a covariant discretization scheme in continuous time, Eq.~\eqref{eq:covKinPi} suggests to write the infinitesimal propagator under the following form 
\begin{equation} \begin{split}\label{eq:propTdis1}
&
 (2\pi \Delta t)^{-\frac d 2} \, \K_{\Delta t}(\bx+\Delta \bx,t+\Delta t | \bx,t) 
\\
&
\qquad
  = \text{exp}
\Bigg\{
-\frac{1}{2}
\Bigg[\bar{g}_{i\mu}(\bx, \Delta \bx)\bar{g}_{j\nu}(\bx, \Delta \bx)\delta^{ij}
\, \frac{\left(\Delta x^\mu - h^{\mu}\Big(\bx + \frac{\Delta \bx}{2}\Big) \Delta t\right)}{\Delta t} 
 \\ & \hspace{48mm}
 \times
 \frac{\left(\Delta x^\nu - h^{\nu}\Big(\bx + \frac{\Delta \bx}{2}\Big) \Delta t \right)}{\Delta t} \; \Delta t + b(\bx) \Delta t\Bigg]\Bigg\} 
 \, ,
\end{split}
\end{equation}
where $b$ is yet to be determined. In order to find $b$, the simplest way to proceed is to start from the already derived Stratonovich-discretized infinitesimal propagator~(\ref{eq:Lstrato}) 
and put by hand the discretization suggested in Eq.~\eqref{eq:propTdis1}. This yields,
\begin{eqnarray}
 &&
 \!\!\!\!\!\!\!\!
 (2\pi \Delta t)^{-\frac d 2} \, \K_{\Delta t}(\bx+\Delta \bx,t+\Delta t | \bx,t) 
\nonumber\\
&&
\quad
 = \, \text{exp}\Bigg\{\!-\frac{1}{2}\left(\bar{g}_{i\mu}(\bx, \Delta \bx) - T_{i\mu\alpha\beta}\Delta x^\alpha \Delta x^\beta\right)
\left(\bar{g}_{j\nu}(\bx, \Delta \bx)-T_{j\nu\rho\sigma}\Delta x^\rho \Delta x^\sigma\right)
\nonumber \\ 
 && 
\hspace{22mm}
 \times \,
\delta^{ij} \,
\frac{\left(\Delta x^\mu - h^{\mu}\Big(\bx + \frac{\Delta \bx}{2}\Big) \Delta t\right)}{\Delta t} 
\frac{\left(\Delta x^\nu - h^{\nu}\Big(\bx + \frac{\Delta \bx}{2}\Big) \Delta t \right)}{\Delta t} \, 
\Delta t
 \nonumber\\ 
 &&
\hspace{22mm}
  - \frac{1}{2}\left[\nabla_\mu h^\mu - \frac{1}{4}R + \frac{1}{4}\left(\omega^{\mu\nu}\Gamma^{\alpha}_{\beta\mu}\Gamma^\beta_{\alpha \nu} + \omega^{\mu\nu}\partial_\nu \Gamma^{\alpha}_{\mu\alpha}\right)\right] \Delta t \Bigg\} \, , 
\end{eqnarray}
which becomes
\begin{eqnarray}
 &&
 \!\!\!\!\!\!\!\!\!\!
 (2\pi \Delta t)^{-\frac d 2} \, \K_{\Delta t}(\bx+\Delta \bx,t+\Delta t | \bx,t) 
\nonumber\\
&&
= \text{exp}\Bigg\{\!-\frac{1}{2}\bar{g}_{i\mu}(\bx, \Delta \bx)\bar{g}_{j\nu}(\bx, \Delta \bx)\delta^{ij} 
\nonumber\\
&& 
 \hspace{15mm}
\times
\left[\dfrac{\Delta x^\mu}{\Delta t}  - h^{\mu}\Big(\bx + \frac{\Delta \bx}{2}\Big) \right]
\left[\dfrac{\Delta x^\nu}{\Delta t} - h^{\nu}\Big(\bx + \frac{\Delta \bx}{2}\Big) \right]  \Delta t
\nonumber\\[6pt]
 &&  
 \hspace{12mm}
  - \frac{1}{2}\left[\nabla_\mu h^\mu - \frac{1}{4}R  + \frac{1}{4}\left(\omega^{\mu\nu}\Gamma^{\alpha}_{\beta\mu}\Gamma^\beta_{\alpha \nu} + \omega^{\mu\nu}\partial_\nu \Gamma^{\alpha}_{\mu\alpha}\right) + \Delta L \right] \Delta t \Bigg\}
\label{eq:second-K}
\end{eqnarray}
with 
\begin{equation} \label{eq:deltaL1} \begin{split}
    \Delta L & = - 2 T_{i \mu \alpha \beta}g_{j \nu}  \delta^{ij} \frac{\Delta x^\mu \Delta x^\nu \Delta x^\alpha \Delta x^\beta}{\Delta t^2} 
     \\ &
      = 2 \omega_{\rho \mu}g_{k \nu}{\mathsf {\B}}^{\rho k}_{\alpha\beta}\frac{\Delta x^\mu \Delta x^\nu \Delta x^\alpha \Delta x^\beta}{\Delta t^2}
       \\[6pt] & \doteq  2 \omega_{\rho \mu} g_{k \nu} {\mathsf {\B}}_{\alpha \beta}^{\rho k}\left(\omega^{\mu\nu}\omega^{\alpha\beta} + \omega^{\mu\alpha}\omega^{\nu\beta}+\omega^{\nu\alpha}\omega^{\mu\beta}\right) .
\end{split}
\end{equation}
The coefficient $b$ can then be deduced by setting Eq.~(\ref{eq:propTdis1}) equal to Eq.~(\ref{eq:second-K})
and reads
\begin{eqnarray}\label{eq:b1}
    && b = \nabla_\mu h^\mu - \frac{1}{4}R + \frac{1}{4}\left(\omega^{\mu\nu}\Gamma^{\alpha}_{\beta\mu}\Gamma^\beta_{\alpha \nu} + \omega^{\mu\nu}\partial_\nu \Gamma^{\alpha}_{\mu\alpha}\right) 
    \nonumber\\
    &&
    \qquad
    + \, 2 \omega_{\rho \mu} g_{k \nu} {\mathsf {\B}}_{\alpha \beta}^{\rho k}\left(\omega^{\mu\nu}\omega^{\alpha\beta} + \omega^{\mu\alpha}\omega^{\nu\beta}+\omega^{\nu\alpha}\omega^{\mu\beta}\right) .
\end{eqnarray}

Before using the particular form of ${\mathsf {\B}}$ displayed in Eq.~\eqref{eq:MTdis}, let us pause for a second and remark that two discretizations of the form given in Eq.~\eqref{eq:propTdis1} together with Eq.~\eqref{eq:barg} and characterized by two different $T'_{i\mu\alpha\beta}$ and $T''_{i\mu\alpha\beta}$ yield the same continuous-time Lagrangian provided that
\begin{eqnarray}
&& 
\omega_{\rho \mu} g_{k \nu} {{\mathsf {\B}}'}_{\alpha \beta}^{\, \rho k}\left(\omega^{\mu\nu}\omega^{\alpha\beta} + \omega^{\mu\alpha}\omega^{\nu\beta}+\omega^{\nu\alpha}\omega^{\mu\beta}\right) =
\nonumber\\
&& \qquad\qquad \omega_{\rho \mu} g_{k \nu} {{\mathsf {\B}}''}_{\alpha \beta}^{\, \rho k}\left(\omega^{\mu\nu}\omega^{\alpha\beta} + \omega^{\mu\alpha}\omega^{\nu\beta}+\omega^{\nu\alpha}\omega^{\mu\beta}\right) ,
\label{eq:equivalence}
\end{eqnarray}
where the relation between ${\mathsf {\B}}'$ and $T'$ (respectively ${\mathsf {\B}}''$ and $T''$) can be read from Eq.~\eqref{eq:Tdisinverse}. Remark that the relevant part of ${\mathsf {\B}}^{\rho k}_{\alpha\beta}$ (only the symmetric part in the down indices interests us) has $d^3(d+1)/2$ degrees of freedom while Eq.~\eqref{eq:equivalence} imposes only one constraint. Accordingly, as soon as $d>1$, there exists a degeneracy in the higher-order discretizations of the form Eq.~\eqref{eq:propTdis1} corresponding to a given continuous-time Lagrangian, similarly to the construction shown in Sec.~\ref{sec:explicit-covariant}.

Let us now specify the result for ${\mathsf {\B}}$ given in Eq.~\eqref{eq:MTdis}. Inserting the latter in Eq.~\eqref{eq:deltaL1} yields,
\begin{equation}\label{eq:DeltaL1}
    \Delta L = \frac{1}{12}g_{i\alpha}\omega^{\mu\nu}\partial_\mu\partial_\nu g^{\alpha i} + \frac{1}{6}g^{\mu i}\partial_\mu\partial_\nu g^{\nu j}\delta_{ij} + \frac{1}{6}\Gamma^{\alpha}_{\mu\alpha}g^{\nu i}\partial_\nu g^{\mu j} \delta_{ij} - \frac{1}{6}g_{\mu i}\partial_\nu g^{\alpha j}\partial_\alpha \omega^{\mu\nu}
\end{equation}
and $b$ becomes
\begin{eqnarray} 
\label{eq:b1_cov} 
  && 
   b =  \nabla_\mu h^{\mu} - \frac{1}{4}R + \frac{1}{4}\left[\omega^{\mu\nu}\Gamma^{\alpha}_{\beta\mu}\Gamma^\beta_{\alpha \nu} + \omega^{\mu\nu}\partial_\nu \Gamma^{\alpha}_{\mu\alpha}\right] +  \frac{1}{12}g_{i\alpha}\omega^{\mu\nu}\partial_\mu\partial_\nu g^{\alpha i} 
    \nonumber \\ 
    & &
    \quad\;
    \; + \, \frac{1}{6}g^{\mu i}\partial_\mu\partial_\nu g^{\nu j}\delta_{ij}
    + \, \frac{1}{6}\Gamma^{\alpha}_{\mu\alpha}g^{\nu i}\partial_\nu g^{\mu j} \delta_{ij} - \frac{1}{6}g_{\mu i}\partial_\nu g^{\alpha j}\partial_\alpha \omega^{\mu\nu} \, .
\end{eqnarray}
As expected, $b$ can be put in a manifestly covariant form by noting that
\begin{equation}\begin{split}
    \Delta L = & \, \frac{1}{12}g_{i\alpha}\omega^{\mu\nu}\nabla_\mu\nabla_\nu g^{\alpha i} + \frac{1}{6}g^{\mu i}\nabla_\mu\nabla_\nu g^{\nu j}\delta_{ij} 
       \\ & 
    - \left(\frac{1}{12}\omega^{\mu\nu}\partial_{\mu}\Gamma_{\nu\alpha}^\alpha 
    + \frac{1}{6}\omega^{\mu\nu}\partial_\alpha \Gamma^{\alpha}_{\mu\nu}+ 
        \frac{1}{12}\omega^{\mu\nu}\Gamma^{\alpha}_{\mu\beta}\Gamma^{\beta}_{\nu\alpha}  + \frac{1}{6}\omega^{\mu\nu}\Gamma_{\alpha \beta}^{\alpha}\Gamma^{\beta}_{\mu\nu}\right) ,
\end{split}
\end{equation}
so that $b$ reads
\begin{equation}
    b = \nabla_\mu h^\mu - \frac{5}{12}R + \frac{1}{12}g_{i\alpha}\omega^{\mu\nu}\nabla_\mu\nabla_\nu g^{\alpha i} + \frac{1}{12}g_{i\alpha}\omega^{\mu\nu}\nabla_\mu\nabla_\nu g^{\alpha i} + \frac{1}{6}g^{\mu i}\nabla_\mu\nabla_\nu g^{\nu j}\delta_{ij} \, .
\label{eq:bcovarform}
 \\[6pt]
\end{equation}
Note that while the expression of $b$ is manifestly covariant, the rotational symmetry in the space of matrix $g^{\mu i}$ is not manifest in the continuous-time Lagrangian that we derived. This symmetry is indeed broken at the level of the discrete-time Langevin equation Eq.~\eqref{eq:Tdis} to order $O(\Delta t^{3/2})$.

\subsection{Higher-order discretization point}
\label{sec:higher-order-disc-point-B}

To provide an alternative time-discretized covariant path-integral (compared to the one of Sec.~\ref{sec:explicit-covariant}),
the question we now ask  is whether it is possible to cast the previous discretization into the form of a higher-order one as described in Eq.~\eqref{eq:higher_order}, \textit{i.e.}~we look for $B^\mu_{\alpha\beta}$ such that discretizing the path integral at the point
\begin{equation}
    \bar{x}^\mu = x^\mu + \frac{\Delta x^\mu}{2} + B^\mu_{\alpha\beta}\Delta x^\alpha \Delta x^\beta \, ,
\end{equation}
gives the same continuous-time Lagrangian as the one inferred from Eqs.~\eqref{eq:propTdis1} and~\eqref{eq:bcovarform}.
 In the notations of the previous section, this amounts to having
\begin{equation}
    T_{i\mu\alpha\beta} = B^\nu_{\alpha\beta} \partial_\nu g_{i\mu} \, ,
\end{equation}
or equivalently
\begin{equation}
    {\mathsf {\B}}^{\mu i}_{\alpha\beta} = B^\nu_{\alpha\beta}\partial_\nu g^{\mu i}  \, .
   \label{eq:BBgprime}
\end{equation}
From Eq.~\eqref{eq:equivalence} we thus require that
\begin{equation}\label{eq:constraintB}
    \omega_{\rho \mu} g_{k \nu} \left(\omega^{\mu\nu}\omega^{\alpha\beta} + \omega^{\mu\alpha}\omega^{\nu\beta}+\omega^{\nu\alpha}\omega^{\mu\beta}\right)B^\sigma_{\alpha\beta}\partial_{\sigma}g^{\rho k} = \Delta L 
\end{equation}
where $\Delta L$ is given in Eq.~\eqref{eq:deltaL1}. This equation imposes only one constraint so that the solutions are in general degenerate. We look for a solution of the form
\begin{equation}\label{eq:solB}
    B_{\alpha\beta}^{\sigma} = \chi \, \Gamma^{\sigma}_{\mu\nu}\omega^{\mu\nu}\omega_{\alpha\beta} \, .
\end{equation}
Equation~\eqref{eq:constraintB} thus becomes a scalar equation over the parameter $\chi$,
\begin{equation}\label{eq:solLambda}
    \chi  = \frac{1}{\Gamma_{\sigma\rho}^{\rho} \Gamma^{\sigma}_{\mu\nu}\omega^{\mu\nu}}\left(\frac{\Delta L}{d + 2}\right) .
\end{equation}
Equation~\eqref{eq:solLambda} together with Eq.~\eqref{eq:solB} extend the results of~\cite{cugliandolo2019building} to multidimensional processes. Note that in the case in which
 the Riemannian manifold defined from the metric $\omega_{\mu\nu}$ is locally flat around some point $\bx_0$, \textit{i.e.}~with vanishing first derivatives of the metric but non-vanishing second ones, then we expect Eq.~\eqref{eq:constraintB} to be singular and $\chi$ defined in Eq.~\eqref{eq:solLambda} to diverge. The points in space at which this can occur are however expected to be isolated. Indeed if the metric has locally vanishing first and second derivatives then Eq.~\eqref{eq:constraintB} is automatically satisfied for any $B$ at that point. 
 
 We have therefore constructed a higher-order discretization scheme generalizing the one 
 of~\cite{cugliandolo2019building} that allows  us
to blindly use the chain rule at the level of continuous-time path-integral weights. In this discretization, all function are evaluated at
\begin{equation}
    \bar{x}^\mu = x^\mu + \frac{\Delta x^\mu}{2} + \frac{1}{\Gamma_{\sigma\rho}^{\rho} \Gamma^{\sigma}_{\zeta\xi}\omega^{\zeta\xi}}\left(\frac{\Delta L}{d + 2}\right) \Gamma_{\eta\delta}^{\mu} \omega^{\eta\delta}\omega_{\alpha\beta}\Delta x^\alpha \Delta x^\beta \, ,
\label{eq:discrschemecovI}
\end{equation}
\\
and the associated continuous-time Lagrangian reads
\begin{eqnarray}
 \label{eq:Lalpha_B}
&& 
\mathcal{L}^{\bx}[\textbf{x},\dot{\textbf{x}}] =  \, \frac{1}{2}\omega_{\mu\nu}\left(\frac{\dd x^\mu}{\dd t} - h^\mu\right)\left(\frac{\dd x^\nu}{\dd t} - h^\nu\right) +  \frac{1}{2}\nabla_\mu h^\mu - \frac{5}{24}R 
\nonumber\\[6pt] &&  
\qquad
+ \frac{1}{24}g_{i\alpha}\omega^{\mu\nu}\nabla_\mu\nabla_\nu g^{\alpha i} + \frac{1}{24}g_{i\alpha}\omega^{\mu\nu}\nabla_\mu\nabla_\nu g^{\alpha i} 
+ \frac{1}{12}g^{\mu i}\nabla_\mu\nabla_\nu g^{\nu j}\delta_{ij} \, .
\end{eqnarray}

These results bring new light on the mathematical subtleties associated to path integrals and the problems these can raise when manipulated improperly. The interest of formula~\eqref{eq:Lalpha_B} lies much more in its very existence, namely in the possibility of deriving such a covariant Lagrangian, than in its practical roll-out. While bringing answers to these questions for Onsager--Machlup path integrals in finite dimension, it certainly raises many questions for path integrals over fields where the internal dimension $d$ is somehow sent to infinity. The fate of covariant derivatives and curvature contributions deserves to be explored.

\section{Martin--Siggia--Rose--Janssen--De$\,\,$Dominicis (MSRJD) formulation}
\label{sec:msrjd}

Since the early path-integral formulation of quantum mechanics, there have been two equivalent 
{expressions} for {the} transition amplitudes. One involves only position operators. An alternative 
one includes additional conjugate momenta operators. 
The latter can be removed or included at will by Gaussian integration. A mirror image of the 
auxiliary momenta exists for stochastic dynamics: the alternative to the original Onsager--Machlup formulation 
that we have discussed at length
{is the Janssen--De~Dominicis path-integral approach~\cite{kubo_fluctuation_1973,dominicis_techniques_1976,Janssen1976,DeDominicis1978}}. It 
adds so-called  response variables or fields. The latter were first identified and introduced, in a (non-commuting) operator language, by Martin, Siggia and Rose~\cite{Martin1973} as an operator canonically conjugate to the field of interest. 

In this Section, we first present the MSRJD path integral in its $\alpha$-discretized form and derive from the previously obtained results about Onsager--Machlup path integrals the transformation properties of such constructions under non-linear changes of variables. We next extend the covariant construction of the path integral to the MSRJD path integral. We adopt the language of stochastic dynamics, but our results equally apply to quantum mechanics. Firstly, we do it for the simpler one-dimensional case, both in continuous and discrete versions. This allows us to sketch a proof of the invariance under time-reversal that allows one to derive the fluctuation-dissipation theorems which relate linear responses to correlation functions in equilibrium and fluctuation theorems out of equilibrium. Lastly, we extend the construction to higher dimensions.

\subsection{The $\alpha$-discretized MSRJD path integral}
We present in this section the construction of the $\alpha$-discretized path MSRJD path integral. Our starting point is the $\alpha$-discretized path integral in the Onsager--Machlup formalism. An additional field, coined as the response field, is introduced through a Hubbard--Stratonovich transformation to linearize the kinetic term. We start with the expression of the scalar invariant propagator with the $\alpha$-discretized path measure given in Sec.~\ref{subsec:new_measure}:
\begin{align}\label{eq:K_discrete_MSR}
& \K(\bxf,\tf|\bx_0,t_0) = \lim_{N \to + \infty} \left(\frac{1}{2\pi\Delta t}\right)^d \int \prod_{k=1}^{N-1}\frac{\dd \bx_k \sqrt{\omega(\bx_{k-1}+\alpha\Delta \bx_{k-1})}}{(2\pi\Delta t)^d}
 \nonumber \\ & \qquad\qquad\qquad\qquad
\prod_{k=0}^{N-1} \exp\left[-\frac{\Delta t}{2}\omega_{\mu\nu}\left(\frac{\Delta x_k^\mu}{\Delta t} - h^\mu\right)\left(\frac{\Delta x_k^\nu}{\Delta t} - h^\nu\right) - \Delta t \, \delta L_k\right],
\end{align}
where all functions are evaluated at $\bx_k + \alpha \Delta \bx_k$ and $\delta L_k$ can be read from Eq.~\eqref{eq:Lalpha_new} as
\begin{align}
\label{eq:deltaLalpha_new_MSR}
\!\!\! \delta L_k
 &
  = 
   \frac{1}{2}\left[\frac{\Delta x_k^\mu}{\Delta t}\left(\left(1-2\alpha\right)\omega_{\mu\nu}\omega^{\rho\sigma}\Gamma_{\rho\sigma}^\nu - 2\alpha \Gamma^\alpha_{\mu\alpha}\right) + 2 \alpha \nabla_\mu h^\mu \right. 
\nonumber\\ 
& \;\;
\left. - (1-2\alpha)\omega_{\mu\nu}\omega^{\rho\sigma}\Gamma^{\nu}_{\rho\sigma}h^\mu + \left(\alpha - \frac{1}{2}\right)^2\omega_{\mu\nu}\omega^{\rho\sigma}\omega^{\alpha\beta}\Gamma_{\rho\sigma}^\mu \Gamma_{\alpha\beta}^\nu  \right. 
\nonumber\\
 &  \;\;
 - \alpha\left(1-\alpha\right)R + \alpha\left(1-\alpha\right)\omega^{\mu\nu}\Gamma_{\beta\mu}^\alpha \Gamma_{\alpha\nu}^\beta + \alpha(2\alpha - 1)\omega^{\mu\nu}\partial_\nu \Gamma^\alpha_{\mu\alpha} \Bigg] \, .
\end{align}
with all functions also evaluated at $\bx_k + \alpha \Delta \bx_k$. The scalar invariant propagator can then be rewritten as
\begin{align}\label{eq:K_discrete_MSR2}
& \K(\bxf,\tf|\bx_0,t_0) = \lim_{N \to + \infty} \int_{-\infty}^{\infty}\int_{-i\infty}^{i\infty} \prod_{k=1}^{N-1}\frac{\dd \bx_k \dd \hat{\bx}_k}{(2i\pi)^d} \int_{-i\infty}^{i\infty}\frac{\dd \hat{\bx}_N}{(2i\pi)^d} \frac{1}{\sqrt{\omega(\bxf)}}
 \nonumber \\ & 
\qquad
\quad
 \exp\left[-\Delta t \sum_{k=1}^{N}\left(-\frac{1}{2}\hat{x}_{k,\mu}\hat{x}_{k,\nu}\omega^{\mu\nu}+\hat{x}_{k,\mu}\left(\frac{\Delta x_{k-1}^\mu}{\Delta t} - h^\mu\right) + \delta L_{k-1}\right)\right],
\end{align}
where all functions are now evaluated at $\bx_{k-1}+\alpha\Delta \bx_{k-1}$ and where the new field $\hat{x}_{k, \mu}$ is integrated along the imaginary axis. In the continuous-time limit, we thus write
\begin{align}\label{eq:K_continuous_MSR}
& \K(\bxf,\tf|\bx_0,t_0) = \int \mathcal{D}\bx \mathcal{D}\hat{\bx} \, \ee^{-\mathcal{S}[\hat{\bx},\bx]} \,,
\end{align}
where the measure over the physical and the response fields is given by
\begin{equation}
\mathcal{D}\bx \mathcal{D}\hat{\bx} = \lim_{N \to + \infty} \prod_{k=1}^{N-1}\frac{\dd \bx_k \dd \hat{\bx}_k}{(2i\pi)^d} \int_{-i\infty}^{i\infty}\frac{\dd \hat{\bx}_N}{(2i\pi)^d} \frac{1}{\sqrt{\omega(\bxf)}}  \, ,
\end{equation}
and the continuous-time action writes
\begin{equation}\label{eq:contMSR}
\mathcal{S}[\hat{\bx},\bx] \stackrel{\alpha}{=} \int_{t_0}^{\tf} \dd t \left(-\frac{1}{2}\hat{x}_{\mu}\hat{x}_{\nu}\omega^{\mu\nu} +\hat{x}_{\mu}\left(\dot{x}^\mu - h^\mu\right) + \delta L[\bx,\dot{\bx}]\right),
\end{equation}
with $\delta L[\bx,\dot{\bx}]$ immediately following from Eq.~\eqref{eq:deltaLalpha_new_MSR}. We are now in a position to use Eqs.~\eqref{eq:new_BT}-\eqref{eq:transfo_lagrange_new} to infer the transformation properties of the continuous-time action \eqref{eq:contMSR} under a non-linear change of variables. Upon introducing $\bu(t) = \textbf{U}(\bx(t))$, we get,
\begin{align}\label{eq:K_continuous_MSR2}
& \K(\buf,\tf|\bu_0,t_0) = \left(\frac{\det \left.\partial X^\mu/\partial u^\nu\right|_{\buf}}{\det \left.\partial X^\mu/\partial u^\nu\right|_{\bu_0}}\right)^{\alpha} \int \mathcal{D}\bu \mathcal{D}\hat{\bu} \, \ee^{-\tilde{\mathcal{S}}[\hat{\bu},\bu]} \,,
\end{align}
where 
\begin{equation}
\mathcal{D}\bu \mathcal{D}\hat{\bu} = \lim_{N \to + \infty} \prod_{k=1}^{N-1}\frac{\dd \bu_k \dd \hat{\bu}_k}{(2i\pi)^d} \int_{-i\infty}^{i\infty}\frac{\dd \hat{\bu}_N}{(2i\pi)^d} \frac{1}{\sqrt{\Omega(\buf)}} 
\end{equation}
and the action $\tilde{\mathcal{S}}[\hat{\bu},\bu]$ is given as usual as the $\alpha$-discretized integral of a Lagrangian $\tilde{\mathcal{L}}_{(\alpha)}^\bu[\hat{\bu},\bu,\dot{\bu}]$. The later can be inferred from the original one $\tilde{\mathcal{L}}_{(\alpha)}^\bx[\hat{\bx},\bx,\dot{\bx}]$ upon using the modified chain rule
\begin{align}
& \hat{\bx} \rightarrow \textbf{X}(\bu) \nonumber  \\
& \dot{x}^\mu \rightarrow \partial_\mu X^\alpha \dot{u}^\mu + \frac{1-2\alpha}{2}\Omega^{\mu\nu}\partial_\mu\partial_\nu X^{\alpha} \, ,
\end{align}
together with the correspondence  
\begin{align}
& \hat{x}_\mu \rightarrow \frac{\partial U^\nu}{\partial x^\mu}\hat{u}_\nu \, ,
\end{align}
and taking into account the systematic corrections given in Eq.~\eqref{eq:transfo_lagrange_new}. Concretely,
\begin{eqnarray}
&& \tilde{\mathcal{L}}_{(\alpha)}^\bu[\hat{\bu},\bu,\dot{\bu}] = 
\nonumber\\
&& 
\qquad
\tilde{\mathcal{L}}_{(\alpha)}^\bx\!\!\left[\frac{\partial U^\nu}{\partial x^\mu}\hat{u}_\nu,\textbf{X}(\bu),\partial_\mu X^\alpha \dot{u}^\mu + \frac{1-2\alpha}{2}\Omega^{\mu\nu}\partial_\mu\partial_\nu X^{\alpha}\right] + \alpha(1-\alpha)\sms\delta\sns\mathcal{L}[\textbf{u}] \,,
\qquad\quad
\end{eqnarray}
where
\begin{align}
\delta\sns\mathcal{L}[\textbf{u}] 
= 
\frac{1}{2}\frac{\partial \omega_{\mu\nu}}{\partial x^\eta}\frac{\partial^2 X^\eta}{\partial u^\alpha \partial u^\beta}\frac{\partial X^\mu}{\partial u^\rho}\frac{\partial X^\nu}{\partial u^\sigma}\Omega^{\alpha\rho}\Omega^{\beta\sigma}- \frac{1}{2}\omega^{\alpha\beta}\frac{\partial U^\nu}{\partial x^\beta}\frac{\partial^2 X^\mu}{\partial u^\phi \partial u^\nu}\frac{\partial^2 U^\phi}{\partial x^\alpha \partial x^\mu} \,.
\label{eq:deltaLu}
\end{align}
As already seen in Sec.~\ref{sec:ItocalculusPI} in the Onsager--Machlup formalism, the correcting term vanishes both for $\alpha = 0$ and $\alpha = 1$.
Also, in dimension one, this term becomes
$
\delta\sns\mathcal{L}[u]
=
-\frac{G(u)^2 X''(u)^2}{2 X'(u)^2}
-\frac{G(u) G'(u) X''(u)}{X'(u)}
$
%
and  corresponds to the last line of Eq.~(E.22) in Ref.~\cite{aron_dynamical_2014_arxiv1}\footnote{
The quantity $\delta\sns\mathcal L$ of Eq.~\eqref{eq:deltaLu} corresponds,
in the arXiv v1 Ref.~\cite{aron_dynamical_2014_arxiv1}, to the integrand of the last line of Eq.~(E.22) multiplied by $-D$ (with $D=\frac 12$ in our settings), which gives $-g^2\,(u''/u')^2/2 - gg'u''/u'$.
This is the same result above but when changing variables from $u$ to $x$.
}.
This shows that the functional action considered in Ref.~\cite{aron_dynamical_2014_arxiv1}, that was built directly in continuous-time,
can also be represented using the time discretization that we made explicit in the current section. 
Beyond such a linear $\alpha$-discretization, and much as in the case of the Onsager--Machlup formulation of path integrals, it is possible to use a modified discretization of the MSRJD action to make the resulting continuous-time action compatible with the usual chain rule, as we now detail.

\subsection{One-dimensional continuous-time covariant action}
\label{sec:response-fields-as}

In the covariant discretization scheme,  the MSRJD path probability measure
$
  \mathcal Dx
  \,  
  \mathcal D\hat x
  \:
  \ee^{-{\mathcal S}[\hat x, x]}
 $
of a one-dimensional process
\begin{equation}
\frac{\dd x}{\dd t} \stackrel{\frac{1}{2}}{=} f(x) + g(x)\eta(t)\,,
\end{equation}
 has an action~\cite{cugliandolo2019building}

\begin{eqnarray}
&& {\mathcal S}[\hat x,x] \stackrel{\betag}{=} \int_0^{\tf} \dd t \, \Big\{\hat x\left(\dot x-f (x) + \frac{1}{2}\,g(x)g'(x)\right) - \frac{1}{2} g(x)^2 \hat x^2 \nonumber\\ && \qquad\qquad\qquad\qquad+\frac{1}{2}f'(x)-\frac{1}{8}g'(x)^2-\frac{1}{2}\frac{g'(x)}{g(x)}\,\dot x\Big\}\; . \label{eq:OMactionalphabetatildeMSRJDalt2}
\end{eqnarray}

Or, up to a translation of the response field:

\begin{eqnarray}
 &&  {\mathcal S}[\hat x,x]\stackrel{\betag}{=}\int_0^{\tf} \dd t \,\Big\{\hat x\left(\dot x-f (x)\right) -\frac 12 g(x)^2 \hat x^2+\frac 12  f'(x)-\frac 12\frac{f(x)g'(x)}{g(x)}\Big\}\; . \qquad
\label{eq:OMactionalphabetatildeMSRJDalt2bis}
\end{eqnarray}
In the corresponding continuous-time 
path integral, one can change variables covariantly, by using the standard chain rule of calculus 
together with the correspondence 
\begin{equation}
\hat x(t)=U'(x(t))\, \hat u(t)
\end{equation}
between response fields. (One notes that there is no Jacobian contribution.)
In contrast, the historically derived MSRJD action in Stratonovich discretization 
misses the two last terms in~(\ref{eq:OMactionalphabetatildeMSRJDalt2}) 
and the application of the chain rule leads to inconsistencies~\cite{aron_dynamical_2016}.

The next subsection discusses the justification of such a covariance property in the discrete time formulation, 
by providing the appropriate discretization of the response field along with the explicit expression of the integration measure.

\subsection{One-dimensional discrete-time covariant action}
\label{sec:discretized-msrjd}
 
The MSRJD representation is constructed rewriting the infinitesimal 
propagator for $x(t)$ with the help of a Hubbard--Stratonovich transformation of the form
\begin{equation}
  \ee^{-\tfrac 12 \tfrac{b^2}{a}}
=
  \sqrt{\tfrac{a}{2\pi}}
\,
  \int_{i \mathbb R}
  d\hat x
  \;
  \ee^{\frac 12 a \hat x^2 -  b \hat x}
\end{equation}
at every time step, 
with the following choice of parameters $a$ and $b$

\begin{eqnarray}
  a={g(\bar x_k)^2}
\,{\Delta t}\; , \qquad\qquad\qquad
\label{eq:HStransfoa}
b=\Big[  {\frac{\Delta x}{\Delta t}-f (\bar x_k )}\Big]\,\Delta t\; . 
\end{eqnarray}
With these transformations, we obtain, for the first time step,
the infinitesimal scalar covariant propagator 
$\K_{\Delta t}(x_1,t_1|x_0,t_0)= g(x_1) \,\mathbb P(x_1,t_1|x_0,t_0)$ which reads, in the $\betag$-discretization, 
%
\begin{equation}
\begin{split}
&\K(x_1, t_1|x_0,t_0) \stackrel{\betag}  = \: g(\bar x_0)\int_{i\mathbb R}\frac{d\hat x_0}{2\pi}\:\ee^{-\delta {\mathcal S}[\hat x_0,\bar x_0]}
\\[2mm]
&\delta {\mathcal S}[\hat x_0,\bar x_0]\stackrel{\betag} =\Delta t\, \left\{\hat x_0\left[\dfrac{\Delta x}{\Delta t}-f (\bar x_0 )\right]
-
\frac{1}{2} g(\bar x_0)^2 \hat x_0^2+\frac 12  f'\!(\bar x_0)-\frac 12\frac{ g'\!(\bar x_0)}{g(\bar x_0)}  f(\bar x_0)\right\}\; .
\end{split}
\label{eq:infinitesimalactionMSRJD}
\end{equation}
It completely encodes the continuous-time expression%
%
\begin{equation}
\label{eq:OMactionalphabetatildeMSRJDalt}
{\mathcal S}[\hat x,x]\stackrel{\betag}=\int_0^{\tf}\!\! \dd t\,\left\{\hat x\,\big(\dot x  -f (x) \big)-\frac 12 g(x)^2 \hat x^2+\frac 12  f'(x)-\frac 12\frac{ g'(x)}{g(x)}  f(x)\right\}\;,
\end{equation}
on condition that, 
when writing 
$
\K(\xf,\tf|x_0,t_0)
=
\int_{x(t_0)=x_0}^{x(\tf)=\xf}  {\mathcal D}\hat{x} \, {\mathcal D}x \,
\ee^{- {\mathcal S}[\hat x,x]}
$,
the following path-integration measure is used
\begin{align}
  {\mathcal D}\hat{x} \, {\mathcal D}x
&= \prod_{k=1}^{N-1}\frac{\dd x_k}{ g(x_k)}\prod_{k=0}^{N-1}\frac{\dd \hat x_k \, g(\bar x_k)}{2 \pi}\nonumber\\&= \,{g(x_0)}  \prod_{k=1}^{N-1}{\dd x_k}\prod_{k=0}^{N-1}\frac{ g(\bar x_k)}{g(x_k)}\frac{\dd \hat x_k}{2\pi}\:.\label{eq:DxDxhat1D}
\end{align}
Up to a translation of the field $\hat x(t)$ by $g'/(2g)$, one 
 recovers the expression~(\ref{eq:OMactionalphabetatildeMSRJDalt2}) of the action
 from Eq.~\eqref{eq:OMactionalphabetatildeMSRJDalt}.
In Eq.~\eqref{eq:OMactionalphabetatildeMSRJDalt}, the symbol $\betag$ over the equality sign means that functions of the variable $x$ 
are $\betag$-discretized.
A variable $\hat x_k$ is introduced at each time step $t_k=k\Delta t$ and merely associated to $\bar x_k$. In other words, at this stage there is no discretization issue with the response field which is evaluated at the beginning of the time-slice.
The proof of the covariance presents more intricate passages 
than for the Onsager--Machlup action (notably because each response field $\hat x_k$ scales as $\Delta x/\Delta t$, \textit{i.e.}~$\Delta t^{-1/2}$), and is sketched in App.~A.3 of Ref.~\cite{cugliandolo2019building}.

\subsection{The time-reversal symmetry for one-dimensional processes}
\label{sec:symmetry-one-dim}

We now follow similar steps to the ones in~\cite{aron_dynamical_2016} to show the invariance of the
propagator of equilibrium processes under time-reversal. We work in the covariant continuous 
time formulation interpreted in the $\betag$ discretization scheme and as we have already shown,  
we are then allowed to use the usual rules of calculus. 

The time-reversal operation on a generic time interval $[t_0, \tf]$ is
\begin{equation}
\tR=\tf+t_0-t 
\end{equation}
and we define the time-reversed variables
%
\begin{equation}
\begin{split}
& {\rm T}x(t) = x(\tR)
\\
& {\rm T}\hat x(t) = \hat x(\tR) + \frac{2}{(g(x(\tR)))^2} \frac{\dd}{{\rm d}t} x(\tR)
\; . 
\end{split}
\end{equation}
The action~(\ref{eq:OMactionalphabetatildeMSRJDalt2}) evaluated in these new variables reads
\begin{eqnarray}
&&\hspace{-1.25cm}{\mathcal S}[ {\rm T}\hat x(t) , {\rm T} x(t) ]\stackrel{\betag}{=}\int_{t_0}^{\tf} \!\! \dd t \,\Bigg\{\left[ \hat x(\tR)+ \frac{2}{(g(x(\tR)))^2}  \frac{\dd}{{\rm d} t} x(\tR) \right]\nonumber\\&& \qquad\qquad  \qquad \qquad \qquad\left[\frac{\dd}{{\rm d} t} x(\tR)-f(x(\tR)) + \frac 12 \,g(x(\tR))g'(x(\tR))\right]\nonumber\\&& \qquad\quad  \qquad   \qquad \qquad-\frac 12 (g(x(\tR)))^2 \left[\hat x(\tR) + \frac{2}{(g(x(\tR)))^2} \frac{\dd}{{\rm d} t} x(\tR)\right] ^2\nonumber \\&&\qquad\quad  \qquad\qquad \qquad +\frac 12  f'(x(\tR))-\frac{1}{8} g'(x(\tR))^2-\frac 12\frac{g'(x(\tR))}{g(x(\tR))}\,\frac{\dd}{{\rm d}t} x(\tR)\Bigg\}. \nonumber\end{eqnarray}
Changing now the time integration argument $t \leftrightarrow \tR$, 
and not writing it explicitly, 
\begin{eqnarray}
&& {\mathcal S}[ {\rm T}\hat x(t) , {\rm T} x(t) ]\nonumber\\&& \qquad\quad\stackrel{\betag}{=} \int_{t_0}^{\tf} \!\! \dd t \,\Bigg \{\left[ \hat x - \frac{2}{g^2(x)}  \frac{\dd}{{\rm d} t} x \right]\left[- \frac{\dd}{{\rm d} t} x-f (x) + \frac 12 \,g(x)g'(x)\right]\nonumber\\&&\qquad\qquad \qquad -\frac 12 g^2(x) \left[\hat x - \frac{2}{g^2(x)}  \frac{\dd}{{\rm d} t} x\right] ^2+\frac 12  f'(x)-\frac{1}{8}g'(x)^2+\frac 12\frac{g'(x)}{g(x)}\,\frac{\dd}{{\rm d}t} x\Bigg \}.\nonumber\end{eqnarray}
 Several terms can already be 
identified as invariant under ${\rm T}$; others need some work. After expanding the product and square, and 
performing some trivial cancellations
\begin{eqnarray}
{\mathcal S}[ {\rm T} \hat x(t) , {\rm T} x(t) ]
 \!\!  & \!\! \stackrel{\betag}{=} \!\! & \!\!
 S[ x(t), \hat x(t) ] + 2 \int_{t_0}^{\tf} \dd t \, \frac{{\dot x} f}{g^2} 
 \; . 
 \label{eq:symmetry-proof}
 \end{eqnarray}
Note that the above equation only holds because the higher-order discretization scheme of Eq.~(\ref{eq:defMdiscreti}) that we proved to be manifestly covariant is also left invariant under time reversal, as is the Stratonovich one. In conservative cases, $f$ obeys Eq.~\eqref{eq:equil-choice-f} with $\alpha = 1/2$, \textit{i.e.}~$f=-\beta g^2 V'/2 + g g'/2$, which guarantees that the steady state distribution is the Boltzmann distribution in a potential $V$ at inverse temperature $\beta^{-1}$. 
The last term of Eq.~\eqref{eq:symmetry-proof} is just the integral of a total derivative and yields 
\begin{eqnarray}
& 
2 
\displaystyle{
\int^{\tf}_{\substack{
 t_0 \\
 \scriptstyle{\beta_g}
  }}} 
 \; \dd t \;\, \dfrac{{\dot x} f}{g^2} \, =  
-\beta\big[V(x(\tf))-V(x(t_0))\big] + \ln \left( \dfrac{g(x(\tf))}{g(x(t_0))}\right) \,.
\label{eq:TRint}
\end{eqnarray}
These terms, which look odd at this stage, are just what is needed to guarantee that detailed balance holds. We can indeed express the steady state probability density to be at $x_0$ at time $t_0$ and $\xf$ at time $\tf$ in terms of the scalar invariant propagator as,
\begin{equation}
\mathbb{P}(\xf, \tf | x_0,t_0) = \frac{e^{-\beta V(x_0)}}{g(\xf)}\,  \mathbb{K}(\xf, \tf \vert x_0, t_0) \,,
\end{equation}
and the steady state probability density to be at $\xf$ at time $t_0$ and $x_0$ at time $\tf$ as,
\begin{equation}
\mathbb{P}(x_0, \tf | \xf, t_0) = \frac{e^{-\beta V(\xf)}}{g(x_0)} \mathbb{K}(x_0, \tf \vert \xf, t_0) \,.
\end{equation}
Furthermore, the two scalar invariant propagators are related by Eqs.~\eqref{eq:symmetry-proof}-\eqref{eq:TRint},
\begin{equation}
\mathbb{K}(x_0, \tf \vert \xf, t_0) = \mathbb{K}(\xf, \tf \vert x_0, t_0)
\exp\left[\beta[V(x(\tf))-V(x(t_0))] - \ln \left( \frac{g(x(\tf))}{g(x(t_0))}\right)\right] \,,
\end{equation}
thus yielding the desired equality 
\begin{equation}
\mathbb{P}(\xf, \tf | x_0,t_0) = \mathbb{P}(x_0, \tf | \xf, t_0) \,.
\end{equation}
 
The application of this symmetry to the calculation of correlation functions between the physical and the auxiliary variables provides simple proofs of the celebrated fluctuation-dissipation theorem linking linear response and correlation functions in equilibrium~\cite{Kubo66}. Moreover, it also allows one to derive the more recent fluctuation theorems valid out of equilibrium~\cite{evans_probability_1993,gallavotti_dynamical_1995-1,gallavotti_dynamical_1995,kurchan_fluctuation_1998,lebowitz_gallavotticohen-type_1999,Evans02,Sevick08,seifert2012stochastic}.
 
\subsection{Higher dimensions}
\label{sec:higher-dim-MSRJD}

We now pick Eq.~(\ref{eq:covStrato}) as our starting point and we linearize the quadratic term by a Hubbard--Stratonovich transformation. In the 
${\B}$-discretization of Eq.~\eqref{eq:defMdiscreti}, the MSRJD higher dimensional action is 
\begin{eqnarray}
{\mathcal S}[\hat {\bf x}, {\bf x}] 
\stackrel{{\B}}{=}
\int_{t_0}^{\tf} \dd t
\; 
\left\{
-\frac{1}{2} \hat x_\mu \, {\omega}^{\mu\nu} \, \hat x_\nu + \left( \dfrac{\dd x^\mu}{\dd t} - h^\mu \right) \hat x_\mu + \nabla_\mu h^\mu - \lambda R
\right\}
. 
\end{eqnarray}
The path-integration measure, similarly to Eq.~\eqref{eq:DxDxhat1D} in one dimension, takes the form
\begin{align}
  {\mathcal D}\hat{\bx} \, {\mathcal D}\bx
  &= 
  \frac{1}{\sqrt{\omega(x_0)}}
  \prod_{k=1}^{N-1}
{\dd^d \bx_k}
  \prod_{k=0}^{N-1}
 \sqrt{\frac{\omega(\bx_k)}{ \omega(\bar \bx_k)}}
 \frac{\dd^d \hat \bx_k}{(2\pi)^d}
\label{eq:DxDxhatanyd}
\end{align}
when writing 
\begin{equation}
\K(\bxf,\tf|\bx_0,t_0)
=
\int_{\bx(t_0)=\bx_0}^{\bx(\tf)=\bxf}  {\mathcal D}\hat{\bx} \, {\mathcal D}\bx 
\,
\ee^{- {\mathcal S}[\hat \bx,\bx]}\,.
\label{eq:MSRJDexpressionK}
\end{equation}
The time-reversal symmetry of Sec.~\ref{sec:symmetry-one-dim} generalizes to 
\begin{equation}
\begin{split}
& {\rm T}x^\mu(t) = x^\mu(\tR)
\\
& {\rm T}\hat x_\mu(t) = \hat x_\mu(\tR) + \omega_{\mu\nu} \frac{\dd}{{\rm d}t} x^\nu(\tR)
\; . 
\end{split}
\end{equation}

\section{Conclusions and open questions}

The quest for a path-integral formulation of stochastic and quantum mechanical problems amenable to such natural manipulations as the change of integration path is almost as old as path integrals themselves. This work has reviewed the progress achieved by DeWitt~\cite{dewitt_dynamical_1957} and later by Graham~\cite{graham_covariant_1977,graham_path_1977}, 
but it has also presented an alternative way to construct manifestly covariant path integrals for generic  
$d$-dimensional Langevin evolutions with Gaussian noises. 

\begin{figure}
\centerline{
\includegraphics[width=.9\columnwidth]{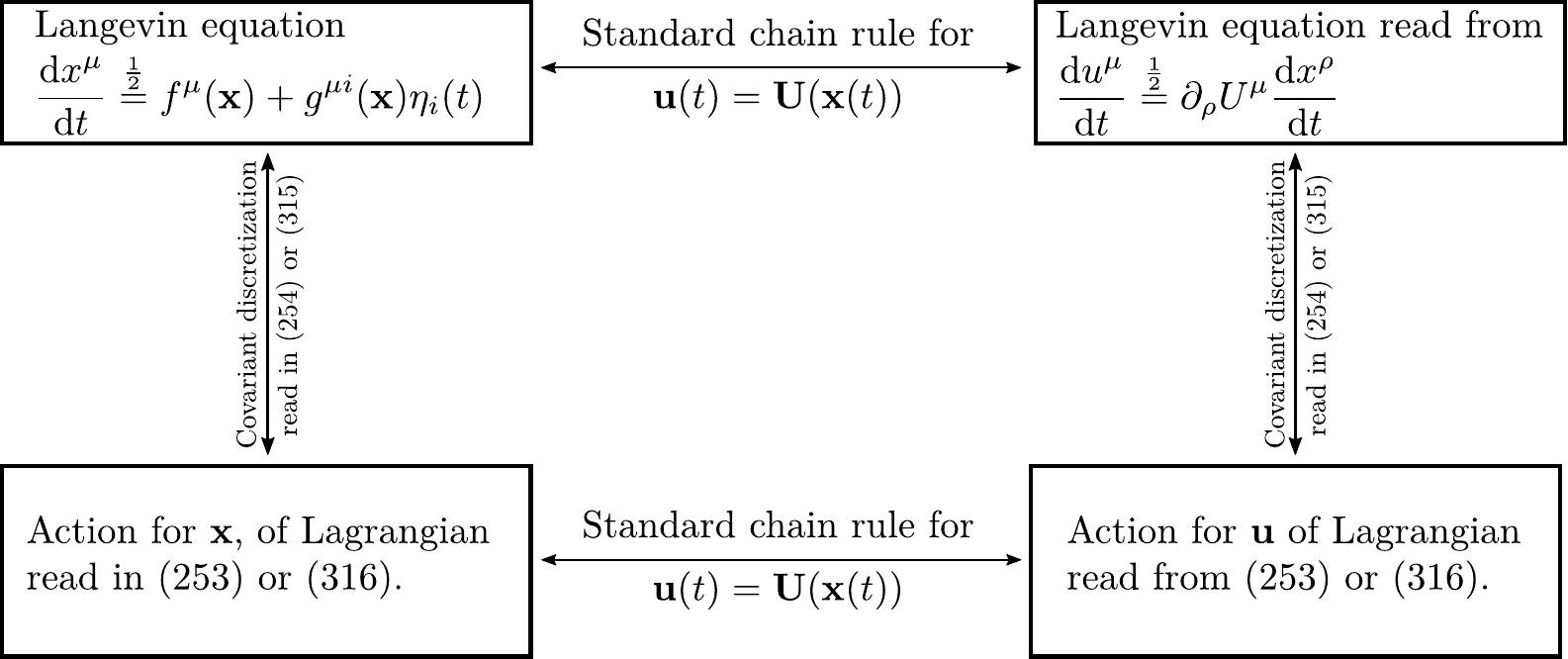}
}
\caption{A graphical summary of the path-integral covariance. We propose two methods to build a non-linear discretization scheme that renders the path-integral construction covariant.
In Sec.~\ref{sec:explicit-covariant}, we showed how the discretization of Eq.~(\ref{eq:defMdiscreti}) leads to a Lagrangian, given in Eq.~(\ref{eq:covStrato}), that generalizes the ones of DeWitt and Graham.
In Sec.~\ref{sec:covLangevin}, we showed how, starting from a discrete-time covariant Langevin equation, one arrives at the discretization scheme of Eq.~\eqref{eq:discrschemecovI},
which yields the Lagrangian~(\ref{eq:Lalpha_B}). The two Lagrangians~(\ref{eq:covStrato}) and~(\ref{eq:Lalpha_B}) 
present different continuous-time expressions but both are covariant, equally valid, 
and define the \textit{same} trajectory probability, because 
they have to be read with their own discretization scheme.
}
\label{fig:conclusion}
\end{figure}

In DeWitt's and Graham's approaches, path integrals are constructed by means of an \textit{implicit} 
discretization resulting from solving the least action principle on infinitesimally small time intervals, eventually stitched one after the other. What is implicit is the dependence of the paths on the discretization time step. %
However, the resulting action is indeed covariant upon a change of integration path, in the sense that one can manipulate the path as if it were differentiable and change variables by applying the usual chain rule in the Lagrangian.
The action involves intrinsic properties of the underlying metric built from the noise correlation matrix, thereby endowing the path-integral construction with a clear geometrical interpretation in terms of geodesics on a well-defined manifold. 
Why then, after all, would there be any need for alternative formulations? The answer is quite simple: at least in the physics of stochastic processes, no one constructs path integrals based on a least-action principle. Instead, paths themselves are directly discretized, and the action evaluated at a given path is \textit{explicitly} expressed in terms of a sequence of values taken by that path. This construction is, of course, much more straightforward, and its appeal lies in the fact that for a given trajectory measured at equally spaced time intervals, one could directly determine its probability of occurrence. 
The surprise is that 
none of the linear discretization schemes that are used at the level of stochastic differential equations is however compatible with a covariant change of path. 
The present work fills this gap by proposing two methods of achieving this goal (corresponding to the results of Secs.~\ref{sec:explicit-covariant} and~\ref{sec:covLangevin}), which we have pictorially encapsulated in Fig.~\ref{fig:conclusion}. 

Alternatively to these covariant constructions, if path integrals are built using a linear $\alpha$-discretization, the usual rule of stochastic calculus for changing variables has to be, generically, abandoned. A notable exception are Itō-discretized path integrals provided that the path measure is itself Itō-discretized~\cite{ding2021timeslicing}. Beyond this special case, this work also presents the non-trivial transformation properties of $\alpha$-discretized path integrals in any dimension $d$, both in the Onsager--Machlup and the MSRJD formalisms. \\ 
	
Neither our review nor our new results have addressed other important issues that may come to mind. The rest of this conclusion section briefly discusses some of these.


A first set of questions has to do with the MSRJD dynamical actions, dual to the Onsager--Machlup ones, that involve, on top of the expected trajectory field, a so-called response field.
It is well-known~\cite{janssen_renormalized_1992} that the Itō discretization, because it makes causality explicit, is the most convenient one for perturbative expansions.
%
%
Yet, for the study and identification of symmetries and invariances, especially under non-linear transformations (such as canonical ones~\cite{tailleur2008mapping}), one may also be after a covariant discretization scheme.
We have presented in Sec.~\ref{sec:msrjd} a scheme that is covariant upon a transformation of the physical variable $\bx$ that is decoupled from the response variable $\hat\bx$.
For a generic canonical transformations (such as the Cole--Hopf one), the added technical difficulty is that $\bx$ and $\hat\bx$ are not discretized according to the same scheme, 
and performing a change of fields will further scramble discretization issues. 
This brings us to a closely related question. While path integrals \textit{{à la}} Onsager--Machlup cannot be written when the noise statistics is not Gaussian, 
the response-field formulation 
allows one to write the action for any process driven by a white possibly non-Gaussian noise (namely a Poisson point process), as long as the generating functional of the latter is known~\cite{phythian1977functional}. For such processes, it is well-known~\cite{falsone2018stochastic} that stochastic calculus requires a discretization rule of infinite order. It is almost certain that the same requirement (working with an infinite order discretization rule) will hold at the path-integral level, but again the interplay with the response field is yet to be understood.

 The fate of the geometrical concepts that pervade the Gaussian white noise case is also an open question.
Of physical interest is the case of time-correlated (so-called colored) Gaussian noise, where the regularity of the solution of the Langevin equation is stronger than in the white noise case,
which eases its manipulation upon differentiation~\cite{hanggi_path_1989,hanggi1993path}.
Furthermore,
supersymmetries are known to hold for stochastic actions of 
equilibrium~\cite{parisi_supersymmetric_1982,feigelman_hidden_1982} and out-of-equilibrium~\cite{marguet_supersymmetries_2021}
stochastic processes
when the underlying Langevin process is additive. The extension to multiplicative noise remains an open problem, and is connected to the covariance of the path integral (since the discretization in the supersymmetric action is Stratonovich).
Finally, interesting connections could be made to the recent approach of Ref.~\cite{kappler_stochastic_2020}, based on the analysis of the probability tubes enclosing trajectories of the stochastic action and that converge to a smooth path as the diameter of the tube is sent to 0 (in the spirit of the approach of~\cite{capitaine_onsager-machlup_2000} in mathematics).
~\\

Path integrals are ubiquitous in statistical and quantum field theory. There, the fields are not simple functions of time. They may depend on time, but above all they depend on space variables, which can be discrete (for lattice field theories) or continuous. In the language of this review, field theory corresponds to working in infinitely many dimensions (a countable infinity on a lattice). Consider for the purpose of discussion a static field theory living in a one-dimensional continuous space, with action
\begin{equation}
  {\mathcal S}[\phi]=\int\dd x \, (\p_x\phi)^2 \, .
\end{equation}
This corresponds, for instance, to the potential energy of a one-dimensional elastic medium. It is not hard to realize that a non-linear change of field will inevitably raise the same questions as those taken up in this work. It is possible, however, that the embedding space dimension plays an important role regarding the smoothness of the typical field fluctuations.
A direct analogy between $x(t)$ and $\phi(x)$ tells us, that in dimension $1$, the elastic field is not differentiable (its spatial gradient behaving as one over the root of the spatial discretization scale).
The case of higher dimensions, ${\mathcal S} [\phi]=\int\dd^d x\,(\nabla\phi)^2$ certainly deserves further attention since
the divergence appears even more severe;
indeed, from power counting, the scaling of the increment $\Delta \phi$ of the field is $\Delta x^{1-d/2}$,
which is singular in dimensions $d>2$.
The case of actions containing higher-order derivative is also of interest, both in quantum field theory or for stochastic processes~\cite{dean_path_2019}.

In fact, difficulties occurring at the field theoretic level have been known for a while. A recent report of the failure of a (canonical) change of fields was presented in~\cite{wilson2011breakdown} (where the authors study the Bose--Hubbard model either in the phase-amplitude variables or using the coherent states of the creation and annihilation operators)
and analyzed in~\cite{bruckmann_rigorous_2018,Rancon20}.
For the purpose of our discussion, we present the issue in a somewhat different context (with an almost identical field-theoretic formulation).
 Consider a one-site pair annihilation process occurring at rate $k$. The master equation for the probability of observing $n$ particles at time $t$ has an evolution operator that can be written~\cite{grassberger1980fock,peliti1986renormalisation,cardy1996renormalisation} in terms of Hermitian conjugate operators $a$ and $a^\dagger$ with a bosonic commutation rule $[a,a^\dagger]=\mathbf{1}$~as
\begin{equation}
\mathbb{W}=k(a^\dagger{}^2-1)a^2
\end{equation}
which one could also express in terms of the two operators $\rho$ and $\rho^+$ defined by $\rho=a^\dagger a$ and $a^\dagger=\ee^{\rho^+}$ that also verify $[\rho,\rho^+]=\mathbf{1}$ (without being Hermitian conjugate),
\begin{equation}
\mathbb{W}=k(1-\ee^{-2\rho^+})\rho(\rho-1)
\; . 
\end{equation}
The corresponding Itō-discretized field theories (with fields $a$, $\bar a$, $\rho$, $\bar \rho$ corresponding to the operators $a$, $a^\dagger$, $\rho$, $\rho^\dagger$), based on a coherent state construction~\cite{doi_second_1976,peliti_path_1985} or on the alternative approach of~\cite{andreanov2006field}, read
\begin{equation}
\begin{aligned}
&
S_1[\bar{a},a] =\int\dd t\left[\bar{a}\dot{a}+k(\bar{a}^2-1)a^2\right] 
\\
&
S_2[\bar{\rho},\rho]=\int\dd t\left[\bar{\rho}\dot{\rho}+k(\ee^{-2\bar{\rho}}-1)\rho(\rho-1)\right]
,
\end{aligned} 
\end{equation}
whereas a naive substitution of the fields $\bar{a}$ and $a$ in $S_1$ by $\bar{a}\to\ee^{\bar{\rho}}$,  $a\to\ee^{-\bar{\rho}}\rho$ (as first introduced in~\cite{grassberger_phase_1982}, leads to an incorrect
\begin{equation}
S_3[\bar{\rho},\rho]=\int\dd t\left[\bar{\rho}\dot{\rho}+k(\ee^{-2\bar{\rho}}-1)\rho^2\right]
\end{equation}
($S_3\neq S_2$). Of course, in hindsight, the appearance of an exponential in the response field $\bar{\rho}$ expresses the strongly non-Gaussian nature of the noise acting on field $\rho$, and this brings us back to our questions on the extension of the MSRJD
formalism to non-Gaussian noises. High energy physics, condensed matter, or stochastic processes, are areas of physics where field-theoretic methods are heavily relied on. Not all situations can be phrased back in terms of an effective Langevin equation. Learning how to deal with fields with confidence, regarding both their time and space discretizations, seems to us to be of prime importance in establishing path integrals on even more solid grounds.

\section*{Acknowledgements} We thank C.~Aron, D.~Barci, R.~Chetrite, R.~Cont, P.-M. D\'ejardin, H.~W.~Diehl, P.~Drummond, Z.~Gonz\'alez-Arenas, H.~J.~Hilhorst, H.~K.~Janssen, 
G.~S.~Lozano, A.~Rançon, S.~Renaux-Petel,  and 
F.~A.~Schaposnik for very helpful discussions. 
LFC \& FvW acknowledge financial support from the ANR-20-CE30-0031 grant THEMA 
and
VL  from the ANR-18-CE30-0028-01 grant LABS.

\bibliographystyle{tfq}
\bibliography{biblio-paths}

\newpage

\appendix 

\section{Bath oscillator models}
\label{app:oscillators}

Although it is conventional to write the Langevin equation in the form in Eq.~(\ref{eq:x-eom-drifted}), 
it can be illuminating to rewrite it as
\begin{equation}
 k^2(x) \frac{\dd x}{\dd t}  =  f(x) -  2D  (1-\alpha) \frac{\dd \ln|k(x)|}{\dd x}   + k(x) \eta\;,
\label{eq:eq-a-la-Aron-Biroli-LFC}
\end{equation}
where we re-parametrized $g(x) \equiv 1/k(x)$. Equation~(\ref{eq:eq-a-la-Aron-Biroli-LFC}) is the one derived for 
the Markovian overdamped dynamics of a particle subject to a force $f$ and interacting with a bath of oscillators \textit{via} a non-linear coupling $\sum_a q_a K(x)$ with $K'(x)\equiv k(x)$ and $q_a$ the coordinates of the 
oscillators labelled $a$~\cite{zwanzig_nonequilibrium_2001, AronLeticia2010}. The exact  integration over the degrees of freedom of the bath gives rise to a viscous friction force, here in the l.h.s., as well as the  multiplicative noise in the r.h.s. 
We can therefore re-interpret the time derivative, $\dd x/\dd t $, in the l.h.s.~of Eq.~(\ref{eq:x-eom-drifted}) as originating from the dissipative interaction with the same bath that is responsible for the time-dependent random noise $\eta$.

\section{Derivation of the Fokker--Planck equation}
\label{app:FP}

The Kramers or Fokker--Planck (FP) approach is useful to prove that a Langevin process 
takes the system to Gibbs--Boltzmann equilibrium at the working temperature. It is a deterministic partial differential equation on the probability distribution 
for the stochastic variable to take a given value, say $y$, at  the measuring time, $t$.

In order to derive the FP equation, let us start from the Markov process identity
\begin{equation}
P(y,t_{k+1}) = \int dx_0 \ \mathbb{P}_{\Delta t}(y, t_{k+1} |x_0, t_k) \ P(x_0,t_k)
\; , 
\label{eq:recursion-P}
\end{equation}
where  $\mathbb{P}_{\Delta t}(y, t_{k+1} |x_0, t_k)$ is the conditional probability of finding $y$ at
the discretized time $t_{k+1}$ provided the system was in the state $x_0$ at the previous instant $t_k$
(note that $x_0$ is not necessarily the initial value here). The integral
runs over all accessible values of $x_0$. This equation holds for
any time increment but we focus on infinitesimal ones here.
It is also-called the Chapman--Kolmogorov equation.

Focus now on the conditional probability for the Langevin process
\begin{equation}
\mathbb{P}_{\Delta t}(y, t_{k+1}|x_0, t_k)
=
\langle \delta(y-x_{k+1}) \rangle
\label{eq:def-Px}
\end{equation}
where the mean value is taken over the noise $\{ \eta_k\}$, and
$x_{k+1}$  is determined
by the Langevin  equation with the ``initial condition''
$x_k=x_0$. Expanding Eq.~(\ref{eq:def-Px})
 in powers of $\Delta x_k \equiv  x_{k+1} - x_k=y-x_0$
we immediately obtain
 \begin{equation}
\mathbb{P}_{\Delta t}(y,t_{k+1}|x_0, t_k)
=
\delta(y-x_0) - \partial_y \delta(y-x_0) \langle \Delta x_k\rangle
+
\frac{1}{2} \partial^2_y \delta(y-x_0) \langle \Delta x_k^2\rangle + \dots
\label{eq:cond-prob-x}
\end{equation}
where the ellipsis indicate terms involving higher-order
moments of $\Delta x_k$.
The idea is to compute the averages  $\langle \Delta x_k\rangle$ and 
$\langle \Delta x_k^2\rangle$ to 
leading order in $\Delta t$ and then take the limit
$\Delta t\to 0$. To do this, we need to use the Langevin equation 
and it is at this point that its discretized form 
plays a role.

Up to corrections $O(\Delta t^{3/2})$ that we neglect, the $\alpha$-discretized Langevin equation reads
\begin{equation}
\Delta x_k =  f(x_k)  \Delta t +  g(x_k)   \Delta\eta_k  + g'(x_k) \alpha \Delta x_k \Delta\eta_k 
\; ,
\end{equation}
where we have deliberately chosen to evaluate the functions $f, \ g$ and $g'$ at the pre-point 
$x_k$,
and the $\eta_k$'s are zero-mean Gaussian variables with variance $\langle \Delta\eta_k  \Delta\eta_{k'}\rangle = 2D\delta_{kk'}\Delta t$.
Replacing $\Delta x_k$ in the last term by this very same equation and  
keeping all terms that contribute to a noise average up to ${\cal O}(\Delta t)$
we get
\begin{eqnarray}
 \Delta x_k =  f(x_k)  \Delta t +  g(x_k)   \Delta\eta_k 
+ \alpha g(x_k) g'(x_k)   \Delta\eta_k^2
\; . 
\end{eqnarray}
If we fix $x_k$ to take the value $x_0$ in the expansion for 
$\mathbb{P}(y, t_{k+1}|x_0,t_k)$, $x_k$ is not correlated with the noise $ \Delta\eta_k $.
Therefore, under  the noise average  the second term in the r.h.s.~vanishes.
Using $\langle   \Delta\eta_k^2\rangle = 2D\Delta t$, 
\begin{equation}
\langle \Delta x_k\rangle = 
f(x_k) \Delta t + 2D  \alpha g(x_k) g'(x_k) \Delta t
\; . 
\end{equation}
The next case to consider, also to ${\cal O}(\Delta t)$, is 
\begin{eqnarray}
&&\langle \Delta x_k^2 \rangle \simeq \langle [g(x_k)    \Delta\eta_k  ]^2 \rangle 
= 2D g^2(x_k) \Delta t 
\; . 
\end{eqnarray}
Replacing now in Eq.~(\ref{eq:cond-prob-x}), and next in Eq.~(\ref{eq:recursion-P}), 
\begin{eqnarray}
&&
P(y,t_{k+1}) =  P(y,t_k) - \Delta t \, \partial_y \! \int dx_0 \ [f(x_0) + 2 D \alpha g(x_0)  g'(x_0)] 
\, \delta(y-x_0) \,  P(x_0,t_k)
\nonumber\\
&&
\qquad\qquad\quad\;\;
+ D\Delta t \, \partial^2_y \int dx_0 \  \delta(y-x_0) \ g^2(x_0) \ P(x_0,t_k)
\; , 
\end{eqnarray}
performing the integrals over $x_0$, and after some rearrangements, in the $\Delta t\to 0$ limit,
\begin{equation}
\partial_t P(y,t) =  - \partial_y \{ [f(y) + 2D \alpha g(y) g'(y)]  \ P(y,t) \}
+ D \partial^2_y  [g^2(y) P(y,t) ] 
\; . 
\label{eq:FP-multiplicative}
\end{equation}
For $g(x)=1$ we recover the well-known FP equation for an additive Gaussian white noise process.

\section{Inverse function in dimension one}
\label{app:inverse}

In the $\alpha$-discretization
\begin{eqnarray}
\bar x_k = x_k + \alpha \Delta x_k = x_{k+1} + (\alpha-1) \Delta x_k
\; . 
\label{eq:app:barxk}
\end{eqnarray}
Take the function $U$ such that $u_k = U(x_k)$,  $\forall \, k=0,\dots, N$. After Taylor expansion of
$\bar u_k = \alpha U(x_{k+1}) + (1-\alpha) U(x_k) = 
\alpha U(\bar x_k + (1-\alpha)  \Delta x_k) 
+ (1-\alpha) U(\bar x_k - \alpha \Delta x_k)$ around $\bar x_k$, and similarly for the inverse function $U^{-1}=X$, 
\begin{eqnarray}
\begin{array}{ll}
& \bar u_k = U(\bar x_k) + \frac12 U''(\bar x_k) \alpha(1-\alpha) \Delta x_k^2 + \dots
\\[6pt]
&
\bar x_k = X(\bar u_k) + \frac12 {X}''(\bar u_k) \alpha(1-\alpha) \Delta u_k^2 + \dots
\end{array}
\end{eqnarray}
and, due to  $\Delta x_k^2 \approx \Delta u_k^2  \approx \Delta t$, the functional relation is translated from 
$\{x_k,u_k\}$ to $\{\bar x_k, \bar u_k\}$ up to $O(\Delta t)$ corrections:
\begin{eqnarray}
&& 
\bar u_k = U(\bar x_k) + O( \Delta t) 
\; , 
\qquad\qquad\qquad
\bar x_k = X(\bar u_k) + O( \Delta t) 
\; . 
\end{eqnarray}
The $\Delta x_k^2$ correction to~(\ref{eq:app:barxk}) introduced in the $(\alpha,\beta)$ discretisation
does not change this conclusion, since it brings other terms of order $O(\Delta t)$ or higher.

\section{Fixing the $(\alpha=1/2, \betag)$ discretization in one dimension}
\label{app:fixing-alpha-beta}

In this Appendix, we fix the parameter $\beta$ 
in the higher order discretization scheme of Eq.~\eqref{eq:higher_order} of one dimensional processes
by imposing that the increment $\Delta x_k$ of the discretized Langevin equation matches the one coming from 
the $\mathbb{T}_{\bff, \bg}$-discretization discussed in Sec.~\ref{sec:covLangevin} (which is covariant at all orders in $\Delta t$).

On the one hand, we expand  $\Delta x_k$, as determined by the one-dimensional Langevin equation
with the $(\alpha, \beta)$ discretization scheme of Eq.~\eqref{eq:higher_order}, keeping all terms up to $O(\Delta t^{3/2})$ and 
evaluating $f$, $g$ and $\beta$ at the  pre-point $x_k$.  We note that one does not have to worry about 
the argument in $\beta$ since it is always multiplied by 
$\Delta x_k^2 = O(\Delta t)$. We then safely set it to $x_k$ and we 
derive
\begin{eqnarray*}
&&
\Delta x_k 
=
f(x_k +\alpha \Delta x_k + \beta(x_k) \Delta x_k^2) \Delta t +g(x_k +\alpha \Delta x_k + \beta(x_k) \Delta x_k^2) \eta_k \Delta t 
\qquad
\nonumber\\[4pt]
&&
=
 O(\Delta t^2) +
f(x_k) \Delta t + f'(x_k) \, [\alpha \Delta x_k + \beta(x_k) \Delta x_k^2] \, \Delta t
\qquad
\nonumber\\
&& 
\quad
+
\left\{ g(x_k) + g'(x_k) \, [\alpha \Delta x_k + \beta(x_k) \Delta x_k^2]  + 
\frac{1}{2} g''(x_k) [\alpha \Delta x_k + \beta(x_k) \Delta x_k^2]^2 \right\}
\eta_k \Delta t 
\qquad
\nonumber\\
&& 
=
 O(\Delta t^2) +
f(x_k) \Delta t + \alpha f'(x_k)  g(x_k)\eta_k \Delta t^2  +g(x_k)\eta_k \Delta t 
\qquad
\nonumber\\[4pt]
&& 
\quad  +
\left\{ \alpha g'(x_k) \, [f(\bar x_k) \Delta t + g(\bar x_k) \eta_k \Delta t] + g'(x_k)  \beta(x_k) \Delta x_k^2  + \frac{\alpha^2}{2} g''(x_k)  \Delta x^2_k  \right\}
\eta_k \Delta t 
\nonumber\\[4pt]
&&
=
 O(\Delta t^2) +
 f(x_k) \Delta t + \alpha f'(x_k)  g(x_k)\eta_k \Delta t^2  +g(x_k)\eta_k \Delta t  + \alpha g'(x_k) f(x_k) \eta_k \Delta t^2 
\nonumber\\[4pt]
&& 
\quad
+
\alpha g'(x_k) g(x_k) \eta^2_k \Delta t^2 + 
\textcolor{black}{\alpha^2} g(x_k) g'(x_k) g'(x_k)\eta^3_k \Delta t^3 
\nonumber\\
&&
\quad
+
g'(x_k)  \beta(x_k) g(x_k)^2 \eta_k^3 \Delta t^3 + \frac{\alpha^2}{2} g''(x_k)   g(x_k)^2 \eta_k^3 \Delta t^3
\; . 
\end{eqnarray*}
Now, we notice that these terms can be grouped differently if one focuses on their
scaling with  $\Delta t$. Indeed, using the substitution rules
$\eta_k^2 \Delta t \mapsto 2D$ and $ \eta_k^3 \Delta t \mapsto 6D \eta_k$, 
\begin{eqnarray}
\Delta x_k 
\!\!\! & \! \doteq  \! & \!\!\!
O(\Delta t^2) +
 \left[ f(x_k) + 2D \alpha g'(x_k) g(x_k) \right] \Delta t +g(x_k)\eta_k \Delta t 
 \nonumber\\[4pt]
&&
 \!\!  \!\! \!\!
 +  \alpha 
 \left[  f'(x_k)  g(x_k)+  g'(x_k) f(x_k) \right]    \eta_k \Delta t^2
\label{eq:app1} \\
&&
\!\! \!\! \!\!
+
\left[
\textcolor{black}{\alpha^2} g(x_k) g'(x_k) g'(x_k)
+ g'(x_k)  \beta(x_k) g(x_k)^2  + \frac{\alpha^2}{2} g''(x_k)   g(x_k)^2 
\right] 
  6D \eta_k \Delta t^2 
  \, .
 \nonumber
\end{eqnarray}

On the other hand, 
the expansion of the generic expression~(\ref{eq:Tdis}) in the $\mathbb{T}_{\bff, \bg}$ discrete Langevin equation 
up to order $\Delta t^{3/2}$ yields (in one dimension)
\begin{eqnarray}
\Delta x_k 
\!\! & \! = \! & \!\!
O(\Delta t^2) + f(x_k) \Delta t  + g(x_k) \eta_k \Delta t  \nonumber\\
&& 
+ \frac{1}{2} 
\left[ 
f(x_k) g'(x_k) \eta_k \Delta t^2 +  g(x_k) f'(x_k) \eta_k \Delta t^2 
\textcolor{black}{+ g(x_k) g'(x_k) \eta^2_k \Delta t^2 }
\right]
\nonumber\\
&&
+ \frac{1}{6} \, g(x_k) \left[ (g'(x_k))^2 +  g(x_k)g''(x_k) \right] \eta^3_k \Delta t^3
\nonumber\\
\!\! & \! \doteq  \! & \!\!
O(\Delta t^2) + [f(x_k) \textcolor{black}{+ D g(x_k) g'(x_k)} ] \Delta t  + g(x_k) \eta_k \Delta t  \nonumber\\
&& 
+ \frac{1}{2} 
\left[ 
f(x_k) g'(x_k)  +  g(x_k) f'(x_k) \right] \eta_k \Delta t^2 
\nonumber\\
&&
+  g(x_k) \left[ (g'(x_k))^2 +  g(x_k)g''(x_k) \right] D \eta_k \Delta t^2
\label{eq:app2}
\end{eqnarray}
where, in the second equality,  we used the substitution rules and we regrouped terms
according to their scaling with $\Delta t$. This expression assumes $\alpha=1/2$.

The comparison of the terms proportional to $\eta_k \Delta t^2$ in
Eqs.~(\ref{eq:app1}) and~(\ref{eq:app2}) implies
\begin{eqnarray}
&&
6 \left[ 
\textcolor{black}{\frac{1}{4}} g (g')^2
+ g'  \beta g^2  + \frac{1}{8} g''   g^2 
\right]
=
g (g')^2  +  g^2 g''  
\end{eqnarray}
which yields that the parameter $\beta$ takes the value
\begin{eqnarray}
\betag = \frac{1}{g g'} 
\left[
-\frac{1}{12} (g')^2 + \frac{1}{24} g g''
\right]
. 
\end{eqnarray}
This is the same result as derived in Sec.~\ref{sec:explicit-covariant} from Eq.~(\ref{eq:infi_prop_general0}) (see Eq.~(\ref{eq:Mbeta1d})),
and as in Sec.~\ref{sec:covLangevin} from the general expressions~(\ref{eq:MTdis}) and~\eqref{eq:BBgprime} when the dimension is set to $d=1$ (see the remark after Eq.~\eqref{eq:BgprimeB}).

\section{List of notations \& results}
\label{sec:appListNotationsRes}

In this Section we list the main definitions, notations and results in equation format.

\subsection{Symbols}

We write the partial derivatives in a compact form, 
$\dfrac{\partial}{\partial t} = \partial_t$ and $\dfrac{\partial}{\partial x} = \partial_x$.

\vspace{0.2cm}

\noindent
The prime in $U'(x(t))$, etc. denote derivative with respect to the argument.

\vspace{0.2cm}

\noindent
The convolution of two functions $f(g( \dots))$ is also written $(f \circ g)(\dots)$ in the text.

\subsection{White noise Langevin equations}

\subsubsection{One-dimensional processes}

The one-dimensional stochastic equation in continuous-time notation reads
\begin{equation}
\begin{split}
& \frac{\dd x}{\dd t} 
\stackrel{\mathfrak{d}}{=} 
f(x) + g(x) \eta 
\\
& \langle \eta(t) \rangle =0 
\; , 
\qquad\qquad
\langle\eta(t)\eta(t')\rangle=\, \delta(t'-t)
\:,
\end{split}
\label{eq:Langevin-app}
\end{equation}
where the symbol $\mathfrak{d}$ denotes the discretization scheme.
The discrete-time stochastic equation is
\begin{equation}
\begin{split}
& \Delta x(t) = x(t + \Delta t) - x(t) = f(\bar{x})\Delta t + g(\bar{x})\Delta \eta(t) 
\\
& \langle \Delta\eta(t) \rangle =0 
\; , 
\qquad\qquad
  \langle \Delta\eta^2(t)  \rangle= \Delta t
\; .
\end{split}
\end{equation}
The quadratic $\alpha,\beta(x)$ discretization scheme is defined as
\begin{equation}
\bar{x} = x = \alpha \Delta x + \beta(x) \Delta x^2
\; . 
\end{equation}
The variable increments satisfy the following scalings with $\Delta t$:
\begin{equation}
\Delta \eta^2 \sim \Delta t \, , \qquad\quad \Delta x^2 \sim \Delta t  \, ,
\qquad\quad  \Delta x^{(\alpha)} - \Delta x^{(\alpha')} \sim \Delta t \, .
\end{equation}
A choice of $\alpha, \beta(x)$  which makes the difference between the discrete and the continuous Langevin processes equal  up to 
order $\Delta t^{3/2}$ (and the path-integral construction covariant, see Eq.~(\ref{eq:requirement-covariance})),~is 
\begin{equation}
\alpha = \frac{1}{2} \; , 
\qquad\qquad
\beta = \betag = - \frac{1}{12} \frac{g'}{g}  + \frac{1}{24} \frac{g''}{g'} 
\; . 
\label{eq:beta-1d}
\end{equation}
For other discretization schemes, the difference between the discrete and the continuous Langevin processes is of order $\Delta t^{1/2}$.
The Fokker--Planck equation for Eq.~(\ref{eq:Langevin-app}) in the $\alpha$-discretization reads
\begin{align}
&
\partial_t P(x,t)
 = 
- \partial_x\big[ (f(x) +\alpha g(x) g'(x)) P(x,t) \big] 
+ \frac{1}{2} \, \partial_x^2 \big[ g^2(x) P(x,t) \big] 
\end{align}
and, with the drift written as  $f(x) = - \frac{1}{2} \, g^2(x) \beta V'(x) + (1-\alpha)  g(x) g'(x) $, it becomes
\begin{align}
\partial_t P(x,t)
=
 \partial_x \left\{ g^2(x) \left[ \frac{1}{2} \, \beta V'(x)  P(x,t) + \frac 12 \, \partial_x P(x,t) \right]
\right\}
. 
\end{align}
The quadratic contribution to the discretization plays no role at the level of the 
Fokker--Planck equation. 
The equivalence between $\alpha$ and $\alpha'$ processes is established through
\begin{align} 
\frac{\dd x}{\dd t} 
\stackrel{\alpha}{=} f(x) + g(x) \eta  \stackrel{\alpha'}{=} f(x) + (\alpha - \alpha') g(x) g'(x)  + g(x) \eta \, .
\end{align}
The three lowest order substitution rules (valid in prefactor of the exponential of the propagator)  are
\begin{align}
\Delta x_k^2 \doteq  g^2(x_k) \Delta t \; , \qquad
\Delta x^3  \doteq 3  g^2(x_k) \Delta x  \Delta t \; , \qquad
\Delta x_k^4 \doteq  3 g^4(x_k) \Delta t^2 
\, . 
\end{align}
We stress that the cubic substitution rule has a very different meaning compared to the quadratic and quartic ones. While the latter hold in a $L^2$ sense at the level of integrated observables, see Eq.~\eqref{eq:proof_convL2}, the cubic one only represents a valid substitution rule within the infinitesimal propagator: an infinitesimal propagator with a cubic term (provided the cubic term is found outside the exponential) describes the same process in the continuous-time limit as an infinitesimal propagator where the cubic term is replaced according to the above equation.  In the exponential of the propagator, the cubic one becomes (see Sec~3.3.1 in~\cite{Cugliandolo-Lecomte17a} and Eq.~(\ref{eq:3rdordersubstanyd}) in the present paper):
\begin{align}
A^{(3)}\, \Delta x^3 
 \doteq 
3 A^{(3)} g^2(x_k) \Delta x  \Delta t 
+
3 \big (A^{(3)}\big)^2 g^6(x_k) \Delta t^2
\, . 
\end{align}
After the change of variables $u=U(x)$ and $x=X(u)$ the Langevin equation becomes
\begin{equation}
\begin{split}
& 
\frac{\dd u}{\dd t}
\stackrel{\alpha}{=}
  F(u) + \left(\frac{1}{2} - \alpha\right)g^2(X(u)) U''(X(u)) + G(u + \alpha \Delta u) \eta  
\\
& F = \left(f U'\right) \circ X \, , \qquad\qquad
G = \left(g U'\right) \circ X \, .
\end{split}
\end{equation}

\subsubsection{Higher dimensional processes}

Consider the generic  $d$-dimensional contravariant vector $\bx$ with components 
$x^\mu$, with $\mu=1,\dots, d$.
The continuous-time stochastic equation reads
\begin{equation}
\label{eq:Langevin-high-app}
\begin{split}
& \frac{\dd x^\mu}{\dd t} \stackrel{\mathfrak{d}}{=} f^\mu(\bx) + g^{\mu i}(\bx) \eta_i 
\\
& \left\langle \eta_i(t) \right\rangle = 0 \, , \qquad\qquad \left\langle \eta_i(t)\eta_j(t') \right\rangle = \delta_{ij}\delta(t-t') 
\end{split}
\end{equation}
and its discrete-time version is
\begin{equation}
\begin{split}
& \Delta x^\mu_k = x^{\mu}((k+1)\Delta t) - x^{\mu}(k\Delta t) = f^\mu(\bx_k) + g^{\mu i}\left(\bx_k + \alpha \Delta \bx_k\right)\Delta \eta_{i,k} 
\\
&
\left\langle \Delta \eta_{i,k} \right\rangle = 0 \, ,  \qquad\qquad \left\langle \Delta \eta_{i,k} \Delta \eta_{j,k'} \right\rangle = \Delta t \, \delta_{ij}\delta_{kk'} \, .
\end{split}
\end{equation}
A covariant Langevin equation to all orders in $\Delta t$ is
\begin{equation}
\Delta x^\alpha = \mathbb{T}_{\bff, \bg} \left( f^{\alpha} \Delta t + g^{\alpha i} \Delta \eta_i(t) \right) 
\end{equation}
with the operator $\mathbb{T}_{\bff, \bg}$ defined by its action on a generic function $h (\bx) $:
\begin{equation}
\mathbb{T}_{\bff, \bg}  \, h  = 
\frac{\exp{\left(f^\mu \Delta t + g^{\mu i}\Delta\eta_i(t)\right)\partial_{x^\mu}}-1}{\left(f^\mu \Delta t + g^{\mu i}\Delta\eta_i(t)\right)\partial_{x^\mu}} h
\; .
\end{equation}
The quadratic discretization is defined as 
\begin{align}
\bar{x}_k^\mu = x_k^\mu + \alpha \Delta x_k^\mu + {\B}_{\alpha\beta}^\mu(\mathbf{x}_k)  \Delta x_k^\alpha \Delta x_k^\beta \, .
\end{align}
The discrete-time Langevin equation is covariant under a non-linear change of variables  
$\bu(t) = \textbf{U}(\bx(t))$ up to order $\Delta t^{3/2}$ if ${\B}^{\mu}_{\alpha\beta}$ is chosen 
to satisfy
\begin{equation}
   \partial_\nu g^{\mu i} {\B}^\nu_{\alpha\beta}
  = \frac{1}{24}\left(\partial_\alpha\partial_\beta g^{\mu i} - 2 g_{\beta j}\partial_\alpha g^{\gamma j}\partial_\gamma g^{\mu i}\right) 
\end{equation}
which boils down to~(\ref{eq:beta-1d}) for $\betag$ in $d=1$.
The $d \times d$ matrix ${\boldsymbol \omega}$ has elements
\begin{align}
\omega^{\mu\nu}(\bx) =  g^{\mu i}(\bx)g^{\nu j}(\bx)\delta_{ij} \, .
\end{align}
The substitution rules (valid in prefactor of the propagator) are extended to
\begin{equation}
\begin{split}
& 
\Delta x^\mu \Delta x^\nu\doteq  \omega^{\mu\nu}(x) \, \Delta t \, ,  \qquad\qquad \Delta \eta_i \Delta \eta_j\doteq  \Delta t \, \delta_{ij} 
\\
& \Delta x^\mu \Delta x^\nu \Delta x^\rho \Delta x^\sigma\doteq  
\left[\omega^{\mu\nu}(x)\omega^{\rho\sigma}(x) + \omega^{\mu\rho}(x)\omega^{\nu\sigma}(x) + \omega^{\mu\sigma}(x)\omega^{\nu\rho}(x)\right]
\Delta t^2 
\\
&
\frac{\Delta x^\mu \Delta x^\nu \Delta x^\rho}{\Delta t} 
\doteq
 \omega^{\mu\nu}\Delta x^{\rho} +
 \omega^{\mu\rho}\Delta x^{\nu} +
 \omega^{\nu\rho}\Delta x^{\mu} 
\, . 
\end{split}
\end{equation}
In the exponential of the propagator, the substitution rule for the cubic term becomes dependent on its prefactor.
Assuming a fully symmetric prefactor $A^{(3)}_{\mu\nu\rho}$ one has:
\begin{equation}
A^{(3)}_{\mu\nu\rho}\frac{\Delta x^\mu \Delta x^\nu \Delta x^\rho}{\Delta t} \doteq 3 A^{(3)}_{\mu\nu\rho} \omega^{\mu\nu}\Delta x^{\rho} + 3 A^{(3)}_{\mu\nu\rho}A^{(3)}_{\alpha\beta\sigma} \omega^{\mu\alpha}\omega^{\nu\beta}\omega^{\rho\sigma}\Delta t 
\, . 
\end{equation}
The relation between $\alpha$ and $\alpha'$ discretized stochastic differential equations~is
\begin{align}
\frac{\dd x^\mu}{\dd t} & \stackrel{\alpha}{=} f^\mu + g^{\mu i} \eta_i 
 \stackrel{\alpha'}{=} f^\mu + \left(\alpha - \alpha'\right) g^{\nu i} \, \partial_\nu g^{\mu i} +  g^{\mu i} \eta_i \, .
\end{align}
Under a change of variables, $\bu(t) = \textbf{U}(\bx(t))$ with 
$\textbf{U}: \mathbb{R}^d \to \mathbb{R}^d $ an invertible transformation, one has, denoting $\mathbf{X}(\bu)=\mathbf{U}^{-1}(\bu)$ the inverse function of $\mathbf{U}(\bx)$
\begin{align} 
\frac{\dd u^\mu}{\dd t} & 
\stackrel{\alpha}{=} \left[f^\rho \, \partial_\rho U^\mu\right]\circ \textbf{X}(u) + \left(\frac{1}{2}-\alpha\right) \left[ \omega^{\rho \sigma} \, \partial_\rho \partial_\sigma U^\mu \right]\circ \textbf{X}(u) 
\nonumber\\
& 
\quad
+ \left[g^{\rho i} \, \partial_\rho U^\mu \right]\circ \textbf{X}(u) \, \eta_i \, .
\end{align}
The Fokker--Planck equation associated to 
Eq.~\eqref{eq:Langevin-high-app} in the $\alpha=1/2$ discretization
\begin{align}
\partial_t P = - \partial_\mu \left[ \left( f^\mu + \frac{1}{2} g^{\nu i} \, \partial_\nu g^{\mu i} \right) P \right] + \frac{1}{2}\partial_\mu\partial_\nu \left(\omega^{\mu\nu}P\right) .
\end{align}
The probability density
$
K(\bx, t) = P(\bx, t)/\sqrt{\omega(\bx)} 
$,
with $\omega$ the determinant of $\omega_{\mu\nu}$,
is governed by  the covariant evolution equation
\begin{equation}
\begin{split} 
\partial_t K = - \nabla_\mu \left( h^\mu K \right) + \frac{1}{2}\nabla_\mu \nabla_\nu \left( \omega^{\mu\nu} K \right) ,
\\
h^\mu = f^\mu - \frac{1}{2} \, \Gamma^\rho_{\nu \rho}\omega^{\nu\mu} - \frac{1}{2} \partial_\nu g^{\nu i} g^{\mu j} \, \delta_{ij}  \, .
\end{split}
\end{equation}
The notation for vectors, tensors and metric is the following
\begin{align}
& f^\mu \;, \; h^\mu \;  \& \; g^\mu \quad\;\; \mbox{contravariant vectors}
\nonumber\\
& \omega^{\mu\nu}  \qquad\quad \qquad\;\; \mbox{$\omega^{\mu\nu} =  g^{\mu i} g^{\nu j} \delta_{ij}$ rank-2 contravariant metric tensor} 
\nonumber\\
& 
\qquad\qquad\;\;\qquad\;\; 
\mbox{inverse $\omega_{\mu\nu}$ such that}  \;\;
\omega^{\mu\rho}\omega_{\rho\nu}=\delta^\mu_{\,\rho} \;\;
\nonumber\\
& 
\qquad\qquad\;\;\qquad\;\; 
\mbox{and of determinant} \;\; \omega=\det  (\omega_{\mu\nu})  \, 
\nonumber\\
& \dd^d \bx \sqrt{\omega(\bx)}  \qquad \;\; \mbox{invariant volume element}
\nonumber\\
& 
\nabla_\mu \qquad\quad\; \qquad\;\;  \mbox{covariant derivative:}
\nonumber\\
& \qquad\qquad \;\; \qquad\;\; \,
\nabla_\mu \phi = \partial_\mu \phi \, , 
\\
& \qquad\qquad  \;\; \qquad\;\; \,
\nabla_\mu A^\nu = \partial_\mu A^\nu + \Gamma^{\nu}_{\mu\rho}A^\rho \, , 
\nonumber\\
& \qquad\qquad  \;\; \qquad\;\; \,
\nabla_\mu T^{\rho\sigma} = 
\partial_\mu T^{\rho\sigma} + \Gamma^{\rho}_{\mu\nu}T^{\nu\sigma} + \Gamma^{\sigma}_{\mu\nu}T^{\rho\nu} 
\;  
\nonumber\\
& 
\Gamma^{\mu}_{\rho\sigma}  \qquad\;\;  \qquad\quad\; \mbox{Christoffel symbols}
\nonumber\\
& \qquad\qquad  \;\; \qquad\;\;  = \frac{1}{2}\omega^{\mu \nu}\left(\partial_\rho \omega_{\nu\sigma} + \partial_\sigma \omega_{\nu\rho} - \partial_\nu \omega_{\rho\sigma}\right) 
\, 
\nonumber\\
& 
R  \qquad\quad\;\; \qquad\;\;\;\; \mbox{Ricci curvature}
\nonumber\\
&  \qquad\qquad  \;\; \qquad\;\; 
=
 \omega^{\mu\nu}\left(\partial_\eta \Gamma^{\eta}_{\mu\nu} - \partial_\mu \Gamma^{\eta}_{\eta \nu} + \Gamma^{\eta}_{\mu\nu}\Gamma^{\rho}_{\eta \rho} - \Gamma^{\eta}_{\mu\rho}\Gamma^{\rho}_{\eta \nu}\right) 
\nonumber
\end{align}

\noindent
In one dimension, one has
\begin{align}
 & 
 g^{\mu i } \mapsto g, \quad 
 g_{\mu i } \mapsto 1/g, \quad  
 \omega^{\mu\nu}\mapsto g^2, \quad  
 \omega_{\mu\nu}\mapsto g^{-2}, \quad  
 \omega\mapsto g^{-2},
\nonumber\\
&
\Gamma^{\mu}_{\rho\sigma}\mapsto - g^2 (g^{-3} g') = - g'/g, \quad R=0, \quad h^\mu \mapsto f 
\nonumber
\end{align}
with the \textit{caveat} that, since $h^\mu$ is a vector, $\nabla_\mu h^\mu \mapsto f'-f g'/g$.

\subsection{Path integrals}

\subsubsection{Definition of covariance}

Two requirements are imposed
\begin{equation}
\begin{split}
&
{\mathcal D} \bx = {\mathcal D} \bu 
\; , 
\\
& 
\mathcal{L}^{\bx}_{\mathfrak{d}}[\textbf{x},\dot{\textbf{x}}] 
= 
\mathcal{L}^{\bu}_{\mathfrak{d}}[\textbf{u}, \dot{\textbf{u}}]
=
\mathcal{L}^{\bu}_{\mathfrak{d}}[\textbf{U}(\textbf{x}),(\partial \textbf{U}/\partial \textbf{x}) \cdot \dot{\textbf{x}}]
\; .
\end{split}
\label{eq:requirement-covariance}
\end{equation}

\subsubsection{One-dimensional processes}

The Chapman--Kolmogorov expression for the transition probability of a Markovian processes is
\begin{align}
\mathbb{P}(x_{\rm f}, \tf | x_0,t_0) = \lim_{N \rightarrow + \infty} \int \prod_{k=1}^{N-1} \dd x_k\prod_{k=0}^{N-1} \mathbb{P}_{\Delta t}(x_{k+1},t_{k+1} | x_{k},t_{k})
P(x_0,t_0)~.
\end{align}
with  $\tf = t_N$ and $x_{\rm f}=x_N$.
The conservation of probability implies 
\begin{equation}
{\dd} x \,  {P}_x(x,t) = {\dd u} \,  {P}_u(u,t) 
\quad 
\Rightarrow 
\quad 
 {P}_x(x,t) = \dfrac{\dd u}{\dd x} \,  {P}_u(u,t) 
\; . 
\end{equation}
The transition probability of an additive noise process written in terms of a linear non-covariant $\alpha$-discretized 
 path integral reads
\begin{align}
& \mathbb{P}(\xf,\tf | x_0,t_0) 
= \lim_{N\rightarrow +\infty} \frac{1}{\sqrt{2\pi \Delta t}}\int \prod^{N-1}_{k=1} \left(\frac{\dd x_k}{\sqrt{2\pi \Delta t}}\right) 
                                            \nonumber\\ 
& \hspace{3.5cm} 
\exp\left\{\!\! -\frac{\Delta t}{2} \sum_{k=0}^{N-1} 
\left[ 
\left(\frac{\Delta x_k}{\Delta t} - f(x_k + \alpha \Delta x_k)\right)^2  + 2 \alpha f'(x_k)\right]
\right\} 
\nonumber \\
& \hspace{3.5cm} 
\stackrel{\alpha}{=}\int_{x(t_0)=x_0}^{x(\tf)=\xf} \mathcal{D}x ~ \exp\left\{- \int_{t_0}^{t} \dd t \, 
\mathcal{L}_{\alpha}[x,\dot{x}] 
\right\}
\end{align}
with 
\begin{align}
\mathcal{L}_{\alpha}[x,\dot{x}]  = \frac{1}{2} \left[ \left(\dot{x}-f(x)\right)^2 + 2 \alpha f'(x) \right]
\; .
\end{align}

\noindent
Covariance is achieved with the $\betag$ discretization
\begin{equation}
\alpha = \frac{1}{2} \; , \qquad\qquad 
\betag = - \frac{1}{12} \frac{g'}{g}  + \frac{1}{24} \frac{g''}{g'} 
\; . 
\end{equation}
The covariant Onsager--Machlup action is 
\begin{equation}
{\cal S}[x,\dot x]  \stackrel{\betag}{=}  \int_{t_0}^{\tf} \dd t \, \left[ \frac{1}{4D g^2} \big(\dot{x}-f(x)\big)^2 + \frac{1}{2} f'(x) - \frac{1}{2} \frac{f g'}{g} \right] 
\end{equation}
and the Martin--Siggia--Rose--Janssen--De Dominicis one reads
\begin{equation}
 {\mathcal S}[\hat x,x]
  \stackrel{\betag}{=}
  \int_0^{\tf} \dd t \,
   \Big\{
  \hat x
  \big(
  \dot x-f (x)
  \big)
   -
  D g(x)^2 \hat x^2
  +
  \frac 12  f'(x)
  -
  \frac 12
   \frac{f(x)g'(x)}{g(x)}
\Big\}
\; .
\end{equation}

\subsubsection{Higher dimensions}

The Onsager--Machlup covariant continuous-time action reads
\begin{equation}
\label{eq:covStratoAppndix}
\mathcal{S}[\bx, \dot \bx] 
\stackrel{{\B}}{=}
\int_{t_0}^{\tf}
\dd t 
\
\left\{ 
 \frac{1}{2} \left[\omega_{\mu\nu}\left(\frac{\dd x^\mu}{\dd t}-h^\mu\right)\left(\frac{\dd x^\nu}{\dd t}-h^\nu\right) + \nabla_\mu h^\mu - \lambda R \right]
 \right\}
\end{equation}
and it should be interpreted in the discretization scheme
\begin{equation}
 \bar x^\alpha =   x^\alpha + \frac{1}{2}\Delta x^\alpha + {\B}^\alpha_{\rho\sigma}\Delta x^\rho \Delta x^\sigma 
 \, 
\end{equation}
where the tensor ${\B}$ verifies 
\begin{equation}
\label{eq:Mapp}
{\B}^\alpha_{\rho\sigma} \left(2 \omega^{\rho\sigma} \Gamma^\beta_{\alpha \beta} + 4 \omega^{\rho \mu} \Gamma^\sigma_{\alpha \mu}\right) 
=
 \frac{1}{4}\omega^{\mu\nu}\left(\partial_\nu \Gamma^{\rho}_{\rho\mu} + \Gamma^\rho_{\mu\sigma}\Gamma^{\sigma}_{\rho\nu}\right) + \Big(\lambda - \frac{1}{4}\Big)R
\,.
\end{equation}
The MSRJD version is 
\begin{eqnarray}
{\mathcal S}[\hat {\bf x}, {\bf x}] 
\stackrel{{\B}}{=}
\int_{t_0}^{\tf} \dd t
\; 
\left\{
-\frac{1}{2} \hat x_\mu \, {\omega}^{\mu\nu} \, \hat x_\nu + \left( \dfrac{\dd x^\mu}{\dd t} - h^\mu \right) \hat x_\mu + \nabla_\mu h^\mu - \lambda R
\right\}
. 
\end{eqnarray}

\subsubsection{Stochastic-quantum correspondence}

\begin{eqnarray}
\mbox{Discretization} &  & \mbox{Operator ordering}
\nonumber
\\
\mbox{Itō} & \leftrightarrow& \mbox{Normal}
\nonumber
\\
\mbox{Stratonovich} & \leftrightarrow& \mbox{Weyl}
\\
\mbox{Hänggi--Klimontovich} & \leftrightarrow& \mbox{Anti-normal}
\nonumber
\end{eqnarray}

\end{document}